\documentclass[11pt]{article}
\usepackage{natbib,amsmath,amsthm,amssymb,color,graphicx}
\usepackage{times}
\usepackage{epstopdf}

\usepackage[usenames,dvipsnames]{xcolor}
\newcommand{\rstar}{r^\star}

\newcommand{\TS}{\mbox{\bf TS}}
\newcommand{\tumS}{S}
\newcommand{\GLTAR}{\mbox{\bf GLT-AR}}
\newcommand{\GLTTS}{\mbox{\bf GLT-TS}}
\newcommand{\GLT}{\mbox{\bf GLT}}
\newcommand{\leadset}[2]{{\cal L}_{#1} (#2)}
\newcommand{\leadsetn}{n}
\newcommand{\AR}{\mbox{\bf AR}}

\newcommand{\NC}{\mbox{\bf NC}}
\renewcommand{\NC}{\mbox{\bf CR}}
\newcommand{\Wishart}[2]{ \mathcal{W} _{#1}\left(#2\right)}
\newcommand{\addb}{,b}
\newcommand{\ji}{{j_i}}

\newcommand{\GIG}[1]{\mathcal{GIG} \left(#1\right)}

\newcommand{\iid }{\mbox{\rm i.i.d.}}
\newcommand{\rankl}{z}
\newcommand{\SSR}{{\mbox{\rm SSR}}}
\newcommand{\oddpost}{O^{\mbox{\rm \tiny post}}}
\newcommand{\Bodd}{D}
\newcommand{\Count}[1]{\#\{#1\}}
\newcommand{\CountAR}{3-5-7-9-\ldots}

\newcommand{\nfacsp}{r_{\footnotesize sp}}
\newcommand{\psplit}{p_{\footnotesize split}}
\newcommand{\pmerge}{p_{\footnotesize merge}}
\newcommand{\newsp}{^{\mbox{\rm \tiny sp}}}
\newcommand{\Km}{{\mathbf K}}

\newcommand{\Gammainv}[1]{\mathcal{G}^{-1} \left(#1\right)}
\newcommand{\Gammad}[1]{ \mathcal{G}\left(#1\right)}

\newcommand{\Normal}[1]{ N\left(#1\right)}
\newcommand{\Normult}[2]{ N _{#1}\left(#2\right)}

\newcommand{\Betafun}[1]{B (#1)}
\newcommand{\Betadis}[1]{\mathcal{B}\left(#1\right)}

\newcommand{\Gamfun}[1]{\Gamma (#1)}
\newcommand{\Poi}[1]{\mathcal{P}(#1)}
\newcommand{\Uniform}[1]{\mathcal{U}\left[#1\right]}
\newcommand{\rvY}{Y} %notation for a random variable
\newcommand{\Real}{\mathbb{R}}

\newcommand{\Sm}{{\mathbf S}}
\newcommand{\Rm}{\mathbf{R}}

\newcommand{\const}{\mbox{\rm c}}
\newcommand{\hyp}{\phi} % true  factor loading matrix
\newcommand{\hypv}{\boldsymbol{\hyp}} % iprob matrix
\newcommand{\pshift}{p_{\mbox{\rm \footnotesize shift}}}
\newcommand{\ym}{{\mathbf y}}             % multivariate observation  y
\newcommand{\cm}{{\mathbf c}}
\newcommand{\om}{\Omega}                  % notation used for the covariance matrix of y_t
\newcommand{\Vary}{{\mathbf \om}}         % covariance matrix of y_t
\newcommand{\dimy}{m}                     % dimension of y_t
\newcommand{\load}{\beta}                 % notation used for factor loadings
\newcommand{\facload}{\boldsymbol{\load}} % factor loading matrix
\newcommand{\facloadsc}{\tilde{\boldsymbol{\load}}} % factor loading matrix
\newcommand{\nfac}{k}                     % number of factors
\newcommand{\nspu}{s}
\newcommand{\fac}{f}                      % notation used for the factor
\newcommand{\facm}{{\mathbf \fac}}        % the latent (multivariate) factor
\newcommand{\error}{\epsilon}             % notation used for the error
\newcommand{\errorm}{\boldsymbol{\error}} % multivariate  errors
\newcommand{\Vare}{{\mathbf \Sigma}}      % covariance matrix of the errors
\newcommand{\Sigmatilde}{\tilde{{\mathbf \Sigma}}}
\newcommand{\Varetrue}{\Vare_0}      % covariance matrix of the errors
\newcommand{\Cov}[1]{\mbox{\rm Cov}(#1)}
\newcommand{\idiov}{\sigma^2}             % noation used for the idiosyncratic variances
\newcommand{\Pm}{{\mathbf P}}             % transformation matrix P
\newcommand{\Qm}{{\mathbf Q}}             % transformation matrix Q
\newcommand{\Mm}{{\mathbf M}}             % transformation matrix M
\newcommand{\Dm}{{\mathbf D}}             % diagonal matrix D
\newcommand{\lm}{{\mathbf  l}}             % diagonal matrix D
\newcommand{\Bm}{{\mathbf B}}             % diagonal matrix D
\newcommand{\Cm}{{\mathbf C}}             % diagonal matrix D
\newcommand{\Gm}{{\mathbf G}}             % diagonal matrix D
\newcommand{\Tm}{{\mathbf T}}             % diagonal matrix D
\newcommand{\deltav}{\boldsymbol{\delta}} % indicator matrix
\newcommand{\deltavtilde}{\tilde{\boldsymbol{\delta}}} % indicator matrix
\newcommand{\deltavlam}{\boldsymbol{\delta}^\Lambda} % indicator matrix
\newcommand{\deltacol}[1]{\deltav_{\cdot,#1}} % indicator matrix
\newcommand{\bm}{{\mathbf b}}             % mean of the prior/posterior of the factor loading
\newcommand{\bV}{{\mathbf B}}             % meancovariance of the prior/posterior of the factor loading
\newcommand{\Xb}{{\mathbf X}}             % regressor matrix
\newcommand{\Psiv}{\boldsymbol{\Psi}}     % covariance of factors in PX-model
\newcommand{\Fm}{{\mathbf F}}             % notation used for  factor matrix (f_1 ... f_T)'
\newcommand{\Pim}{\boldsymbol{\Pi}}       % row selection matrix
\newcommand{\bP}{\Pm}
\newcommand{\jtwo}{l}

\newcommand{\traces}{\mbox{\rm tr}}

\newcommand{\trace}[1]{\traces (#1 )}

\newcommand{\ones}{{\mathbf{1}}}

\newcommand{\facloadtilde}{\tilde{\facloadtrue}} % true  factor loading matrix
\newcommand{\betatilde}{\tilde{\facload}} % true  factor loading matrix
\newcommand{\Varetilde}{\tilde{\Vare}_0}      % covariance matrix of the errors
\newcommand{\facloadstar}{\facload ^\star} % true  factor loading matrix

\renewcommand{\facloadtilde}{\facload} % true  factor loading matrix
\renewcommand{\Varetilde}{\Vare}
\newcommand{\Omegav}{{\mathbf P}}     % info matrix of the joint distribution
\newcommand{\cv}{{\mathbf m}}         % corresponding covector
\newcommand{\Lv}{{\mathbf L}}         % Cholesky factor of info matrix of the joint distribution
\newcommand{\Am}{{\mathbf A}}         % Cholesky factor of info matrix of the joint distribution
\newcommand{\Asub}[1]{{\mathbf A}^{(#1)}}         % Cholesky factor of info matrix of the joint distribution
\newcommand{\Bsub}[1]{{\mathbf B}^{(#1)}}         % Cholesky factor of info matrix of the joint distribution
\newcommand{\Bsubr}[1]{k_{#1}}
\newcommand{\xm}{{\mathbf x}}             % multivariate observation  y
\newcommand{\zm}{{\mathbf z}}             % multivariate observation  y

\newcommand{\nfacr}{r_+}                     % number of nonzero columns
\newcommand{\Diag}[1]{\mbox{\rm Diag}\!\left(#1\right)} % diagonal matrix,
\newcommand{\dimmat}[2]{#1\times #2}  % Dimension of a matrix
\newcommand{\bfz}{{\mathbf{0}}}         % vector of zeros
\newcommand{\bfzmat}{{\mathbf{O}}}      % matrix of zeros
\newcommand{\identm}{{\mathbf I}}       % symbol used for identity matrix
\newcommand{\identy}[1]{{\identm}_{#1}} % identity matrix with dimension
\newcommand{\unit}[1]{{\mathbf{1}}_{#1}} % identity matrix with dimension
\newcommand{\Probsym}{\mbox{\rm Pr}}    % symbol used for probability of an event
\newcommand{\Prob}[1]{\Probsym (#1)}    % Probability of event, use \Prob{event}
\newcommand{\Ew}[1]{\mbox{\rm E}(#1)}   % Expectation of a rv
\newcommand{\rank}[1]{\mbox{\rm rg}\,(#1)}  % rank of a matrix
\newcommand{\Scov}[1]{{\mathbf S}_{#1}} % sample covariance
\newcommand{\trans}[1]{#1 ^{'}}         % transposed sign in linear algebra
\newcommand{\Var}{\mbox{\rm V}}         % symbol used for variance
\newcommand{\V}[1]{\Var (#1 )}          % variance of variance-covariance matrix
\newcommand{\indic}[1]{\mathbb{I} (#1)}

\newcommand{\odd}{O}
\newcommand{\oddpr}{O^p}

\newcommand{\loadtrue}{\Lambda}                 % notation used for true factor loadings
\newcommand{\loadtruetilde}{\tilde{\Lambda}}                 % notation used for true factor loadings
\renewcommand{\loadtruetilde}{\beta}
\newcommand{\facloadtrue}{\boldsymbol{\loadtrue}} % true  factor loading matrix
\newcommand{\nfactrue}{r}                     % number of factors
\newcommand{\nfacind}{\tilde{r}}
\newcommand{\new}{^{\mbox{\rm \tiny new}}}
\newcommand{\old}{^{\mbox{\rm \tiny old}}}

\newcommand{\nullmod}{n}

\newenvironment{Figure}[4]%arguments: caption/label/name of eps-file without .eps extension/scaling factor
{\begin{figure}[t!]
\begin{center}
\scalebox{#4}{\includegraphics{#3.eps}}
\caption{#1}\label{#2}}
{\end{center}\end{figure}}

\newenvironment{Figure2}[5]%arguments: caption/label/2 times name of eps-file without .eps extension
{\begin{figure}[t!]
\begin{center}
\scalebox{#5}{\includegraphics{#3.eps}}
\hspace*{1cm}
\scalebox{#5}{\includegraphics{#4.eps}}
\caption{#1}\label{#2}
}{\end{center}\end{figure}}

{\begin{figure}[t!]
\begin{center}
\begin{tabular}{cc}
\scalebox{#5}{\includegraphics{#3.pdf}}
& % \hspace*{1cm}
\scalebox{#6}{\includegraphics{#4.pdf}}
\end{tabular}
\caption{#1}\label{#2}
}{\end{center}\end{figure}}

{\begin{figure}[t!]
\begin{center}
\begin{tabular}{ccc}
\scalebox{#6}{\includegraphics{#3.pdf}}
& % \hspace*{1cm}
\scalebox{#6}{\includegraphics{#4.pdf}}
&
\scalebox{#6}{\includegraphics{#5.pdf}}
\end{tabular}
\caption{#1}\label{#2}
}{\end{center}\end{figure}}

\newenvironment{Figure3}[6]%arguments: caption/label/2 times name of eps-file without .eps extension
{\begin{figure}[t!]
\begin{center}
\begin{tabular}{ccc}
\scalebox{#6}{\includegraphics{#3.eps}}
& % \hspace*{1cm}
\scalebox{#6}{\includegraphics{#4.eps}}
&
\scalebox{#6}{\includegraphics{#5.eps}}
\end{tabular}
\caption{#1}\label{#2}
}{\end{center}\end{figure}}

\newenvironment{Tabelle}[2]%arguments: caption/label
{\begin{table}[t!]
\begin{center}
\caption{#1}\label{#2}}
{\end{center}\end{table}}
\theoremstyle{plain}
\newtheorem{thm}{Theorem}%[subsubsection]
\newtheorem{cor}[thm]{Corollary}
\newtheorem{lem}[thm]{Lemma}

\theoremstyle{definition}
\newtheorem{alg}{Algorithm}

\begin{document}

\title{Sparse Bayesian Factor Analysis when the Number of Factors is Unknown\footnote{Several  research report versions of this paper were circulated that did not address variance identification. A  June 2009 Chicago Booth School of Business Research Report selected  the number of factors  in a sparse Bayesian factor models under the positive lower triangular constraints. In \citep{fru-lop:par}, we  introduced  sparse Bayesian factor models with the generalized lower triangular constraints. This final version of the paper extends the later work by fully addressing variance identification.  The first author would like to thank James J. Heckman for many inspiring discussions about this subject.  The paper in its various forms was presented on many occasions, such as  the
 2010 SBIES Meeting at UT Austin, %Seminar on Bayesian Inference in Econometrics and Statistics
the 25th Anniversary Celebration of the Department of Statistical Science at Duke University (2012),
the  2014 ESOBE Meeting in Paris, the 30$^{\rm rd}$ International Workshop on Statistical Modelling in  Linz (2015) and  the 2016 CFE  Meeting in Seville,
and  we acknowledge helpful comments from many people, in particular Remi Piatek und Sylvia Kaufmann.}}

\author{Sylvia Fr\"uhwirth-Schnatter\footnote{Department of Finance, Accounting, and Statistics, WU Vienna University of Economics and Business, Austria. Email: {\tt sfruehwi@wu.ac.at}} \ \ and \ \ Hedibert Freitas Lopes\footnote{Insper Institute of Education and Research, S\~ao Paulo, Brazil. Email: {\tt hedibertfl@insper.edu.br}}}

\maketitle

\begin{abstract}

%It is overwhelming to notice the increasing popularity of factor models, and its variants, over the last three decades across the whole spectrum of applied sciences, in particular the class of sparse factor models with unknown number of factors.  Nonetheless,
Despite the  popularity of sparse factor models,
little attention has been given to formally address  identifiability of these models beyond standard rotation-based identification such as the positive lower triangular constraint.
To fill this gap,  we provide a counting rule on the number of nonzero factor loadings
that is sufficient for achieving  uniqueness of the variance decomposition in the factor representation.
Furthermore, we introduce the   generalised lower triangular representation to resolve rotational invariance
and show that within this model class the unknown number of common factors can be recovered in an overfitting sparse factor model.
By combining point-mass mixture priors with a highly efficient and customised MCMC scheme,
% with overfitting sparse factor modelling  in a generalised lower triangular representation,
we obtain posterior summaries regarding  the number of common factors  as well as the factor loadings   via postprocessing.
Our methodology  is illustrated for  monthly exchange rates of 22 currencies with respect to the euro over a period of  eight years and  for monthly log returns of 73 firms from the NYSE100 over a period of 20 years.
\end{abstract}

\vspace{0.5cm}

\vspace{0.5cm}

\centerline{JEL classification: C11, C38, C63}

%\newpage
%
%\tableofcontents
%
%\newpage
%
%\listoffigures
%
%\newpage
%
%\listoftables

%\setcounter{page}{0}

\section{Introduction}

For many decades, factor analysis has been a popular method  to
model the covariance matrix $\Vary$ of correlated, multivariate observations $\ym_t$ of dimension $\dimy$,
see  e.g. \citet{and:int} for a comprehensive review.
Assuming  $\nfactrue$  uncorrelated factors,
a  factor model yields the representation  $\Vary=  \facloadtrue  \trans{\facloadtrue } + \Varetrue $,
with a   $\dimmat{\dimy}{\nfactrue}$ factor loading matrix   $\facloadtrue$  and  a diagonal matrix  $\Varetrue $.
The considerable reduction of the
  number of   parameters  compared to  an unconstrained  covariance matrix
  %the  $\dimy(\dimy+1)/2$ elements  in  $\Vary$,
is a main  motivation for the application of factor models  in economics and finance, especially, if $\dimy$ is large,
see e.g. \citet{fan-etal:hig_je} and \citet{for-etal:ope}.
% Since the number of factors $\nfactrue$ is typically much smaller than $\dimy$, the basic factor model  defined above yields by definition a parsimonious, low dimensional representation of $\Vary$.
 Beyond that, the goal of  factor analysis is often  to estimate  the  loading matrix $\facloadtrue$  to understand the driving
  forces behind the correlation between the features observed through $\ym_{t}$.

The recent years  have seen   considerable  research in the area of sparse Bayesian factor analysis which
achieves additional sparsity beyond the  natural parsimonity of  factor models in two different ways.
One strand of literature considers sparse factor models through continuous shrinkage priors on the factor loadings,  see e.g. \citet{bha-dun:spa}, \citet{roc-geo:fas} and \citet{kas:spa}, among others.
Alternatively, following the pioneering paper by   \citet{wes:bay_fac},  many authors considered  sparse factor models with point mass mixture
   priors on the factor loadings, including   basic factor models  \citep{car-etal:hig},  dedicated factor models with correlated
 (oblique) factors \citep{con-etal:bay}  and
   dynamic factor models \citep{kau-sch:bay}.

 Sparse Bayesian factor analysis with point mass mixture
   priors  assumes that (many) elements of  the factor loading matrix $\facloadtrue$ are 0, without being specific as to which elements are
 concerned. Inference with respect to  zero loadings is considered as a variable selection problem and
there are several reasons, why variable selection is of interest in sparse Bayesian factor analysis. First of all, sparse Bayesian factor analysis  allows to identify \lq\lq simple structures\rq\rq\ where in each row only a few nonzero loadings are present \citep{and-rub:sta}. %, i.e. $q_i=\sum_{j=1}^\nfactrue\delta_{ij}<\nfactrue$.
 Identifying simple structures has been a long standing issue in factor analysis,  in particular in psychology,  and was implemented recently through sparse Bayesian factor analysis in \citet{con-etal:bay}.
 A second motivation is identifying irrelevant variables $y_{it}$  in $\ym_t$ which are uncorrelated
with the remaining variables, meaning that for these variables the entire  row of
the factor loading matrix $\facloadtrue$ is zero.   The possibility to identify such variables  within the framework  of  sparse Bayesian factor analysis is
of high relevance  in economic analysis, given the recent practice to include as many variables as possible \citep{sto-wat:mac,boi-ng:are},
 and was implemented through sparse Bayesian factor analysis in  \citet{kau-sch:ide}.\footnote{Identifying irrelevant variables also of importance in areas such as bioinformatics, where typically only a few out of potentially ten thousands of genes may be related to a certain physiological outcome \citep{luc-etal:spa}.}
% \comment{Within the framework of sparse Bayesian factor analysis using point mass mixture priors, both issues are easily addressed.
  % by choosing a prior $p(\deltav)$  that implies a certain prior distribution for  the number of nonzeros rows as well as for the degree of simplicity $q_i$ in each nonzero row.
% A~posteriori,  irrelevant variables can be identified, by exploring the 0/1 pattern of the indicator matrix $\deltav$ with respect to  zero rows, and for all nonzero row one can learn which individual loadings are zero or not, hence how simple the factor loading matrix is, given the data.}\marginpar{Delete?}

The present paper contributes to the literature on sparse Bayesian factor models using  point mass mixture
   priors in several ways.   As a first major contribution, we explicitly address identifiability issues that arise in sparse
  Bayesian factor analysis.   In the econometrics  literature, identifiability  is often reduced  to solving
  rotational  indeterminacy, see e.g.~\citet{gew-sin:int}. However, for sparse Bayesian factor models identification goes
  beyond this problem and  concerns uniqueness of the variance decomposition in the covariance matrix $\Vary$.
  This problem which
  has been known for a long time \citep{and-rub:sta} went largely unnoticed  in the literature on sparse Bayesian factor analysis, both in bioinformatics as well as in  econometrics,   and was addressed only recently by
   \citet{con-etal:bay} in the context of   dedicated sparse factor models.   %, where equation   (\ref{fac1}) is combined with correlated (oblique) factors, $ \facm_t \sim  \Normult{\nfactrue}{\bfz,\Rm}$. and the factor loading matrix $\facloadtrue$  has a  perfect simple structure, i.e. each observations loads on at most one factor. Under the assumption that  $\facm_t$ and  $\errorm_t$ are independent,  this yields the representation
 %\begin{eqnarray*} % \label{fac4} $\Vary=      \facloadtrue  \Rm \trans{\facloadtrue } + \Varetrue$.
% \end{eqnarray*}Since the factors are correlated, this approach yields a sparse structure for the factor loading matrix $\facloadtrue$, but not for $\Vary$.
 Our paper provides a major achievement in this respect.
We reverse the two-step identification strategy of \cite{and-rub:sta} and  first force a structure on the loading matrix that solves  rotational invariance up to trivial rotations. To this aim, we introduce the class of  generalized lower triangular (GLT) factor models where the loading matrix is a generalized lower triangular matrix.
Given a GLT structure, we  introduce in a second step  %the \CountAR\
 a simple counting rule for the nonzero factor loadings as a sufficient  condition for verifying variance identification.
  % and develop an efficient algorithm for doing so.

  As a second contribution, we operate in a sparse overfitting Bayesian factor model  to yield inference with respect to  the number of unknown  factors.
   Selecting the number of factors has been known since long to be a very difficult issue. \citet{bai-ng:det2002} define  information criteria to choose the number of factors.  \citet{lee-son:bay} and  \citet{lop-wes:bay} were among the first   to address this issue in a careful Bayesian manner  using marginal likelihood.
   More recently,  \citet{con-etal:bay} use Bayesian variable selection in an overfitting model to determine the number of factors in a dedicated factor model.
   However, the recent econometric literature on Bayesian factor analysis,
including \citet{ass-etal:bay}, \citet{cha-etal:inv}, and \citet{kau-sch:bay},  does not provide any intrinsically Bayesian solution for determining the
number of factors.
 %, but rely  on asymptotic, non-Bayesian criteria such as \citet{bai-ng:det2002}.
  In the present paper, we discuss identification in an overfitting sparse factor model from a  formal viewpoint. We gain very useful insights into the structure of the
loading matrix in an overfitting model, if we confine ourselves to the class of  GLT factor models.
Using  a point-mass mixture  prior in an overfitting sparse factor model, we are able to identify the number of factors by postprocessing posterior draws and exploiting \lq\lq column sparsity\rq\rq, i.e.~by counting the number of   nonzero columns  among the variance identified factor loading matrices.

As a final contribution,  we design an efficient Markov chain Monte Carlo (MCMC)  procedure that delivers posterior draws from  an overfitting sparse factor model under point mass priors which is know to be particularly challenging, see e.g. \citet{pat-etal:pos}.
In addition, we carefully discuss prior specifications on all levels of the model, including a prior for the idiosyncratic variances that
   avoids the well-known Heywood problem and  a fractional prior for the unrestricted factor loadings.

   The rest of the paper  is organized as follows.  Section~\ref{secide} discusses identification issues for sparse factor models and introduces the class of
   GLT factor models.  Section~\ref{secbayes} discusses Bayesian inference  and selecting the number of factors for GLT factor models.  Section~\ref{secalpp} considers applications to exchange rate data and NYSE100 returns.  Section~\ref{secconcluse} concludes. Mathematical proofs and technical details  are summarized in a  comprehensive Web-Appendix.

\section{Identification issues in sparse Bayesian factor analysis} \label{secide}

  % \subsection{The sparse basic factor model}\label{basic}

   A basic factor model relates each observation $\ym_t=\trans{(y_{1t}, \ldots,  y_{\dimy t})}$ in a random sample $\ym=\{ \ym_t, t=1,\ldots,T\}$ of $T$ observations to a latent  $\nfactrue$-variate random variable $\facm_t=\trans{(\fac_{1t} \cdots \fac_{\nfactrue t})}$, the so-called common factors, through:
\begin{eqnarray}  \label{fac1}
 \ym_t =  \facloadtrue \facm_t + \errorm_t,
\end{eqnarray}
where $\facloadtrue$ is the unknown $\dimmat{\dimy}{\nfactrue}$ factor loading matrix with
factor loadings $\loadtrue_{ij}$.  $\nfactrue$ is called the number of factors.
Throughout the paper, the common factors are assumed to be orthogonal:
\begin{eqnarray}
 \facm_t  \sim  \Normult{\nfactrue}{\bfz,\identy{\nfactrue}} \label{fac2} .
\end{eqnarray}
A basic assumption in factor analysis is
that   $\facm_t$, $\facm_s$, $\errorm_t$, and  $\errorm_s$ are pairwise independent for all $t \neq s$.
 Furthermore, the following assumption is made concerning the idiosyncratic errors $\errorm_t$:
 \begin{eqnarray}
 \errorm_t \sim \Normult{\dimy}{\bfz,\Varetrue} ,  \qquad  \Varetrue=\Diag{\idiov_1,\ldots,\idiov_{\dimy}}.  \label{fac3}
\end{eqnarray}
 Assumption (\ref{fac3}) implies that conditional on   $\facm_t$   the $\dimy$ elements of $\ym_t$ are independent, hence all dependence among these variables is explained through the common factors.
 For the basic factor model, assumption (\ref{fac3}) together with (\ref{fac2}) implies that the observations $\ym_t$  arise  from a multivariate normal distribution,
$\ym_t  \sim \Normult{\dimy}{\bfz,\Vary}$, with zero mean and a covariance matrix  $\Vary$
with the following constrained  structure:
 \begin{eqnarray}
\Vary=     \facloadtrue  \trans{\facloadtrue } + \Varetrue .  \label{fac4}
\end{eqnarray}
For a  sparse Bayesian factor model,   a binary indicator $\delta_{ij}$ is introduced for each element $\loadtrue_{ij}$ of the factor loading matrix $ \facloadtrue$ which takes the value $\loadtrue_{ij}=0$, iff  $\delta_{ij}=0$, and $\loadtrue_{ij} \in \Real$ is unconstrained otherwise. This yields a binary indicator matrix $\deltav$ of 0s and 1s of the same dimension as  $\facloadtrue$.
   In sparse Bayesian factor analysis, the indicators $\delta_{ij}$  are  unknown and are inferred from the data, using point-mass mixture priors (also called spike-and-slab priors), see Subsection~\ref{priordelta} for more details.

 \subsection{Identification of sparse basic factor models}\label{onefactro}

 In the present paper, we explicitly address identifiability issues that arise in sparse
  Bayesian factor analysis with respect to uniqueness of the variance decomposition.
 Assume that  $\facloadtrue$ is of full column rank ($\rank{\facloadtrue}=\nfactrue$) and let $\nfactrue$ be the smallest number  compatible with representation (\ref{fac4}).  %For the remainder of this subsection, we will  assume that $\nfactrue$ is known.
Identification means that for any %pair $(\facloadtrue ,\Varetrue)$ and
$(\facloadtilde ,\Varetilde)$  satisfying (\ref{fac4}),   that is:
 \begin{eqnarray}
\Vary %=     \facloadtrue  \trans{\facloadtrue } + \Varetrue
= \facloadtilde \trans{\facloadtilde} +
\Varetilde ,   \label{facide1}
\end{eqnarray}
 where
$\Varetilde$ is a diagonal matrix and  $\facloadtilde$ a $\dimy \times \nfactrue$ loading matrix,
it follows that  $\facloadtilde=\facloadtrue$ and   $\Varetilde =\Varetrue$.

  Well-known identification problems arise for factor models, meaning that additional structure is
necessary to achieve identifiability.  A rigorous approach toward identification of factor models was first offered by \citet{and-rub:sta}.
They considered  identification as a two-step procedure, the first step being
 identification of the variance decomposition, i.e.~identification of $\Varetrue$  from (\ref{fac4}),
 which implies identification of   $\facloadtrue  \trans{\facloadtrue }$,
 and the second step being subsequent identification of $\facloadtrue$ from
 $\facloadtrue  \trans{\facloadtrue }$, also know as solving the rotational identification problem.

 The econometric literature typically   reduces identification of factor models to the second problem
 %, i.e.  uniqueness of the factor loading,
 and focuses on  rotational  identification,
taking variance identification for granted, see e.g. \citet{gew-zho:mea}.
However,  uniqueness of the factor loading matrix
of $\facloadtrue$ given  $\facloadtrue  \trans{\facloadtrue }$ does not imply identification. Variance identification is easily violated  in
particular for sparse factor analysis,  as following considerations  illustrate.
 Consider a sparse one-factor model for $\dimy \geq3 $ measurements, for which rotational invariance is not an issue, with two different
loading matrices. In the first case all but two factor loadings are 0  (e.g. $\lambda_1\neq 0$, $\lambda_2\neq 0$),
 %whereas $\lambda_i= 0$ for $i=3,\ldots, \dimy$)
 whereas in the second case
all but three factor loadings are 0 (e.g. $\lambda_i\neq 0$, $i=1,2,3$),  %, whereas $\lambda_i= 0$, for $i=4,\ldots, \dimy$),
implying, respectively,  the following covariance matrices $ \Vary$:
{\begin{eqnarray*} \small
 & \left(
\begin{array}{ccccc}
    {\bf  \lambda_1^2 + {\idiov_1} } & {\bf \lambda_1  \lambda_2}    & && \\
    {\bf \lambda_1   \lambda_2} & {\bf \lambda_2^2 + {\idiov_2} }   & &&\\
    &&    {\idiov_3} &  &   \\
   && & \ddots &    \\
   & &&&   {\idiov_\dimy} \\
  \end{array}\right), \,
   \left(
  \begin{array}{cccccc}
    {\bf  \lambda_1^2 + {\idiov_1} } & {\bf \lambda_1   \lambda_2} & {\bf \lambda_1  \lambda_3}  & && \\
    {\bf \lambda_1  \lambda_2} & {\bf \lambda_2^2 + {\idiov_2} } & {\bf \lambda_2  \lambda_3}  & &&\\
    {\bf \lambda_1   \lambda_3} & {\bf \lambda_2   \lambda_3} & {\bf \lambda_3^2 + {\idiov_3} }  & &&  \\
   & &&    {\idiov_4} &  &   \\
  & && & \ddots &    \\
  & & &&&   {\idiov_\dimy} \\
  \end{array}\right). &
\end{eqnarray*}}
%If $\dimy\geq 3$, then we  have $\dimy(\dimy+1)/2$\geq 2\dimy$ sample moments %$\Cov{y_{it}, y_{lt}}=\Omega_{il}$ to identify the $2\dimy$ unknwon parameters $(\lambda_i, \idiov_i)$, $i=1,\ldots,\dimy$ in $\Vary$.
As only the diagonal elements  % $\om_{ii}$
 of $\Vary$  depend on $\idiov_i$, the
 factor loadings can be identified only via the off-diagonal elements of
 $\Vary$. For the first model,  only $\Cov{y_{1t}, y_{2t}}=\Omega_{12}$ is nonzero, whereas all remaining covariances are equal to zero, hence, only the three sample moments $\V{y_{1t}}=\Omega_{11}$, $\V{y_{2t}}=\Omega_{22}$, and $\Cov{y_{1t}, y_{2t}}=\Omega_{12}$ are available to identify  the four parameters $\idiov_1$, $\idiov_2$, $\lambda_1$, and $\lambda_2$.
Therefore, a sparse factor model with only two nonzero factor loadings is not identified,
since infinitely many different parameters $\idiov_1$, $\idiov_2$, $\lambda_1$, and $\lambda_2$
 imply the same distribution for the observed data $\ym_t$.
For the second model the  three covariances $\Cov{y_{1t}, y_{2t}}=\Omega_{12}$, $\Cov{y_{1t}, y_{3t}}=\Omega_{13}$, and $\Cov{y_{2t}, y_{3t}}=\Omega_{23}$  are nonzero and in total six sample moments are available to identify the six parameters $(\lambda_i, \idiov_i)$, $i=1,2,3$.
From these considerations, it is evident that a one-factor model is identifiable only, if
at least 3 factor loadings are nonzero, which has been noted as early as \citet{and-rub:sta}.

% \subsubsection{Uniqueness of  variance decomposition for sparse basic factor models} \label{uniquevar}

For a   basic factor model with at least two factors,
 uniqueness of the variance decomposition, i.e.~the identification of the idiosyncratic variances $\idiov_1, \ldots, \idiov_\dimy$  in
 $\Varetrue$ from the variance decomposition (\ref{fac4}) of $\Vary$ has to be verified in addition to   solving rotational invariance. More precisely, given any pair $(\facloadtrue ,\Varetrue)$ and
$(\facloadtilde ,\Varetilde)$ satisfying (\ref{fac4}) and (\ref{facide1}), under which condition does this imply  that
 $\Varetilde =\Varetrue$ and $ \facloadtilde \trans{\facloadtilde} = \facloadtrue  \trans{\facloadtrue } $?
 In the present paper, we rely on the row deletion property
 of \citet{and-rub:sta} to ensure variance identification. \citet[Theorem~5.1]{and-rub:sta} prove that the following condition is sufficient  for
the identification of $\facloadtrue  \trans{\facloadtrue} $ and  $\Varetrue$ from the marginal covariance matrix $\Vary$  given in (\ref{fac4}):
\begin{itemize}
  \item[\AR .] Whenever an arbitrary row is deleted from $\facloadtrue$, two disjoint submatrices of rank $\nfactrue$ remain.
\end{itemize}
In standard factor analysis, where all rows of $\facloadtrue$ are nonzero  and the factor loadings $\loadtrue_{ij}$ are unconstrained  except for dedicated zeros that are introduced to resolve the rotation problem (see Subsection~\ref{uniqload}),  condition \AR\   is typically satisfied, if the following
upper bound for the number of factors $\nfactrue $ holds:
\begin{eqnarray}
   \nfactrue  \leq \frac{\dimy-1}{2}, \label{boundAR}
\end{eqnarray}
i.e.  $\dimy \geq 2 \nfactrue +1$.
 From  condition \AR\  it is  apparent that for a sparse factor model
   a minimum number of three nonzero elements has to be preserved in each column, despite variable selection,
   to guarantee uniqueness of the variance decomposition and identification of $\Varetrue$. Hence, too many zeros in a sparse factor loading matrix may lead to non-identifiability of
$\Varetrue$ and $  \facloadtrue \trans{\facloadtrue}$, and subsequently to a failure to identify $\facloadtrue$.
This issue is hardly ever addressed  in the literature on sparse Bayesian factor analysis.
    In Theorem~\ref{rule357}  in Subsection~\ref{varidesp}, we   introduce  a  counting rule
    (which will be called the \CountAR\  rule for obvious reasons)    that provides a sufficient condition to verify  the row deletion property \AR\ for sparse Bayesian factor models.\footnote{A less restrictive bound than (\ref{boundAR}) which is widely used
     in psychological research is the Lederman bound \citep{led:ran}. However,  for the time being we did not succeed in formulating a sufficient counting rule within this class of factor models.}

% \subsubsection{Uniqueness of the factor loadings - solving rotational invariance} \label{uniqload}

The identifiability of  $\Varetrue$ guarantees  that   $\facloadtrue  \trans{\facloadtrue }$ is
identified.  % Given T $\facloadtrue  \trans{\facloadtrue }$,
The second step of identification is then
 to ensure  uniqueness of the factor loadings, i.e.~unique identification of $\facloadtrue$ from $\facloadtrue  \trans{\facloadtrue }$.
  As is well-known, without imposing  constraints on $\facloadtrue$,  the model is invariant under  transformations of the form
  $\facloadtilde =  \facloadtrue \Pm$ and  $\facm_t^{\star} =   \trans{\Pm} \facm_t$, where $\Pm$ is an arbitrary
 $\dimmat{\nfactrue}{\nfactrue}$  orthogonal matrix  (i.e. $\Pm \trans{\Pm}= \identy{\nfactrue}$), since evidently,
  \begin{eqnarray}
   \facloadtilde  \trans{\facloadtilde}  =   \facloadtrue \Pm \trans{\Pm}  \trans{\facloadtrue } = \facloadtrue  \trans{\facloadtrue }. \label{rotation}
\end{eqnarray}
 A special case of rotational invariance is the following trivial rotational invariance,
    \begin{align}\label{eq:Ralpha}
   \facloadtilde   =      \facloadtrue  \bP _{\pm} \bP _{\rho} ,
\end{align}
where the permutation matrix $\bP_{\rho}$ corresponds to one of   the $\nfactrue$! permutations %  of the $\nfactrue$ columns of $\facloadtrue$
 and the reflection matrix
 $\bP_{\pm}=\Diag{\pm 1, \ldots, \pm 1}$  % corresponds
 to one of  the $2^\nfactrue$ ways to  switch the signs of the  $\nfactrue$  columns of $\facloadtrue$.
 Often, identification rules  are  employed
 that guarantee identification of $\facloadtrue $ only up to such column and sign switching, % as in (\ref{eq:Ralpha}),
see e.g. \citet{con-etal:bay}.  Any  structure $\facloadtrue$ obeying such an identification rule represents a whole equivalence class of matrices $\facloadtilde$  given by all possible $2^\nfactrue \nfactrue !$ trivial rotations of $\facloadtrue$ defined in (\ref{eq:Ralpha}).

The usual way of dealing with rotational invariance is to constrain $\facloadtrue$  in such a
way that the only possible rotation  in  (\ref{rotation}) is the identity $\Pm=\identy{\nfactrue}$.
For orthogonal factors as defined in (\ref{fac2}),  at least $\nfactrue(\nfactrue-1)/2$ restrictions on the elements of
 $\facloadtrue$  are needed to eliminate rotational  indeterminacy \citep{and-rub:sta}.
The common  constraint both in econometrics \citep{gew-zho:mea} and statistics \citep{wes:bay_fac,lop-wes:bay} is to consider positive lower triangular (PLT) matrices,
i.e.~to constrain the upper triangular part of $\facloadtrue$ to be zero and
     to assume that the main diagonal elements $\loadtrue_{11},\ldots, \loadtrue_{\nfactrue \nfactrue}$  of $\facloadtrue$ are strictly positive.
     % , i.e. $\loadtrue_{jj}>0$ for all $j=1,\ldots,\nfactrue$.
 Although the PLT  constraint is pretty popular,
 it is often  too restrictive in practice.
  It induces an order dependence among the responses,  making  the appropriate choice of the first
  $\nfactrue$ response variables an important modeling decision  \citep{car-etal:hig}.
Difficulties arise in particular, if one of the true  factor loadings  $\loadtrue_{jj}$ is equal
 or close to 0, see e.g. \citet{lop-wes:bay}.

 Alternative strategies have been suggested,  for instance by \citet{kau-sch:ide} who
  exploit the single value decomposition of
  $\facloadtrue  \trans{\facloadtrue }$ to  solve rotational invariance.
  %  \comment{ADD comments on \citet{ass-etal:bay}.}
  In Subsection~\ref{secGLT}, we introduce a new  identification rule based on generalized lower triangular (GLT) structures.
%
%%%
 It should be emphasised that constraints imposed on $\facloadtrue$  to solve rotational invariance do not necessarily guarantee  uniqueness of the variance decomposition.\footnote{Consider, for instance, a PLT  loading matrix where in some column $j$ only two factor loading are nonzero:
the diagonal element $\loadtrue_{jj}$ which is nonzero by definition and  a second factor loading  $\loadtrue_{n_j,j}$ in some row $n_j>j$.
Such a loading matrix obviously violates the necessary  condition for variance identification that each column contains at least three nonzero elements.}
This issue  is  hardly  ever addressed explicitly in  the econometric literature, an exception being \citet{con-etal:bay}.\footnote{\citet{con-etal:bay} investigate identification of a dedicated factor model,  where equation (\ref{fac1}) is
  combined with correlated (oblique) factors, $\facm_t \sim  \Normult{\nfactrue}{\bfz,\Rm}$,
and the factor loading matrix $\facloadtrue$  has a  perfect simple structure, i.e. each observation
loads on at most one factor. They prove
a condition  that implies uniqueness of the variance decomposition as well as uniqueness of the
factor loading matrix and, consequently, the 0/1 pattern of the indicator matrix $\deltav$, namely:
the correlation matrix $\Rm$ is of full rank ($\rank{\Rm}=\nfactrue$) and
 each column of $\facloadtrue$ contains at least three nonzero loadings.} Variance identification for sparse Bayesian factor models is discussed in detail in  Subsection~\ref{varidesp}.

 %
 %typically  implies uniqueness of the variance decomposition, since condition {\bf AR} is fulfilled except for a set of Lebesgue measure 0.

\subsection{Solving rotational invariance through GLT structures} \label{secGLT}

 In this paper, we relax the PLT  constraint
  by allowing $\facloadtrue$ to be a generalized lower  triangular  (GLT) matrix:
  \hspace*{2mm}    \begin{itemize}
     \item[\GLT .]   Let $\facloadtrue$ be a $\dimmat{\dimy }{\nfactrue}$ factor loading matrix and let (for each $j=1,\ldots,\nfactrue$) $l_j$  denote
          the row index of the top nonzero entry in the $j$th column of  $\facloadtrue$ (i.e.  $ \loadtrue_{ij}=0, \forall \, i<l_j$).
     $\facloadtrue$  is a {\em generalized lower triangular} matrix, if  $l_1 < \ldots <  l_\nfactrue$ and   $\loadtrue_{l_j,j} > 0$ for $j=1,\ldots,\nfactrue$.
   \end{itemize}
  %Note that  $\facloadtrue$  has  full column-rank,  i.e. $\nfactrue=\rank{\facloadtrue }$.

   For a GLT matrix $\facloadtrue$,  the leading indices
    $l_1 , \ldots,  l_\nfactrue$ satisfy  $l_j\geq j$ and  need not lie on the main diagonal. Obviously, the class  of GLT matrices  contains PLT matrices as that special case where $l_j= j$
   for $j=1,\ldots,\nfactrue$. %, but allows for more general forms  of triangular  matrices.
This generalization is particularly useful, if the ordering of the response variables is in conflict with the PLT assumption.
 Since  $\loadtrue_{jj}$ is allowed to be 0, response variables  different from the first
  $\nfactrue$  ones  may lead the  factors. Indeed, for each factor $j$, the  leading variable is the
   response variable $y_{l_j,t}$ corresponding to the  leading index $l_j$.
   % top nonzero element $\loadtrue_{l_j, j}$.
%
  An example of such a  GLT matrix is displayed in the left-hand side of Figure~\ref{figsparseglt}.
Evidently, all loadings \emph{above}  the leading element $\loadtrue_{l_j,j}$ are zero by definition. A \emph{sparse GLT matrix} results, if in addition some factor loadings \emph{below} the leading  element  $\loadtrue_{l_j,j}$ are zero as well.
  The  condition $\loadtrue_{l_j,j}>0$ prevents  sign switching and can be substituted by the condition
  $\loadtrue_{i_j,j}>0$ for any row $i_j \geq l_j$ with a nonzero factor loading in column $j$.
   Condition \GLT\  resolves rotational invariance, provided that the leading
    indices  $l_1 < \ldots <  l_\nfactrue$ are ordered: evidently,   for any two GLT matrices
     $\facloadtilde$ and $ \facloadtrue $  with identical leading indices % $l_1 < \ldots <  l_\nfactrue$
     the identity      $\facloadtilde =  \facloadtrue \Pm$ holds, iff $\Pm=\identy{\nfactrue}$.

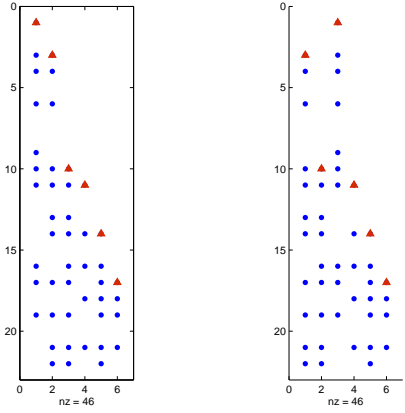
    \begin{Figure}{An example of  a sparse GLT matrix with  leading indices
$(l_1, \ldots, l_6)=(1,3,10,11,14,17)$ marked by triangles:  the ordered GLT structure (left-hand side) and one of the $2^6 \cdot 6$! corresponding unordered  GLT structures (right-hand side).}{figsparseglt}{glt}{0.4}
\end{Figure}

Any  GLT structure $\facloadtrue$ represents a whole equivalence class of
unordered GLT matrices $\facloadtilde$  given by all possible $2^\nfactrue \nfactrue !$ trivial rotations
of $\facloadtrue$ defined in (\ref{eq:Ralpha}).
Any unordered GLT structure $\facloadtilde$  has (unordered) leading
indices $l_1 , \ldots , l_{\nfactrue}$,
occupying different rows, see the right-hand side of Figure~\ref{figsparseglt}.
%, and   is  identified up to sign and column switching.
  The corresponding (ordered) GLT structure
    is recovered  % $l_1 < \ldots <  l_\nfactrue$
  from the order statistics   $l_{(1)} , \ldots ,  l_{(\nfactrue)}$ of $l_1, \ldots , l_ \nfactrue$ by a trivial rotation  and has leading indices
   $l_{(1)} <  \ldots  < l_{(\nfactrue)}$.  %, which obviously fulfills .
 %

%\subsubsection{Identifying the leading indices of a GLT structure}

In practice, the leading indices $l_1 , \ldots , l_\nfactrue$ of a GLT structure are unknown and need to be identified from the data for a given number of factors $\nfactrue$. This is achieved in sparse Bayesian factor analysis by introducing an indicator matrix $\deltav$ that obeys a GLT structure.
     Hence, we need to identify the entire 0/1 pattern in $\deltav$ from  $\Vary$, including the leading indices.
      Given variance identification, i.e. assuming that $\facloadtrue \trans{\facloadtrue}$ is identified,
 a particularly important issue for the identification of a sparse factor model
  is whether the 0/1 pattern in
 $\deltav$ is uniquely identified. In general,   $\deltav$ is not uniquely identified
 from $\facloadtrue \trans{\facloadtrue}$, because
    non-trivial   rotations $\Pm$ might exist that change  the zero pattern in
     $\facloadtilde =  \facloadtrue \Pm$.

     In the context of GLT structures, assume  that an  unordered  GLT  matrix $\facloadtilde$   exist with leading indices
      $\tilde{l}_1 , \ldots ,   \tilde{l}_\nfactrue$ being possibly
       different from the leading indices $l_1 , \ldots ,  l_\nfactrue$ of the
      loading matrix $\facloadtrue$ and both matrices  solve   $ \facloadtilde  \trans{\facloadtilde}  = \facloadtrue  \trans{\facloadtrue }$. % (\ref{rotation}).
     Then, Theorem~\ref{theGLT}  shows that the entire GLT structure $\facloadtrue$ including the leading indices and all zero loadings  is uniquely identified from $\facloadtrue  \trans{\facloadtrue }$, up to trivial rotations, i.e.
   $  \facloadtilde =  \facloadtrue \bP _{\rho} \bP _{\pm}$, meaning in particular that  the sets of leading indices  $\{\tilde{l}_1 , \ldots ,   \tilde{l}_\nfactrue\}$  and  $\{l_1 , \ldots ,  l_\nfactrue\} $ are identical.
   %The loading matrices $ \facloadtilde $ constitute exactly the equivalence class of all unordered GLT structures corresponding to $\facloadtrue$.
 %  Hence, also the  (unordered) leading indices $\tilde{l}_1 , \ldots , \tilde{l}_{\nfactrue}$ is identified. \\[1mm]
  % Hence,  the  leading indices $l_1 < \ldots  < l_{\nfactrue}$ of $\facloadtrue$ are  identified. \\[1mm]

\begin{thm}\label{theGLT}
    For a sparse GLT structure,
    $\deltav$ is uniquely identified, provided that uniqueness of  the variance decomposition holds, i.e.:
      if     $\facloadtrue$ and $\facloadtilde$ are sparse GLT matrices, respectively,  with
     leading indices  $l_1 < \ldots <  l_\nfactrue$ and  $\tilde{l}_1 < \ldots <  \tilde{l}_\nfactrue$
      that  satisfy
      $\facloadtilde  \trans{\facloadtilde}  =
      \facloadtrue  \trans{\facloadtrue }$, then  $\facloadtilde = \facloadtrue$. Hence,
      the leading indices as well as the entire 0/1 pattern
      of $\facloadtilde$ and $\facloadtrue$ are identical.
      %\begin{itemize}
      %  \item the leading indices $l_1 < \ldots <  l_\nfactrue$ as well as the entire 0/1 pattern
      %of $\facloadtilde$ and $\facloadtrue$ are identical;
      %  \item furthermore, $\facloadtilde = \facloadtrue \bP _{\pm}$,
      %  where  $\bP_{\pm}=\Diag{\pm 1, \ldots, \pm 1}$ switches the sign of  column $j$, if
      %  $(\bP_{\pm})_{jj}=-1$.
      %\end{itemize}}
\end{thm}

\noindent See Appendix~\ref{app:proof} for a proof. %\marginpar{Include rotations into the Theorem?}
While the assumption of a GLT structure resolves the rotational invariance, it does not
guarantee  uniqueness  of the variance decomposition.\footnote{Consider, for instance,   a GLT matrix with
the leading index in column  $\nfactrue$ being equal to $l_\nfactrue= \dimy -1$.  The loading matrix has at most  two nonzero elements in column $\nfactrue$ and violates the necessary  condition for variance identification that each column contains at least nonzero three elements.}
%  Given  $\nfactrue$ and $\dimy$, %% and condition such as \NC{1} and  \AR\ have to be verified.
  In particular,  an upper bound on the leading indices is necessary for    % \NC\ and
 \AR\ to hold.

 \begin{enumerate}
  \item[\GLTAR .] Let $\facloadtilde$ be an unordered GLT  structure with %  (unordered)
  leading indices $l_1, \ldots , l_{\nfactrue}$. %$\tilde{l}_1 , \ldots , \tilde{l}_{\nfactrue}$.
 The following condition % on the position of the leading indices  $l_1, \ldots , l_{\nfactrue}$
  is necessary  for  condition  \AR :
      \begin{eqnarray} \label{condlj}
 %        && \comment{OLD}  l_j \leq \dimy  -  2(\nfactrue-\rankl_j +1), \qquad j=1,\ldots,\nfactrue,\\
       \dimy-   l_j \geq   2(\nfactrue-\rankl_j +1), \qquad j=1,\ldots,\nfactrue,
 \end{eqnarray}
 where $\rankl_j$ is the rank  of $l_j$ in the ordered sequence $ l_{(1)} <  \ldots < l_{(\nfactrue)}$.
 % $\lm= (l_{(1)}, \ldots , l_{(\nfactrue)})$.
 %, which obviously fulfills $l_1< \ldots < l_ \nfactrue$.
  For an ordered GLT structure, (\ref{condlj})  reduces to $\dimy - l_j\geq  2(\nfactrue - j+1)$.
 %, i.e.  $ l_1 \leq \dimy - 2\nfactrue, \ldots, l_\nfactrue \leq \dimy - 2$.
\end{enumerate}
 For sparse GLT structures $\facloadtilde$ with zeros below the leading elements,  \GLTAR\ is only a necessary, but not  a sufficient condition for  {\AR }\footnote{A GLT structure obeying (\ref{condlj})  with $l_\nfactrue = \dimy - 2$ and  $\delta_{m \nfactrue}=0$, for instance,  contains only
 two  nonzero loadings  in column $\nfactrue$ and violates the necessary  condition for variance identification that each column contains at least nonzero three elements.}    % For a sparse GLT structure, where elements below $l_j$
and  variance identification has to be  verified  explicitly.  An efficient procedure for  dealing
with this   challenge  is introduced in the following  subsection.
%  Subsection~\ref{varidesp}.
 % and many measurements   load only on a few factors.

\subsection{Verifying  the row deletion property  for sparse factor loading matrices} \label{varidesp}

% \subsubsection{The \CountAR\ rule for verifying the row deletion property}

For  sparse Bayesian factor analysis, conditions for verifying directly  from the zero pattern in the factor loading matrix, whether
the row deletion property \AR\ holds, would be very useful, but so far only  necessary conditions have been provided.
\citet{and-rub:sta}, for instance,  prove  the following necessary conditions for  \AR : for every nonsingular   $\nfactrue$-dimensional square matrix  $\Gm$,
the matrix $\facload=\facloadtrue \Gm$ contains  in each column \emph{at least 3}  % nonzero factor loadings
and in each pair of columns   \emph{at least 5}   nonzero factor loadings.
\citet[Theorem~3.3]{sat:stu} extends  these necessary conditions in the following way:  every subset of $1\leq q \leq  \nfactrue$  columns
of $\facloadtrue$ contains \emph{at least $2q+1$}  nonzero factor loadings.

Extending the results of  \citet{sat:stu}, we  prove in the following Theorem~\ref{rule357}  that  for unordered GLT factor matrices
it is {\em sufficient} (and not only necessary)  for \AR\ that  such a  counting rule  holds  for the indicator matrix $\deltav$ for
a  {\em single  trivial rotation}  $\Gm= \bP _{\pm} \bP _{\rho}$
of the factor loading matrix  $ \facloadtrue $ % $\facloadtilde =  \facloadtrue \Gm$
 (and not for every nonsingular matrix $\Gm$).

    \begin{thm}[{\bf The \CountAR\ counting rule}]\label{rule357}
    % For  a GLT factor matrix  $\facloadtrue$,  a   is
  Consider   the following counting rule  for an unordered GLT structure $\facloadtilde =  \facloadtrue \bP _{\pm} \bP _{\rho}$ corresponding to
   an ordered GLT structure $\facloadtrue$:
   \begin{itemize}
  \item[\NC\ ] For  each $q =1,\ldots,\nfactrue$ and for each submatrix consisting of $q$ column of $\facloadtilde$, the  number  of nonzero rows in this sub-matrix is at least equal to $2q+1$.
  \end{itemize}
  Condition  \NC\ is both necessary and sufficient for the  row deletion property \AR\ to hold for  $\facloadtrue$.
 \end{thm}

    \noindent See Appendix~\ref{app:proof} for a proof.    Theorem~\ref{rule357} operates on the indicator matrix $\deltav$
    which is very convenient for verifying variance identification in sparse Bayesian factor analysis.
  Most importantly,   condition \NC\ extends the 3-5 counting rule of \citet{and-rub:sta}  to a more general \CountAR$\cdots$ rule for  the indicator matrix $\deltav$ corresponding to the factor loading matrix. % $\facloadtilde$.
    Obviously, if   \NC\ is violated for a single subset of   $q$ % \in \{1, \ldots, \nfactrue\} $
    columns of $\deltav$, then  \AR\ is violated  for $\facloadtrue$.
For $q=1,2$ as well as for $q=\nfactrue -1,   \nfactrue$   the corresponding  counting rules  can be easily  verified from simple functionals of  the
indicator   matrix
$\deltav$, see  Corollary~\ref{Lemma1} in  Appendix~\ref{simcount}.
Hence, for factor models with up to 4 factors ($\nfactrue \leq 4$) it is trivial  to  verify, if the \CountAR\ counting rule  and hence variance identification
 holds.

For   models with more than four factors ($\nfactrue > 4$), these simple counting rules are necessary conditions that quickly help  to   identify
indicator matrices  $\deltav$ where  \NC\ (and hence \AR ) is violated.
  If the simple counting rules of Corollary~\ref{Lemma1} hold,  then  \NC\   could be verified  by
 iterating over   all %$\footnotesize{\Bincoefsmall{\nfactrue}{q}}$
 subsets  of  $q=3, \ldots, \nfactrue-2$ columns of $\deltav$; a  number rapidly  increasing  with  $\nfactrue$.
The following Theorem~\ref{Lemma2}   shows that  verifying   \AR\ greatly simplifies, if  the
loading matrix has a block diagonal representation. In this case, \NC\ has to be checked only up to  the maximum
block size, rather than for the entire loading matrix.

\begin{thm}\label{Lemma2}
   Let  $ \facloadsc$ be a $\dimmat{m_n}{\nfacr}$ factor loading matrix   of full column rank,  $\rank{\facloadsc}=\nfacr$
   with $m_n$ nonzero rows. Assume that
 $ \facloadsc$  has following  block diagonal representation after  suitable permutations of rows and columns,
    with $\Pim_r$ and  $\Pim_c$  being the corresponding permutation matrices:
 \begin{eqnarray} \label{blockl}
\Pim_r  \facloadsc \Pim_c =
\left(
  \begin{array}{llll}
   \Asub{1} & \bfzmat &\bfzmat &\bfzmat \\
\times  & \ddots & \bfzmat &\bfzmat \\
 \times &  \times &  \Asub{Q-1} & \bfzmat \\
 \times  & \times  & \times  &  \Asub{Q} \\
  \end{array}
\right),
 \end{eqnarray}
where %,
 $\Asub{q}$,  $q=1,\ldots ,Q$,  are $(\dimmat{m_q}{r_q})$-dimensional  matrices  % and   $\Csub{q}$  is an   $\dimmat{(m-m_q)}{r_q}$ matrix
   such that    $\sum r_q = \nfacr$ and
 $\sum m_q = m_n $.  Assume that $\Asub{1}, \ldots,  \Asub{Q-1}$ are of  full column rank   $r_q=\rank{\Asub{q}}$. Then the following holds:
    \begin{itemize}
       \item[(a)]
       If all  sub matrices $\Asub{1}, \ldots,  \Asub{Q}$ satisfy the row deletion property \AR\ with $r=r_q$, then  the entire loading matrix $\facloadsc$
       satisfies the row deletion property \AR\ with $r=\nfacr$.
          \item[(b)]    If the  submatrix $\Asub{Q}$ violates the row deletion property \AR\ with $r=r_Q$, then the row deletion property \AR\ is violated
           for the entire loading matrix $\facloadsc$.

         %   \item[(c)]   If, for any $q=1,\ldots ,Q-1$,   the  submatrix $\Csub{q}$ is $\bfzmat_{\dimmat{(m-m_q)}{r_q}}$ is a zero matrix and   submatrix $\Asub{q}$   violates the row deletion property \AR\ with $r=r_q$, then the row deletion property \AR\ is violated  for the entire loading matrix $\facloadsc$.\footcomment{Part (c) is MOVED to IDE PAPER.}

         \end{itemize}
             \end{thm}

           \noindent  See Appendix~\ref{app:proof} for a proof. % of  Theorem~\ref{Lemma2}.
           Part~(a) of  Theorem~\ref{Lemma2}
           is useful to verify  that \AR\ holds for sparse loading matrices that have a block diagonal representation as in (\ref{blockl}).
           Part~(b)  of Theorem~\ref{Lemma2} is  useful to quickly identify indicator matrices $\deltav$ where  \AR\ does not hold.
In Appendix~\ref{verpartbig},
           Algorithm~\ref{algARIDE}  is discussed that derives representation (\ref{blockl}) sequentially  and is  useful for verifying variance identification in practice. %,   see Subsection~\ref{subGL} for more details. % e.g. for postprocessing MCMC estimation.

\subsection{Identification of irrelevant variables} \label{uniqload}

Irrelevant variables are observation $y_{it}$ for which the entire row $i$ of
the factor loading matrix $\facloadtrue$ is zero. This implies that  $y_{it}$ is uncorrelated
with the remaining variables. As argued by \citet{boi-ng:are}, it is useful to identify such variables.
Within the framework of sparse Bayesian factor analysis, such irrelevant variables can be identified
by exploring the 0/1 pattern of the indicator matrix $\deltav$ with respect to  zero rows, see \citet{kau-sch:ide}.
In Lemma~\ref{theirr} formal  identification of irrelevant variables from $\deltav$ is proven,
provided that the number of factors  $\nfactrue$ satisfies a more general upper bound than (\ref{boundAR}).
This commonly used upper bound is based on the assumption that all rows of $\facloadtrue$ are nonzero and
 a different upper bound is needed,   if we want to learn the position of the
 zero rows from a sparse factor analysis applied to   all $m$ variables. The corresponding bound
 %(\ref{boundZERO}) in Theorem~2
is derived from the fact that we need at least $2\nfactrue+1$ nonzero rows for the row deletion
property \AR\ to hold.
\\[1mm]

 \begin{lem}\label{theirr} Assume that a $\dimy \times \nfactrue $  factor loading matrix $\facloadtrue$ contains
 $m_0$ zero rows and that the number of factors  $\nfactrue$ satisfies following upper bound:
 \begin{eqnarray}
   \nfactrue  \leq \frac{\dimy-m_0-1}{2}. \label{boundZERO}
\end{eqnarray}
 If  uniqueness of the variance decomposition holds, then the position of the zero rows
 in $\facloadtrue$ is uniquely identified, that is,  any other $\nfactrue$-factor loading matrix $\facloadtilde$ satisfying $  \facloadtilde \trans{\facloadtilde} =
 \facloadtrue  \trans{\facloadtrue }$ has exactly the same set of zero rows.
\end{lem}
\noindent  See Appendix~\ref{app:proof} for a proof.

\subsection{Identification in overfitting factor models} \label{secover}

Assume that the data $\ym=\{\ym_1, \ldots,\ym_T\}$ are generated by the basic factor model (\ref{fac1}) with
 the corresponding variance decomposition in (\ref{fac4}) being unique, however,
  the true number of factors $\nfactrue$ is not known. In this case, a common procedure is to
   perform exploratory factor analysis based on a model with increasing number of factors $\nfac$,
\begin{eqnarray}  \label{fac1reg}   % \label{fac1reg}
 \ym_t =  \facload  \facm_t + \errorm_t,  \qquad  \errorm_t \sim \Normult{\dimy}{\bfz,\Vare} ,
\end{eqnarray}
where   $\facload$ is a  $\dimmat{\dimy}{\nfac}$   loading matrix
 with elements $\load_{ij}$ and $\Vare$ is a diagonal matrix with strictly positive diagonal elements.  As before, we allow the elements
 $\load_{ij}$ of $\facload$ in this potentially overfitting  sparse factor model  to be zero, with the corresponding indicator matrix being denoted by  $\deltav$.
 % \subsubsection{Extended variance identification} \label{secover_extend}
%\comment{MENTION THAT THIS IS REVIEW OF LITERATURE.}
 Factor analysis based on model (\ref{fac1reg}) yields the extended  variance decomposition
 \begin{eqnarray}
\Vary=     \facload   \trans{\facload  } + \Vare,   \label{fac4beta}
\end{eqnarray}
instead of the true variance decomposition (\ref{fac4}).
 If  model (\ref{fac1reg}) is not overfitting, that is $\nfac = \nfactrue$, then variance identification implies that  $\Vare=\Varetrue$ and  $\facload =\facloadtrue \Pm$ for some orthogonal  matrix  $\Pm$.

However,  if $\nfac  > \nfactrue$, then model  (\ref{fac1reg}) is, indeed, overfitting and
additional identifiability issues have to be addressed   for such overfitting factor models.
In particular, identifiability of  $\facload   \trans{\facload  } $ and $\Vare$ from (\ref{fac4beta}) is lost, as infinitely many representations $(\facload,\Vare)$
with $\Vare \neq \Varetrue$ exist that imply the same covariance matrix $\Vary$ as $(\facloadtrue,\Varetrue)$.
This identifiability problem has been noted earlier by \citet{gew-sin:int} and \citet{tum-sat:ide}. %, among others.
 Consider, e.g., a   model that  is  overfitting  with $\nfac  = \nfactrue+1$. Then  infinitely many representations $(\facload,\Vare)$ can be
constructed  that imply the same covariance $\Vary$ as $(\facloadtrue,\Varetrue)$, namely:
\begin{eqnarray} \label{adsp}
&&  \Vare =  \Diag{\idiov_1,\ldots, \idiov_{l_\nfac} - {  \loadtrue_{l_\nfac, \nfac}^2}, \ldots, \idiov_\dimy} ,
\quad \facload=  \left(\begin{array}{cc}
               {\large \bf{ \facloadtrue}}  & \left|\begin{array}{c}
                                           \bfz \\
                                           {  \loadtrue_{l_\nfac,  \nfac}} \\
                                           \bfz
                                         \end{array} \right.
               \end{array} \right) ,
\end{eqnarray}
where    $\loadtrue_{l_\nfac, \nfac}$
is an arbitrary factor loading satisfying  $0< \loadtrue_{l_\nfac, \nfac}^2< \idiov_{l_\nfac}$
and  $l_\nfac$ is an  arbitrary row index  different from the leading indices  $l_1, \ldots, l_\nfactrue $  in $ \facloadtrue$.
The last column of $\facload$ corresponds to a so-called \emph{spurious factor} which loads only on a single
observation.  Hence, factor analysis in an overfitting model with $\nfac=\nfactrue +1 $
may yield  factor loading matrices $\facload$  of   rank  $\nfactrue +1$, containing a spurious factor, rather than loading matrices of rank $\nfactrue$ with a zero column.
  For arbitrary  $\nfac  > \nfactrue$,  \citet{tum-sat:ide}  provide  a
 general  representation of the  factor loading matrix  in an overfitting factor model.
 Suppose that $\Vary$ has a decomposition  as in  (\ref{fac4})   with $\nfactrue$ factors   and for some  $\tumS \in \mathbb{N}$  with
 $\dimy \geq 2\nfactrue + \tumS + 1 $, or equivalently,
 \begin{eqnarray}  \label{kbound_extend}
   \nfactrue  \leq \frac{\dimy-\tumS -1}{2},
\end{eqnarray}
   the following extended row deletion property holds:
  \begin{itemize}
  \item[\TS ] Whenever $1+\tumS$  rows are deleted  from $ \facloadtrue$, then two disjoint submatrices of rank $\nfactrue$ remain.
\end{itemize}
  If $ \Vary $ has another decomposition  such that  $\Vary=     \betatilde   \trans{\betatilde} + \Vare$ where    $\betatilde$ is  a $\dimmat{\dimy}{(\nfactrue+s)}$-matrix of rank $\nfactrue+s$ with $ s \leq S$,  then  \citet[Theorem~1]{tum-sat:ide} show that there exists  an orthogonal matrix $\Tm$ of rank $\nfactrue+s$  such that
 \begin{eqnarray}  \label{decover}
\betatilde  \Tm =  \left(\begin{array}{cc}
                \bf{ \facloadtrue} & \Mm
               \end{array} \right) ,    \qquad   \Vare =  \bf{ \Varetrue} -  \Mm  \trans{\Mm },
\end{eqnarray}
where  the off-diagonal elements of  $\Mm \trans{\Mm}$ are zero.
Hence,  $\Mm $ is a  so-called \emph{spurious factor loading matrix} that does not contribute to explaining the correlation in $\ym_t$, since
 \begin{eqnarray*}
  \betatilde  \trans{\betatilde } + \Vare =  \betatilde \Tm \Tm' \trans{\betatilde } + \Vare =
\facloadtrue  \trans{\facloadtrue } + \Mm  \trans{\Mm } + (\Varetrue  -  \Mm  \trans{\Mm } ) =
\facloadtrue  \trans{\facloadtrue } + \Varetrue  =\Vary  .   \label{fac4A}
\end{eqnarray*}
While (\ref{decover}) is an important result,  without imposing further structure on the factor loading matrix  it is of limited use in applied factor analysis,
 as  the separation of $\betatilde$  into the true factor loading matrix $\facloadtrue$ and  the spurious factor loading matrix  $\Mm $  is possible only up to a general rotation  $\Tm$ of  $\betatilde$.

The following Theorem~\ref{theoverGLT} shows that extended identification in overfitting sparse factor models can be achieved
within the class of unordered GLT structures as introduced in this paper.
If  $\betatilde$  in model  (\ref{fac1reg})  is constrained to be an unordered GLT structure,
then  $\facloadtrue$  can be easily recovered  from (\ref{decover}).   First,  all rotations in (\ref{decover})  are equal  to trivial rotations $\Tm = \bP _{\pm} \bP _{\rho}$, only. Hence,  the columns of the spurious loading matrix  $\Mm $  appear  in between the columns  of $\facloadtrue$. Second,  the spurious loading matrix $\Mm $ is easily identified as an \textit{unordered spurious GLT matrix}, where in each column the leading element is the only nonzero loading. This  powerful result  is exploited subsequently  in our MCMC procedure to navigate through overfitting models
with varying the number of factors, by adding  and deleting spurious factors.

% \subsubsection{Extended variance identification for GLT structure} \label{secover_extend}

%In addition to assumption \TS , we require that the nonzero columns $\betatilde$  of $\facload$ in the extended variance identification (\ref{fac4beta}) are  constrained to be

\begin{thm} \label{theoverGLT}
Assume that  $\facloadtrue $ is a GLT factor loading matrix with leading indices $l_{1} < \ldots < l_{\nfactrue}$ that obeys the extended
row deletion property \TS\ for some $\tumS \in \mathbb{N}$. If  $\betatilde$ in the extended variance decomposition  $\Vary=     \betatilde   \trans{\betatilde} + \Vare$ is restricted  to be an  unordered GLT matrix with leading indices $\tilde{l}_1, \ldots,  \tilde{l}_{\nfactrue+s}$, then the following holds:
\begin{itemize}
  \item[(a)]  $\facloadtrue$  and $\Varetrue$  can be represented in terms of $\betatilde$, $\Vare$, and $\Mm $ as in  (\ref{decover})
  up to trivial rotations $\Tm= \bP _{\pm} \bP _{\rho}$.
  \item[(b)] $\Mm $ is a  spurious GLT structure with leading indices ${n_1} , \ldots , {n_s}$
with exactly one nonzero loading in each column. Furthermore, all leading indices
  $\{ {n_1}, \ldots, {n_s} \}$  are different from the leading indices $\{ l_{1} , \ldots , l_{\nfactrue}\}$ of $\facloadtrue $.
  \item[(c)] The leading indices $\{\tilde{l}_{1}, \ldots, \tilde{l}_{\nfactrue+s}\}$ of $\betatilde$ are identical to the leading indices  $\{l_{1} , \ldots , l_{\nfactrue}, {n_1} , \ldots , {n_s}\}$
       %$l_{1} < \ldots < l_{\nfactrue}$  and $ {n_1} < \ldots < {n_s}$ of  $\facloadtrue$ and $\Mm $
      of the matrix $ \betatilde \Tm$.
\end{itemize}
\end{thm}

 \noindent See Appendix~\ref{app:proof} for a proof.
For an unordered GLT structure,  \TS\ implies a constraint on the leading indices of  $\betatilde$ which extends \GLTAR :
 \begin{enumerate}
  \item[\GLTTS .] Let $\facloadtilde$ be an unordered GLT  structure with $\nfacr$ nonzero columns with %  (unordered)
  leading indices $l_1, \ldots , l_{\nfacr}$. %$\tilde{l}_1 , \ldots , \tilde{l}_{\nfactrue}$.
 The following condition   on the leading indices  is necessary  for  condition  \TS : % Each $l_j$ satisfies:
      \begin{eqnarray} \label{condljTS}
       \dimy-   l_j - \tumS\geq   2(\nfacr-\rankl_j +1), \quad j=1,\ldots, \nfacr,
 \end{eqnarray}
 where $\rankl_j$ is the rank  of $l_j$ in the ordered sequence $ l_{(1)} <  \ldots < l_{(\nfacr)}$.
\end{enumerate}

\section{Bayesian inference}  \label{secbayes}

 Bayesian inference is performed in the overfitting sparse factor model   (\ref{fac1reg})   where   $\nfac$
satisfies the upper bound (\ref{kbound_extend}) for  a given degree of overfitting  $\tumS \in \mathbb{N}$.
Both $\nfac$  as well as $\tumS$ are user-selected parameters.
The  maximum number of potential factors  $\nfac$  is chosen large enough that zero and spurious columns will appear during posterior inference.
We found it useful to allow for at least $\tumS \geq 2$ spurious columns.
% If  $\dimy$ is small, this might reduce the maximum number $k$ of potential factors,  but has no consequences for larger models.

%\comment{ A one-sweep procedure is applied, where the unknown number of factors is  estimated jointly with the remaining parameters.     To estimate $\nfactrue$, we  exploit \lq\lq column sparsity\rq\rq\ of the indicator matrix based  on the number of   nonzero columns  $\nfacr$ in variance identified indicator matrices $\deltav$.}

\subsection{Prior specifications} \label{priorel}

Let $\deltav$  be the  $\dimmat{\dimy}{\nfac}$ indicator matrix   corresponding to the $\dimmat{\dimy}{\nfac}$  loading matrix $\facload $ in model   (\ref{fac1reg}).  Within our sparse Bayesian factor analysis, a  joint prior for $\deltav$,  $\facload$  and  the variances $\idiov_1, \ldots, \idiov_\dimy$ is
selected, taking the form
% : \begin{eqnarray*} \label{priall}
 %$p (\deltav, \idiov_1, \ldots, \idiov_\dimy, \facload)=
 $p (\deltav) p(\idiov_1, \ldots, \idiov_\dimy)  p(\facload|\deltav, \idiov_1, \ldots, \idiov_\dimy).$
% \end{eqnarray*}

\subsubsection{The prior on the indicators} \label{priordelta}

Following common hierarchical point mass mixture prior  on the indicator matrix $\deltav$ is applied:
  \begin{eqnarray} \label{prigen}
&& \Prob{\delta_{ij}=1|\tau_{j}}=\tau_{j}, \qquad  \tau_j \sim \Betadis{a_0,b_0}, \qquad j=1,\ldots, \nfac, \\ && \Prob{\beta_{ij}=0|\delta_{ij}=0}=1, \nonumber
 \end{eqnarray}
where all indicators  are  independent {\em a priori} given %the hyperparameter
  $\hypv=(\tau_1, \ldots, \tau_\nfac)$.\footnote{Alternative priors (which are not pursued in the present paper) have been considered e.g.
  by  \citet{con-etal:bay}    and  \citet{kau-sch:bay}.}
  % $\tauv=(\tau_1, \ldots, \tau_\nfac)$.
   Since the true number of factors $\nfactrue$ is unknown, we employ a prior on $\deltav$ that implies column sparsity apriori.   To this goal, the hyperparameters of   prior (\ref{prigen}) % and (\ref{prialt})
   are chosen such that the number of nonzero columns  $\nfacr$ in $\deltav$ is random apriori,  taking values less than $\nfac$ with high probability.  In this case,  %the number of potential factors $\nfac$is larger than $\nfactrue$, hence
   the model is overfitting and we are able to learn the number of factors $\nfactrue$.
  Hyperparameters that exclude zero  columns in $\deltav$ apriori are prone to overfit the number of factors.
 Prior (\ref{prigen})  can be rewritten as:
 \begin{eqnarray} \label{prialt}
 \tau_j \sim \Betadis{a_0,b_0} = \Betadis{b_0 \frac{\alpha}{\nfac},b_0},
 \end{eqnarray}
 where $\nfac$ is the number of potential factors.
 For $\nfac \rightarrow \infty$, prior (\ref{prialt}) converges to the two-parameter Beta prior  introduced by \citet{gha-etal:bay}
 in Bayesian nonparametric latent feature models which can be regarded as a  factor model with infinitely many columns. However, if $k$ exceed the upper bound (\ref{kbound_extend}), variance identification can no longer be achieved. For this reason, we stay within the framework of
 factor models with finitely many columns in the present paper, but exploit column sparsity as explained above.

 Following  \citet{gha-etal:bay},   we choose values $b_0 < 1$   considerably smaller than 1 (a sticky prior) %, e.g. $b_0=0.3$,
 to allow apriori  zero columns for factor models where  the number of factors is unknown.
The choice of $\alpha$ (or  $a_0$) is guided by the apriori expected simplicity $ \Ew{q_i}$ of the factor loading matrix, where $ q_i = \sum_{j=1}^\nfac \delta_{ij}$ is the number  of nonzero   loadings in each row which is typically smaller than $\nfac$. This leads to following choice for $a_0$ and $\alpha$:
   \begin{eqnarray} \label{defqi}
    \Ew{q_i} % = \frac{\nfac \alpha^\tau}{\alpha^\tau + \nfac}
    =   \frac{\nfac a_0 }{a_0 + b_0}= \frac{\alpha}{1 + \alpha/\nfac} \quad   \Rightarrow \quad
     a_0 =  \frac{ b_0\Ew{q_i}}{\nfac -\Ew{q_i}}, \quad \alpha =  \frac{\Ew{q_i}}{1 -\Ew{q_i}/\nfac}.
  \end{eqnarray}
  As common in statistics and machine learning, the prior on  $\deltav$  does not account explicitly for identification.
   To deal with rotational invariance, an unordered GLT structure  as introduced in Subsection~\ref{secGLT}
   is imposed  on $\deltav$ during MCMC estimation,  by sampling only indicator matrices where  the leading indices $l_{1}, \ldots ,  l_{\nfacr}$  of the $\nfacr$ nonzero columns $\betatilde$  of $\facload $
 satisfy condition \GLTTS\ given in (\ref{condljTS}) for the specified value of  $\tumS$, i.e.  prior $p(\deltav)$  is constrained implicitly to  unordered sparse GLT structures.  The  unordered GLT structure enforced during MCMC estimation breaks the invariance of the procedure with respect to the ordering of the data. However, it is less sensitive to the ordering of the data than the PLT constraint.
 % The resulting posterior draws $\deltav$, however, are  not necessarily variance identified and have to be postprocessed.

\subsubsection{The prior on the idiosyncratic variances} \label{priorsi}

 When estimating factor models using classical statistical methods, such as maximum likelihood (ML) estimation,
  it frequently happens that  the optimal solution lies outside the admissible parameter space with one
   or more of the idiosyncratic variances $\idiov_i$s being negative, see e.g.
   \citet[Section~3.6]{bar:lat}. An empirical study in  \citet{joe:som} involving 11 data
   sets revealed that such improper solutions are quite frequent and this difficulty became
 known as the Heywood problem. % and is likely to happen for  samples with  either $\T$ or $\dimy$ being small, for data where the true $\idiov_i$s are very unbalanced and for overfitting models where fitting more factors than present leads to an inflation of the communalities $R_i^2$  defined in (\ref{fac4A}),
% \begin{eqnarray*}
%R_i^2 = \frac{\sum_{j=1}^{\nfac} \load_{ij}^2}{\sum_{j=1}^{\nfac} \load_{ij}^2+\idiov_i} =
%\frac{\sum_{j=1}^{r} \load_{ij}^2 + \sum_{j=r}^{\nfac} \load_{ij}^2}{\sum_{j=1}^{\nfac} %\load_{ij}^2+\idiov_i}  \geq (R_i^2) \true + \frac{\sum_{j=r}^{\nfac} \load_{ij}^2}{\sum_{j=1}^{\nfac} %\load_{ij}^2+\idiov_i},
%\end{eqnarray*}
 % forcing $\idiov_i$ toward 0.
%
 The introduction of a prior on the  idiosyncratic variances $\idiov_1, \ldots, \idiov_\dimy$ within a Bayesian framework, typically chosen from the  inverted Gamma family, that is
 \begin{eqnarray}
\idiov_i \sim \Gammainv{c_0,C_{i0}}, \label{priorsiidg}
\end{eqnarray}
 naturally avoids negative values for  $\idiov_i$.
  Nevertheless, there exists a Bayesian analogue of the Heywood problem which takes the form of
  multi-modality of the posterior of  $\idiov_i$ with one mode lying at 0. This is likely to happen,
  if a small value $c_0$ and fixed hyperparameters  $C_{i0}$ are chosen in (\ref{priorsiidg}),
   as common in Bayesian factor analysis.

  Subsequently, we select $c_0$ and  $C_{i0}$ in such a way that Heywood problems are avoided.
    Heywood problems typically occur, if the  constraint
      \begin{eqnarray}
 \frac{1}{\idiov_i} \geq (\Vary^{-1})_{ii}   \quad \Leftrightarrow \quad   \idiov_i  \leq \frac{1}{(\Vary^{-1})_{ii}} \label{const1}
\end{eqnarray}
is violated, where the matrix $\Vary$ is the covariance matrix of $\ym_t$ defined in (\ref{fac4}),
see e.g. \citet[p.~54]{bar:lat}. It is clear from  inequality (\ref{const1}) that $1/\idiov_i$  has to be bounded away
from 0. For this reason, improper priors on the idiosyncratic variances such as $p(\idiov_i)\propto 1/\idiov_i$
 \citep{mar-mcd:bay,aka:fac} are not able to prevent Heywood problems.
Similarly, proper inverted Gamma prior with small degrees of freedom such as $c_0=1.1$
 \citep{lop-wes:bay} allow values too close to 0.

 As a first improvement, we choose  $c_0$  in (\ref{priorsiidg}) large enough to
 bound the prior away from 0, typically $c_0=2.5$.  Second, we reduce the occurrence probability of a Heywood problem which is equal to
  $\Prob{X\leq C_{i0}(\Vary^{-1})_{ii}}$ where $X \sim \Gammad{c_0,1}$ through
  the choice of $C_{i0}$.  The smaller $C_{i0}$, the  smaller is this probability.   However, since
  $\Ew{\idiov_i}=C_{i0}/(c_0-1)$, a downward bias may be introduced, if $C_{i0}$ is too small.
  We choose $C_{i0}=(c_0-1)/(\widehat{\Vary^{-1}})_{ii}$ as the largest value for which inequality
   (\ref{const1}) is fulfilled by the prior expectation $\Ew{\idiov_i}$ and $\Vary^{-1}$
   is substituted by an estimator  $\widehat{\Vary^{-1}}$.
%  \begin{eqnarray*} C_{i0} =   \frac{(c_0-1)}{(\widehat{\Vary^{-1}})_{ii}}, % \approx \frac{1}{(\Scov{y}^{-1})_{ii}} . %\label{const2} \end{eqnarray*}
 This yields the following prior:
\begin{eqnarray}
\idiov_i \sim \Gammainv{c_0,(c_0-1)/(\widehat{\Vary^{-1}})_{ii}}. \label{priorsiid}
\end{eqnarray}
   % $1-R_i^2 =\idiov_i /\om_{ii} $,  % $1-R_i^2 =\idiov_i /\om_{ii} $,
  Inequality (\ref{const1}) introduces an upper bound for $\idiov_i /\om_{ii} $, the proportion of variance not explained by the common
  factors,
%\begin{eqnarray*}
% \frac{1}{(1-R_i^2) \om_{ii}} \geq (\Vary^{-1})_{ii} \Rightarrow
%   (1-R_i^2) \leq  \frac{1}{(\Vary^{-1})_{ii}  \om_{ii}},  % \label{const3}
%\end{eqnarray*}
which is considerably smaller than 1 for small  idiosyncratic variances $ \idiov_i $.
Hence, our prior is  particularly sensible, if the communalities $R_i^2=1-\idiov_i /\om_{ii}$ are rather unbalanced across variables and the variance of some observations is very well-explained by the common factors, while this is not the case for other variables.
Our case studies illustrate  that this prior usually leads to unimodal
posterior densities for the idiosyncratic variances.

An estimator $\widehat{\Vary^{-1}}$  of the inverse $\Vary^{-1}$ of the marginal covariance matrix
is required to formulate prior (\ref{priorsiid}).
If $T>> \dimy$, then the inverse of the sample covariance matrix $\Scov{y}$ could be used,
i.e. $\widehat{\Vary^{-1}}=\Scov{y}^{-1}$. However, this estimator is unstable, if  $ \dimy$ is not
small compared $T$, and does not exist, if  $\dimy>T$. Hence, we prefer
a Bayesian estimator  which is obtained by combining the
 sample information with the inverted Wishart prior $\Vary^{-1} \sim \Wishart{\dimy}{\nu_o, \nu_o \Sm_o}$:
%$$\widehat{\Vary^{-1}}= \frac{1}{\nu_O  + T/2 - (\dimy+1)/2}
 %(\nu_O \identy{\dimy} + 0.5 \sum_{t=1}^T \ym_t \trans{\ym_t} + ),$$
 \begin{eqnarray}
\widehat{\Vary^{-1}}= (\nu_o + T/2)(\nu_o \Sm_o + 0.5 \sum_{t=1}^T \ym_t \trans{\ym_t})^{-1}. \label{Varyhat}
\end{eqnarray}
If  the variables $y_{jt}, j=1,\ldots,m,$  are standardized over $t$, then  $\Sm_o=\identy{\dimy}$ is a sensible choice.

% If the idiosyncratic variances are not too unbalanced, we found it also useful to consider a hierarchical prior, where $C_{i0}\equiv C_0$ and  $C_0$ is equipped with a $\Gammad{g_0,G_0}$ prior with $g_0$ being a small integer, e.g. $g_0=5$. This prior allows for more shrinkage than the previous one. Once again we use the upper bound defined in (\ref{const1}) to choose $G_0$. We assume that the expected mode of the prior of  $\idiov_i$ which is given by $\Ew{g_0/C_0}=g_0/c_0/G_0$ is smaller than the average of the upper   bounds defined in (\ref{const1}), thus
%\begin{eqnarray}
%\idiov_i \sim \Gammainv{c_0,C_{0}},  \qquad C_{0} \sim \Gammad{g_0,G_0}, \qquad  G_0=\frac{g_0}{c_0 \sum_{i=1}^\dimy 1/(\Scov{y}^{-1})_{ii} }. \label{priorG0}
%\end{eqnarray}

\subsubsection{The prior on the factor loadings} \label{priorfl}

Finally, conditional on $\deltav$ and $\idiov_1, \ldots, \idiov_\dimy$, a prior has to be
 formulated for all nonzero factor loadings.
Since the likelihood function factors into a product over the rows of the loading matrix,   prior independence across the rows  is assumed.
 For a given $\deltav$, let  $\facload_{i\cdot}^{\deltav}$ be
    the vector of unconstrained elements in the $i$th row of  $\facload$.
   The variance of the prior  of $\facload_{i\cdot}^{\deltav}$  is assumed to depend on   $\idiov_i$, because  this allows
joint drawing of   $\facload$ and  $\idiov_1, \ldots, \idiov_\dimy$  and, even more importantly, sampling the model indicators $\deltav$ without conditioning on the model parameters  during MCMC estimation, see   Algorithm~\ref{Algo3} in Subsection~\ref{mcmc}.

For each  row $i$ with  $q_i>0$  nonzero elements,  the standard prior takes  the form
  \begin{eqnarray}
\facload_{i\cdot}^{\deltav}|\idiov_i  \sim \Normult{q_i}{\bfz, \bV_{i0}^{\deltav}\idiov_i},  \label{prior1}
\end{eqnarray}
where, typically,  $\bV_{i0}^{\deltav}=A_0 \identy{q_i}$   % with $A_0$ being fixed
\citep{lop-wes:bay,gho-dun:def,con-etal:bay}.
%\citet{kau-sch:bay} apply a hierarchical prior, where the $\bV_{i0}^{\deltav}=\Diag{A_1 \ldots A_\nfac}$ and $A_j$ follows an inverted Gamma distribution.
%
 In addition, a  fractional prior in the spirit of \citet{oha:fra} is introduced  in this paper for sparse Bayesian factor models which can be interpreted as the posterior of a  non-informative prior and a small fraction $b>0$ of the data. This yields  a conditionally fractional prior for the \lq\lq regression model\rq\rq\
\begin{eqnarray}
\tilde{\ym}_i= \Xb_i ^{\deltav} \facload_{i\cdot}^{\deltav} + \tilde{\errorm}_i,
 \label{regnonp}
\end{eqnarray}
where  $\tilde{\ym}_i=\trans{(y_{i1} \cdots y_{iT})}$ and
$\tilde{\errorm}_i=\trans{(\error_{i1} \cdots \error_{iT})}$.   $\Xb_i ^{\deltav}$ is a regressor matrix
% for  $\facload_{i\cdot}^{\deltav}$
constructed from the latent factors  $\facm_1,  \ldots, \facm_T$ (see Appendix~\ref{postdisfac} for details).
 % If no element  in row $i$ of $\facload$ is restricted to 0, then  $\Xb_i ^{\deltav}=\Fm$.  If some elements are restricted to 0, then $ \Xb_i ^{\deltav}$ is obtained from  $\Fm$  by deleting all columns $j$ where $\delta_{ij}=0$, i.e. $ \Xb_i ^{\deltav}=   \Fm \Pim_i ^{\deltav} $, where $\Pim_i ^{\deltav}$ is a $\dimmat{\nfac}{\sum_{j=1}^{\nfac} \delta_{ij} }$ selection matrix, selecting those columns $j$ of $\Fm$ where $\delta_{ij}\neq 0$.
 The fractional  prior is then defined  as a fraction of the full conditional likelihood, derived from regression model (\ref{regnonp}):
\begin{eqnarray*} p(\facload_{i\cdot}^{\deltav}|\idiov_i \addb ,\facm)
\propto \displaystyle p(\tilde{\ym}_i| \facm, \facload_{i\cdot}^{\deltav} ,\idiov_i)^b
= \left(\frac{1}{2\pi \idiov_i}\right)^{Tb/2}
 \exp\left(-\frac{b}{2\idiov_i}
(\tilde{\ym}_i- \Xb_i ^{\deltav} \facload_{i\cdot}^{\deltav} )'(\tilde{\ym}_i- \Xb_i ^{\deltav} \facload_{i\cdot}^{\deltav})\right).
\end{eqnarray*}
This yields the following fractional  prior:\footnote{Similar conditionally conjugate fractional priors have been applied
 by several authors for variable selection in latent variable models  \citep{smi-koh:par,fru-tue:bay, tue:bay,fru-wag:sto}.} %,
%  Using  $p(\facload_{i\cdot}^{\deltav}|\idiov_i \addb ,\facm )\propto  p(\tilde{\ym}_i| \facm, \facload_{i\cdot}^{\deltav} ,\idiov_i)^b$ we obtain from regression model (\ref{regnonp}):
 \begin{eqnarray}   \label{priorfrac}
 \facload_{i\cdot}^{\deltav} | \idiov_i \addb ,\facm  \sim
\Normult{q_i}{\bm_{iT}  ^{\deltav}  , \bV_{iT}^{\deltav}  \idiov_i /b},
\end{eqnarray}
 where $\bm_{iT}^{\deltav} $ and $\bV_{iT}^{\deltav} $ are the posterior  moments under the non-informative prior
  $p(\facload_{i\cdot}^{\deltav} | \idiov_i) \propto \const$:
\begin{eqnarray}
 \bV_{iT} ^{\deltav} = \left(\trans{(\Xb_i ^{\deltav})} \Xb_i ^{\deltav} \right) ^{-1} , \qquad
 \bm_{iT}  ^{\deltav} = \bV_{iT} ^{\deltav}  \trans{(\Xb_i ^{\deltav})}\tilde{\ym}_i . \label{postmomA_frac}
\end{eqnarray}
Concerning the choice  of the fraction $b$,  in general,
larger values of $b$ extract more information from the likelihood than smaller values,
which reduces the influence of the sparsity prior $p(\deltav)$ as $b$ increases, leading to a  larger
 number of estimated factors.
 Depending on the relation between $\nfac$, $\dimy$, and $T$, small values such as $b=10^{-3}$, $b=10^{-4}$ or $b=10^{-5}$  yield sparse solutions.
In total,   $N=\dimy T$  observations  are available to estimate  $d(\nfac,\dimy) = \nfac \dimy-\nfac(\nfac-1)/2 = \nfac (\dimy- (\nfac-1)/2)$ free elements in the coefficient matrix $\facload$  for a  GLT structure.

 If $d(\nfac,\dimy)$ is considerably smaller than $N$, then the variable selection literature suggests to choose $b_N=1/(T\dimy)$.
This is in particular the case,  if the potential number of factors $k$ is considerably smaller than $T$.
On the other hand,  if $d(\nfac,\dimy)$ is in the order of  $N$, then $b_N$ implies a fairly small penalty and may lead to  overfitting models.  Following \citet{fos-geo:ris}, the risk inflation criterion $b_R=1/d(\nfac,\dimy)^2$ can be applied in this case. For a GLT sructure,  $b_R$ implies a stronger penalty than $b_N$, if $d(\nfac,\dimy)> \sqrt{ T \dimy}$.

\subsection{MCMC estimation} \label{mcmc}

We use MCMC techniques to sample from the posterior $ p(\deltav, \idiov_1, \ldots, \idiov_\dimy, \facload,\hypv, \facm|\ym)$
(with $\facm=(\facm_1,\ldots,\facm_T))$  of the overfitting model (\ref{fac1reg}),
% , where  the maximum number of potential factors $\nfac$ satisfies the upper bound (\ref{kbound_extend}) for  a given value of  $\tumS$.
given  the priors  introduced in Subsection~\ref{priorel}.
As noted by many authors,   e.g. \citet{pat-etal:pos}, MCMC sampling for sparse Bayesian factor models is notoriously difficult,
  since sampling the indicator matrix $\deltav$ corresponds to navigating through an extremely high dimensional model space.
  %, corresponding to all possible values of $\deltav$.
 This is even more challenging, if the sparse factor model is overfitting. % as in our case.

In this paper, a  designer  MCMC scheme is employed which is  summarized in Algorithm~\ref{Algo3}, where several steps have been  designed  specifically   for sparse  Bayesian factor models  under the GLT constraint when  the number of factors is unknown. This designer  MCMC scheme delivers posterior draws of $\facload $ and $\deltav$  with a  varying number $\nfacr$  of nonzero columns. %  $\betatilde $ and  $\deltavtilde$.
  An  unordered  GLT structure is imposed  on the nonzero columns $\betatilde $ and  $\deltavtilde$ % throughout posterior sampling
  by requiring that the  leading indices $l_1, \ldots , l_{\nfacr} $
  obey condition \GLTTS\ given in (\ref{condljTS}).
  % Beyond this condition,    no further constraints  guaranteeing  variance identification    are imposed during MCMC sampling   and variance identification  is verified
  Non-identification  with respect to trivial  rotations introduces column and sign switching during MCMC sampling.
  Hence, the sampler produces draws that fulfill various {\em necessary} conditions for identification, while the more
  demanding {\em sufficient} conditions are assessed through a scanning of the posterior draws
   during postprocessing, see Subsection~\ref{subGL}.

 %Given initial values\footnote{See Appendix~\ref{init} for details} for  $\nfacr \in \{1, \ldots \nfac\}$ and $\deltav, \facload,\idiov_1,\ldots,\idiov_{\dimy},\hypv$, we iterate $M$ times through Algorithm~\ref{Algo3}, after discarding the first $M_0$ draws as burn-in.

 \begin{alg}[\textbf{MCMC estimation for sparse Bayesian factor models with unordered  GLT structures}] \label{Algo3}
 % Given initial values for  $\deltav, \facload,\idiov_1,\ldots,\idiov_{\dimy}$ and $\hypv$,  iterate through following steps:
  Choose initial values\footnote{See Appendix~\ref{init} for details.} for  $(\nfacr, % \in \{1, \ldots \nfac\}$ and
  \deltav, \facload,\idiov_1,\ldots,\idiov_{\dimy},\hypv)$, iterate $M$ times through the following steps and  discard the first $M_0$ draws as burn-in:
  \begin{itemize}
  \item[(F)]  Sample the latent factors $\facm_1,\ldots,\facm_T$ conditional on
the model parameters $\facload$ and $\idiov_1,\ldots,\idiov_{\dimy}$ from $ p(\facm_1,\ldots,\facm_T|\facload,\idiov_1,\ldots,\idiov_{\dimy},\ym) $.
%, see Subsection~\ref{SectionF_fac}.

\item[(A)] Perform a boosting step based either on ASIS or marginal data augmentation. %;  (see Subsection~\ref{accelerate}).

%   \item[(S)]   Sample the indicator matrix $\deltav$ conditional on  the factors  $\facm_1,\ldots,\facm_T$ and the hyper parameters $\hypv$ without conditioning on the model parameters   $  \facload$ and $\idiov_1,\ldots,\idiov_{\dimy}$:  \footcomment{REWRITE! NOTE the  (R) depends on $  \facload$ and $\idiov_1,\ldots,\idiov_{\dimy}$!! We probably don't need (S)}.
 % \begin{itemize}
      \item[(R)] Perform a  reversible jump MCMC step  %discussed in Subsection~\ref{RJMCMC},
  to add or delete spurious columns in $\deltav$ and  $  \facload$. %, see  .

 \item[(L)] Loop over all  nonzero columns $j$  of the indicator matrix $\deltav$ in a
      random order and sample the leading index $l_j$
       conditional on    the remaining columns $\deltacol{-j}$, the factors $\facm_1,\ldots,\facm_T$, and $\hypv$
       without conditioning on the model parameters   $  \facload$ and $\idiov_1,\ldots,\idiov_{\dimy}$. %\footcomment{CHANGE!, SWITCH is no loop!}

 \item[(D)] Loop over all nonzero columns of the indicator matrix $\deltav$ in a random order. Sample for each
 column $j$ all indicators below the leading index  $l_j$ (i.e.~$\delta_{ij}$ with $i \in I_j=\{l_j +1, \ldots, \dimy \}$)
  conditional on the remaining   columns $\deltacol{-j}$, the factors $\facm_1,\ldots,\facm_T$, and $\hypv$
   (without conditioning on the model parameters
   $  \facload$ and $\idiov_1,\ldots,\idiov_{\dimy}$)  jointly using Algorithm~\ref{AlgoInd} in Appendix~\ref{mcmcsmodi}.
  %\end{itemize}

\item[(H)]  Sample    $\tau_j |\deltav  \sim \Betadis{a_0  + d_j,b_0 + \dimy- d_j},  j=1,\ldots,\nfac$,
where   $d_j=\sum_{i=1}^\dimy {\delta_{ij}}$  is  the number  of  nonzero factor loadings  in column $j$.

 %  \item [(BD)] \comment{COMMENT: IS NOT VALID, since $p(d_j)$ is not valid.} With probability $1-p_{sp}$,\comment{\footnote{Currently decision not independent from decision    in step~{RJ}. Is this the right way to do it?}}  perform a birth-death step, see Subsection~\ref{birthdeath}.

 \item [(P)]   Sample   the model parameters  $\facload$ and $\idiov_1,\ldots,\idiov_{\dimy}$   jointly conditional on the indicator matrix   $\deltav$ and the factors $\facm_1,\ldots,\facm_T$ from $p(\facload, \idiov_1,\ldots,\idiov_{\dimy}| \deltav,\facm_1,\ldots,\facm_T,\ym)$.
    % see Subsection~\ref{jointfac}.

 \end{itemize}
\end{alg}

\noindent The most innovative part of this MCMC scheme  concerns sampling the indicator matrix $\deltav$.
 Updating  $\deltav$   for sparse exploratory Bayesian factor analysis  without identification constraints on $\deltav$
is fairly  straightforward, see e.g.  \citet{car-etal:hig} and \citet{kau-sch:bay}, among many others.
However,  a more refined approach is implemented in the present  paper to address the econometric  identification issues for sparse factor models discussed  in Section~\ref{secide}.
The nonzero columns of  $\facload$ and $\deltav$ are instrumental for  estimating  the number of factors  during postprocessing,
     see Subsection~\ref{estr}.
To increase and decrease the number   of nonzero  columns in  $\facload$ and $\deltav$,
    Step~(R)  exploits   Theorem~\ref{theoverGLT}   to add and delete  spurious factors
     through a  reversible jump MCMC step  described  in  Subsection~\ref{RJMCMC}.
     Similarly as in \citet{con-etal:bay},  it is much easier to introduce new latent factors into the model through these spurious factors,
 compared to alternative approaches that would split existing factors or add new ones only under the condition that enough nonzero elements are preserved.
        To  force the unordered  GLT structure on the  $\nfacr$ nonzero columns of  $\facload$ and $\deltav$,     Step~(L) performs  MH steps
 to navigate  through the space of all admissible leading indices $(l_1, \ldots, l_{\nfacr})$ that satisfy \GLTTS , see Subsection~\ref{movelead}.
To implement Step~(D) efficiently, a  method for  sampling an entire set of  indicators $\{ \delta_{ij}, i \in I_j \}$ in a particular column $j$ in one block is  developed  in Appendix~\ref{mcmcsmodi}.

%,  in particular Step~(H), see Appendix~\ref{Sec_hyp}.
Step~(F) and  Step~(P) operate in a \lq\lq confirmatory\rq\rq\  factor model where  certain loadings are
constrained to zeros according to the indicator matrix $\deltav$. Although these steps are standard in Bayesian
factor analysis (see   e.g. \citet{lop-wes:bay} and  \citet{gho-dun:def})
 improvements are suggested   such as  multi-move sampling of all unknown model parameters $\facload$, and $\idiov_1,\ldots,\idiov_{\dimy}$ in Step~(P),
see  Appendix~\ref{SectionF_fac} and \ref{jointfac} for futher details.
     Finally, % since    Step~(F) and  Step~(P)  perform full conditional Gibbs sampling,
      the boosting Step~(A)  is added to improve mixing of the MCMC scheme,  see Subsection~\ref{accelerate} and Appendix~\ref{accelerate_App} for more details.

%  \comment{MODIFY?    Since we allow for trivial rotations,  we loose  identifiability of the signs of the elements of $\facm_t$ and $\facload$ with our approach. For each $j=1,\ldots,\nfac$ all elements $\fac_{j1}, \ldots, \fac_{jT}$ of the latent factor as well as the $j$th column of $\facload$ may be changed by the same sign switch without changing the likelihood. In combination with the prior  introduced in Subsection~\ref{priorfl}  this leads to multimodality of the posterior density. To make  sure that our sampler explores all possible modes,   a random sign switch is performed in   Step~(d-3) for  each column and the columns are randomly permuted:
% \begin{itemize}
  %  \item [(d-3)] Perform a  random sign switch independently for each column $j=1,\ldots, \nfactrue$: substitute the  draws of   $\{ \fac_{jt} \}_{t=1}^T$ and $\{ \load_{ij} \}_{i=l_j}^\dimy$  with probability 0.5 by $\{- \fac_{jt} \}_{t=1}^T$ and $\{- \load_{ij} \}_{i=l_j}^\dimy$,  otherwise leave these draws  unchanged. \comment{ADD all other rotations!}.  Rotate the $\nfactrue$ columns of $\facload$ by one of the $2^\nfactrue \nfactrue !$ randomly selected matrices  $\Pm= \bP _{\rho} \bP _{\pm}$.
% \end{itemize}}

\subsubsection{Special MCMC moves for  unordered GLT structures} \label{movelead}

Step~(L) in Algorithm~\ref{Algo3} implements moves that explicitly change the position of the leading indices  in the $\nfacr$  nonzero columns of $\deltav$ (including
spurious columns), without violating \GLTTS .  Let $\lm =(l_{1} , \ldots , l_{\nfacr})$ be the set of leading indices. Since an unordered GLT structure has to be preserved,  the leading index $l_j$  in column $j$
  is not free to move, but  restricted  to a subset $\leadset{\tumS}{\lm_{-j}} \subseteq \{1,\ldots,\dimy\}$ which
  depends  on  the leading   indices $\lm_{-j}$  of the other  columns and the maximum degree of overfitting  $\tumS$.\footnote{See Subsection~\ref{GLTMCMC} for a definition of   $\leadset{\tumS}{\lm_{-j}}$.} % To guarantee that \GLTTS\ holds, we set  $\leadsetn=\dimy-\tumS$; .}
   We scan all nonzero columns of $\deltav$ in a random order and  propose  to change the position
of  $l_j$  in a selected column $j$ using  one of four local moves, namely
 shifting the leading index,   adding a new leading index, deleting a leading
  index and  switching  the leading elements (and all indicators in between) between column $j$ and a randomly selected column $j'$; see Figure~\ref{figStepL} for illustration and  Subsection~\ref{updatelead} for further details.

  %\begin{Figure2}{Special MCMC moves to change the leading elements of an unordered GLT structure}{figStepL}{lead_change}{switch_lead}{0.4} \end{Figure2}

 %\begin{Figure2pdf}
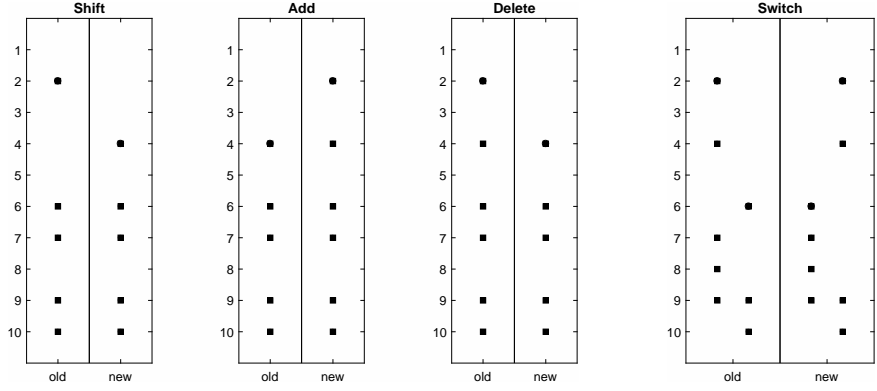
 \begin{Figure2}{MCMC moves to change the leading indices of an unordered GLT structure;
 from left to right: shifting the leading index,   adding a new leading index, deleting a leading
 index and  switching  the leading elements}{figStepL}{lead_change}{switch_lead}{0.4}
 \end{Figure2}% \end{Figure2pdf}

\subsubsection{Split and merge moves for overfitting models}  \label{RJMCMC}

 For overfitting factor models,  Step~(R)    in Algorithm~\ref{Algo3}
is a dimension changing move that explicitly changes  the number $\nfacr$ of nonzero columns in $\deltav$ and $\facload$  by adding  and deleting a spurious column.
 If  a spurious column  $\Mm$ is identified among  the nonzero columns  % $\betatilde$
of $\facload$, then as demonstrated  in Subsection~\ref{secover} it can be substituted by a zero column  without changing the likelihood function,
  by adding $\Mm  \trans{\Mm }$ to $\Vare$.   On the other hand, any zero column in $\facload$  can be turned into an (additional) spurious column  without changing the likelihood function either, see  (\ref{adsp}).  %
  This is the cornerstone of our procedure, however, while the likelihood is invariant to these moves, the prior is not  and
 simply adding or deleting spurious columns  would lead to an invalid  MCMC step.
 A reversible jump MCMC step as implemented in Step~(R) can correct for that.

 The split and  merge moves outlined above form a reversible pair  that operates in the latent variable model (\ref{fac1reg})   conditional on all parameters, except the hyperparameter $\hypv=(\tau_1, \ldots, \tau_\nfac)$ which is integrated out  of prior (\ref{prigen}).
 Split and merge  moves are local moves operating between the two following factor models:
 \begin{eqnarray}
&&  y_{l_j,t}=  \facload_{{l_j},-j}^{\deltav} \facm_{t,-j} + \error_{l_j,t}  ,
\qquad \error_{l_j,t}  \sim \Normal{0,\idiov_{l_j}}, \label{spurious1}\\
&& y_{l_j,t}=  \facload_{{l_j},-j}^{\deltav} \facm_{t,-j} + \load_{l_j,j} \newsp  \fac_{jt}\newsp \delta_{l_j,j}  +  \tilde{\error}_{l_j,t},
\qquad \tilde{\error}_{l_j,t}  \sim \Normal{0,\idiov_{l_j}-\delta_{l_j,j}(\load_{l_j,j}^2) \newsp},   \label{spurious2}
\end{eqnarray}
where model  (\ref{spurious2}) contains a spurious column with $ \load_{l_j,j} \newsp$ being the only nonzero loading in this column.
If $\delta_{l_j,j}=0$ in model (\ref{spurious2}), then model (\ref{spurious1}) results. However, if $\delta_{l_j,j}=1$, then, as discussed in Subsection~\ref{secover},  model (\ref{spurious2}) is not identified and
$\load_{l_j,j} \newsp$  can take  any value such that $ (\idiov_{l_j}) \newsp = \idiov_{l_j} -  (\load_{l_j,j}^2) \newsp >0 $.
By integrating model (\ref{spurious2})  with respect to  the spurious factor  $ \fac_{jt}\newsp$, it can be easily verified that both models imply the same distribution   $p(y_{l_j,t}| \facload_{{l_j},-j}^{\deltav}, \facm_{t,-j},\idiov_{l_j})$.

  % \paragraph{Designing the split move.}
  The split move turns one of the zero columns
$j$ in (\ref{spurious1}) into a spurious column, by selecting a row $l_j$ % (where, obviously, $\delta_{l_j,j}=0$ and $\load _{l_j,j}=0$), defining $\delta\newsp_{l_j,j}=1$,
not occupied by any other leading index and splitting the variance $\idiov_{l_j}$ of the idiosyncratic  error between  the new
variance  $(\idiov_{l_j}) \newsp$ and
the spurious factor loading  $\load \newsp_{l_j,j}$ such that
\begin{eqnarray*}
(\load \newsp_{l_j,j})^2 + (\idiov_{l_j}) \newsp = \idiov_{l_j}.
\end{eqnarray*}
% \comment{COMMENT $L_j$!}
Splitting  is achieved by sampling $U$
 from a distribution with support [-1,1] % and density $g(u)$
 and defining:\footnote{Specific  choices  for the distribution of $U$ are discussed in Appendix~\ref{RJdetails}.  For instance, sampling $U^2$  from a uniform distribution on  [0,1]  worked pretty well in many situation.}
\begin{eqnarray*} \label{prorjitAmain}
\load \newsp_{l_j,j} = U \sqrt{\idiov_{l_j}}   , \qquad (\idiov_{l_j}) \newsp= (1-U^2) \idiov_{l_j} .
\end{eqnarray*}
% All other coefficients of  $\deltav$ and $\facload$ remain unchainged.
  %
Given $\load \newsp_{l_j,j}$ and  $(\idiov_{l_j}) \newsp$, new factors $\fac_{jt} \newsp$ are
proposed for the spurious column $j$, independently for  $t=1, \ldots,T$,  from the conditional density $ p(\fac_{jt}\newsp| \facm_{t,-j},\facload_{{l_j},-j}^{\deltav}, \load \newsp_{l_j,j}, (\idiov_{l_j}) \newsp, y_{l_j,t})$ which takes a very  simple form (see Appendix~\ref{RJdetails} for details):
\begin{eqnarray*} \label{mainonfsp}
\fac_{jt} \newsp | \cdot  \sim \Normal{E_{jt} \newsp,V_{j} \newsp},  \quad
 V_j  \newsp =  1- U^2, \quad
\displaystyle  E_{jt}  \newsp=  U/\sqrt{\idiov_{l_j}} \times  % \tilde{y}_{l_j,t}  ,  &
 \left( y_{l_j,t}- \facload_{l_j,-j} \facm_{t,-j} \right).
\end{eqnarray*}
 By  reversing the split move, the  merge move   sets  the only nonzero factor loading  $\load \newsp_{l_j,j}$  in row $l_j$   of  a  spurious columns  $j$
 in (\ref{spurious2}) to zero, while increasing the idiosyncratic variance $\idiov_{l_j}$ at the same time. Deleting the spurious column   determines $\idiov_{l_j}$ and $U$  in the following way:
\begin{eqnarray*}
 \idiov_{l_j} =  (\load _{l_j,j}\newsp) ^2 + (\idiov_{l_j}) \newsp,\qquad
 U=\load _{l_j,j}\newsp / \sqrt{\load _{l_j,j}\newsp)^2 + (\idiov_{l_j}) \newsp}.
\end{eqnarray*}
% as well as  $\load  _{l_j,j}=0$   and  $\delta_{l_j,j}=0$.
Since column $j$ is turned into a zero column,  new factors  are proposed from the prior, i.e.  $\fac_{jt} \sim \Normal{0,1}$ for  all $t=1, \ldots,T$.

At each sweep of the MCMC scheme,  a decision has to be made  whether a split or  a merge move is performed.
Evidently, no merge  move can be performed,  whenever the current factor loading matrix contains no spurious columns. Similarly,  no split move can be performed, whenever no additional spurious columns can be introduced. This  happens if no more zero columns are present or if the number of spurious columns is equal to  $S$. Otherwise, split and merge move are  selected randomly, see   Appendix~\ref{RJdetails} which also contains details on the acceptance rates both for split and merge moves.

%Naturally, a split move is selected,  whenever the current factor loading matrix contains no spurious columns,  and a merge move is selected, whenever no further  spurious columns can be introduced.  \comment{This happens if no more zero columns are present or if the number of spurious columns is equal to  $S$.} Otherwise, a merge move is selected with  high probability equals $1-p_s$, where $p_s$ is a small probability that  yet another spurious column is introduced.

\subsubsection{Boosting MCMC}  \label{accelerate}

  Step~(F) and  Step~(P)  in Algorithm~\ref{Algo3} perform full conditional Gibbs sampling for a confirmatory factor model corresponding to the current indicator matrix $\deltav$, by sampling the factors  conditional on the loadings and idiosyncratic variances  and sampling  the loadings and idiosyncratic variances  conditional on the factors.  Depending on the signal-to-noise ratio of the latent variable representation,   such full conditional Gibbs sampling   tends to be poorly mixing.
For the basic  factor model  (\ref{fac1reg}), where  ${\facm}_t  \sim  \Normult{\nfac}{\bfz,\identy{\nfac}}$,
the information in the data (the \lq\lq signal\rq\rq ) can be
 quantified by  the matrix $\trans{\facload}  \Vare ^{-1}  \facload  $ in comparison to the identity matrix  $\identy{\nfac}$ (the \lq\lq noise\rq\rq ) in the filter for $\facm_t|\ym_t,\facload, \Vare$
(see Appendix~\ref{SectionF_fac}):
\begin{eqnarray*}
{\facm}_t
|\ym_t,  \facload, \Vare \sim \Normult{\nfac}{(\identy{\nfac} + \trans{\facload} \Vare ^{-1} \facload) ^{-1} \trans{\facload} \Vare ^{-1} \ym_t , (\identy{\nfac} + \trans{\facload}  \Vare ^{-1}  \facload) ^{-1} }.
\end{eqnarray*}
In particular for large
factor models with many measurements, one would expect that the data contain ample information to estimate the factors ${\facm}_t$. However, this is the case only,  if the information matrix $\trans{\facload}  \Vare ^{-1}  \facload $  increases with $\dimy$, hence if most of the factor loadings are nonzero. For sparse factor models  many columns with quite a few  zero loadings  are present, leading to a low signal-to-noise ratio and, as a consequence, to poor mixing of   full conditional Gibbs sampling, as illustrated in the left-hand panel in Figure~\ref{Boostadd} showing posterior draws of  $\trace{\trans{\facload}  \Vare ^{-1}  \facload }$  without boosting Step~(A)  for the exchange data to be discussed in Subsection~\ref{applicEx22}.

\begin{Figure3} %\begin{Figure3pdf}
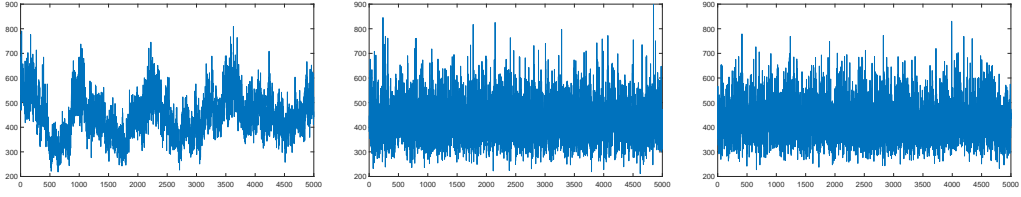
{Exchange rate data;  fractional prior with   $b=b_N$. Posterior draws of  $\trace{\trans{\facload}  \Vare ^{-1}  \facload}$ without boosting (left-hand side), boosting through ASIS based on choosing $\sqrt{\Psi_j}$ as the largest loading (in absolute values) in each nonzero column (middle) and boosting through MDA based on the inverted Gamma working prior $\Psi_j  \sim \Gammainv{1.5,1.5}$ (right-hand side).}{Boostadd}{ex22_noboost}{ex22_boost_asis}{ex22_boost_mda}{0.2}
\end{Figure3}  %\end{Figure3pdf}

Hence, for sparse factor models  it is essential to include boosting steps
to obtain  MCMC scheme with improved mixing properties, while keeping all priors unchanged.
Popular boosting algorithms  are  the ancillarity-suffiency interweaving strategy (ASIS), introduced by \citet{yu-men:cen}, and  marginal data augmentation (MDA), introduced by \citet{van-men:art}.\footnote{ASIS  has been applied to SV models \citep{kas-fru:anc}, TVP  models  \citep{bit-fru:ach}, and factor SV models \citep{kas-etal:eff};  MDA has been applied  to factor models by \citet{gho-dun:def,con-etal:bay,pia-pap:bay}.} %to  discrete choice models \citep{mcc-etal:bay} and
    There are numerous examples in the literature, where boosting enhances mixing at the cost of changing the prior, an example being  the MDA algorithm applied by \citet{gho-dun:def} to the basic factor model.  However, changing the prior of the factor loading matrix $\facload$ in the original model is undesirable in any  variable selection context and  is avoided by the boosting strategies applied in the present paper.

Both for ASIS and MDA, boosting is based on moving from  model  (\ref{fac1reg}) where  ${\facm}_t  \sim  \Normult{\nfac}{\bfz,\identy{\nfac}}$
% $\facm_t$  follow  prior  (\ref{fac2})
to an  expanded model
 with a more general prior:
\begin{eqnarray*}
 \ym_t =  \tilde{\facload}   \tilde{\facm}_t + \errorm_t, \quad \errorm_t \sim \Normult{\dimy}{\bfz,\Vare},  \qquad
  \tilde{\facm}_t   \sim  \Normult{\nfac}{\bfz,\Psiv},
\end{eqnarray*}
where $\Psiv=\Diag{\Psi_1,\ldots,\Psi_{\nfac}}$ is diagonal. The relation between the two systems is given by following transformation:
\begin{eqnarray}
% \tilde{\facm}_t = (\Psiv \old)^{1/2} \facm_t ,  \quad  \tilde{\facload} = \facload  (\Psiv \old)^{-1/2}. \label{fac5px}
 \tilde{\facm}_t = (\Psiv)^{1/2} \facm_t ,  \quad  \tilde{\facload} = \facload  (\Psiv)^{-1/2}. \label{fac5pxmain}
\end{eqnarray}
Note that the nonzero elements in  $\tilde{\facload}  $  have the same position as the nonzero elements in $\facload$.
 An important aspect of applying boosting in the context of  sparse Bayesian factor models is the following.
The transformation (\ref{fac5pxmain}) has to be a one-to-one mapping for any kind of boosting based on parameter expansion to be valid.
For sparse Bayesian factor models, this is true only for the {\em nonzero} columns of  $ \facload$, whereas for any  zero column $j$,   (\ref{fac5pxmain}) would be satisfied for arbitrary values $\Psi_{j}$  and many different expanded systems would map into the original system.\footnote{Applying a boosting step to an unobserved factor $f_{jt}$ has the undesirable effect  that the prior of  $f_{jt}$ is no longer a  normal distribution. Rather, it is a scale mixture of Gaussian distributions with the mixing distribution being equal to the distribution of  $\Psi_{j}$. For instance, if $\Psi_{j}$ follows an
inverted Gamma distribution as in marginal data augmentation, then moving to the expanded model by rescaling the  factors $f_{jt}$ for all $t$ would lead to a model where $f_{jt}$ follows a $t$-prior  rather than a normal distribution with scale $\Psi_{j}$.} Hence, we set  $\Psi_{j}=1$ for all zero columns of  $ \facload$ and, for nonzero columns $j$,  choose $\Psi_j $  in a deterministic fashion for ASIS   and  sample  $\Psi_j $ from a working prior for MDA.

For boosting based on  ASIS, a nonzero factor loading $\load_{n_j,j}$  is chosen in each nonzero  column $j$, to define
the current value of  $\Psi_j$ as $\sqrt{\Psi_j}=\load_{n_j,j}$. This creates a factor loading matrix $\tilde{\facload}$ in the expanded system where
for all  nonzero columns $j$,  $\tilde{\load}_{n_j,j}=1$  whereas $\tilde{\load}_{i,j}= \load_{ij}/\load_{n_j,j}$ for $i\neq n_j$.
 For  MDA,
 $\Psi_j$ is sampled from a working prior  $p(\Psi_j)$,
 which is independent  both  of $\facload$ and $\Vare$.   % , i.e. $\Psiv \sim p(\Psiv) $.
 Our assumption of prior independence  between the working parameter  $\Psiv$ and the remaining parameters   $\facload$ and $\Vare$  guarantees that the prior distribution of  $\facload$  remains unchanged, despite moving between the two models.
 For both boosting strategies, Step~(A) in Algorithm~\ref{Algo3} is implemented as described in detail in Algorithm~\ref{AlgoA} in Appendix~\ref{accelerate_App}.
 For  illustration, Figure~\ref{Boostadd} shows considerable efficiency gain in  the posterior draws of  $\trace{\trans{\facload}  \Vare ^{-1}  \facload }$    for the exchange data, when a boosting strategy is applied,   both for ASIS (middle panel) as well as  MDA (right-hand panel).

\subsection{Bayesian inference through postprocessing posterior draws} \label{exbaykrandom}

MCMC estimation through Algorithm~\ref{Algo3} delivers  draws from the posterior $ p(\deltav, \idiov_1, \ldots, \idiov_\dimy, \facload|\ym)$  that are not
identified in the strict sense discussed in Subsection~\ref{onefactro}.
The only quantity that can be inferred from the posteriors draws,  without caring at all about identification, is the marginal covariance matrix $\Vary=\facload \trans{\facload} +\Vare$. % Functionals of $\Vary$, such as the trace $\trace{\Vary}$,  the  (log) determinant $\log \Det{\Vary}$,  as well as  $ \trans{\unit{\dimy}} \Vary^{-1}  \unit{\dimy} $ can be useful to assess convergence of the MCMC sampler. %
 %
% \subsubsection{Verifying variance identification}
 %
  For  posterior inference beyond $\Vary$ such as estimating the number $\nfactrue$ of factors  and posterior identification of $\Vare$ and $\facloadtrue \trans{\facloadtrue}$, it  is  essential to consider only posterior draws for which
the variance decomposition is unique. While most papers ignore this important aspect,
variance identification for sparse Bayesian factor models is fully addressed  in the present paper during post-processing.
Due to the point-mass mixture prior employed in this paper, the posterior draws of
$\deltav$ contain valuable information both concerning the sparsity and  identifiability
of the factor loading matrix, as the  point-mass
 mixture prior allows  exact zeros in the factor loading matrix both apriori as well as aposteriori.

All posterior draws obtained from Algorithm~\ref{Algo3} are post-processed, to verify  if  the $\nfacr$ nonzero column $\tilde{\facload}$
of $\facload$ satisfy  the row-deletion property condition \AR\ with $r=\nfacr$.
  For draws with $\nfacr \leq 4$,  the simple counting rules outlined in Corollary~\ref{Lemma1} in  Appendix~\ref{simcount}  are applied. For draws with  $\nfacr > 4$, a very efficient procedure is applied that  derives a block diagonal representation
as in Theorem~\ref{Lemma2} for $\tilde{\facload}$  sequentially  and applies the \CountAR\  rule  to  the corresponding subblocks,
%Based on the results of Subsection~\ref{varidesp},
 see  Algorithm~\ref{algARIDE}  in  Appendix~\ref{verpartbig} for more details.
 Any further Bayesian inference is  performed for the  $M_V$ variance identified draws, only.
 % An important number  in this context is   the fraction $p_V$ of variance identified draws  among all $M$ posterior draws. If $p_V$ is very small, then it should also be verified, e.g. by simulations, that the shrinkage prior used for $\deltav$ does not put too much  prior mass  on indicators favouring unidentified models. %, see SFS (2017b) for more details.

 \subsubsection{Identification of the number of factors $r$} \label{estr}

 Given posterior draws of $\facload$ and  $\deltav$, the challenge is to estimate the number of factors $\nfactrue$, if the model is overfitting.
 A common procedure  to identify
  the number of factor  is to apply
   an incremental procedure, by increasing $\nfac$ step by step, and to use model selection  criteria such as  information criteria \citep{bai-ng:det2002}
  or Bayes factors \citep{lee-son:bay,lop-wes:bay}  to choose the number of factors.

Alternatively, a number of authors suggested to estimate  the number of factors in one sweep together with the parameters.
  \citet{car-etal:hig}, for instance,   infer $\nfactrue$ from
   the  columns from $\deltav$, after removing
   columns with a few nonzero elements in a heuristic manner.
   \citet{bha-dun:spa} employ a procedure which increasingly shrinks factor loadings toward zero
  with increasing column number. The number of factors is changed during sampling by setting
  an entire column of the loading matrix to zero, if all factor loadings are close to 0.
  \citet{kau-sch:bay} estimate a sparse dynamic factor model with an increasing number $\nfac$
  of potential factors and use so-called \lq\lq extracted factor representation\rq\rq\ during MCMC post-processing procedure to select the number of factors.

 However, any such heuristic method of inferring the number of factors from the nonzero columns from $\deltav$ in an overfitting model without checking uniqueness of variance decomposition  is  prone to be biased. Instead, our procedure relies on the  mathematically justified representation of
 the loading matrix $\facload$ in an overfitting factor model given by Theorem~\ref{theoverGLT} and provides a new, non-incremental approach for selecting the number of factors.
   We identify $\nfactrue$  through a   one-sweep MCMC procedure  which is based on  purposefully overfitting the number
 $\nfac$ of potential factors  within the framework of sparse Bayesian factor analysis as implemented above.
A related strategy was also applied in  \citet{con-etal:bay} within the framework of dedicated Bayesian Factor analysis.

 Evidently,  zero columns (if any) in $\facload$ can be removed,  since   $\facload   \trans{\facload }=\betatilde  \trans{\betatilde}$, where $\betatilde$ contains the $\nfacr$  nonzero columns of $\facload$.
As outlined in Section~\ref{secover}, the number $\nfacr$ of nonzero columns is the equal to the number of factors $\nfactrue$,  if  the variance decomposition is unique for $\nfactrue=\nfacr$.
  This is no longer true, if uniqueness of the variance decomposition  does not hold for $\nfactrue=\nfacr$.  In an overfitting factor model with $\nfac>\nfactrue$, many draws with   $\nfacr$ nonzero columns will have a representation as in   Theorem~\ref{theoverGLT} and  contain  a submatrix $\Mm $ with  $s$ spurious columns, each of which has exactly one nonzero element. Hence, these draws violate even the most simple condition for variance identification.    For such posterior draws   $\betatilde$,  $\nfacr$ overestimates $\nfactrue$ since,  according to Theorem~\ref{theoverGLT},    $\nfacr=\nfactrue+s$, or equivalently:
% \begin{eqnarray} \label{NNFAC}
$\nfactrue = \nfacr-s$.
% \end{eqnarray}
 Hence,  methods of inferring the number of factors from the nonzero columns $\nfacr$  of the unconstrained posterior draws $\deltav$ in an overfitting factor model with $\nfac>\nfactrue$   are prone to overestimate the number of factors,  in particular, if  many draws violate  simple conditions for variance identification.
%
%Hence, $ \nfacr$ typically overestimates the number the factors $\nfactrue$ in an overfitting factor model which  can  be estimated from (\ref{NNFAC}) as the number of nonzero columns  minus the number of spurious columns,   \comment{ADD? provided that the remaining, non-spurious columns $ \facloadtrue$ obey condition {\bf TS} with (fixed)   maximum degree of overfitting $S$.}

  As opposed to this, we   rely  on uniqueness of variance decomposition and discard draws from the posterior  sample
   that violate  uniqueness of the variance decomposition for $\nfactrue=\nfacr$. %, see  Subsection~\ref{subGL}.
 %
 %Most importantly, identification of $\Vare$ implies identification of  $\facload \trans{\facload}=\betatilde \trans{\betatilde}$. Since $\betatilde= \facloadtrue \Pm $ is one of possibly  many permutations of $\facloadtrue$,  also $  \facloadtrue \trans{\facloadtrue} = \betatilde \trans{\betatilde} $ is identified.
For the remaining draws, the %rank  of $\facload \trans{\facload}$, or more or less equivalently, the
number $\nfacr$   of nonzero columns of  $\betatilde$ can be considered as a posterior draw of the number
of factors  $\nfactrue$.   The entire (marginal) posterior distribution $p(\nfacr|\ym)$ can be estimated
from these draws,  using the empirical pdf of the sampled values for $\nfacr$.
The posterior mode  $\tilde{\nfactrue}$ of $p(\nfacr|\ym)$
 provides a point  estimator of the number of factors $\nfactrue$.
 This inference is valid,  even if  the rotation problem for $\facload$ is not solved,
    as only  uniqueness of the variance decomposition is essential.

     It should be noted that point mass mixture priors are particularly
 useful in identifying  spurious factors, since these priors are able to identify exact zeros in the columns
 corresponding to spurious factors.
Under continuous shrinkage priors, see  e.g. \citet{bha-dun:spa,roc-geo:fas}, it is not straightforward, how
to identify spurious factors.

\subsubsection{Further inference for unordered variance  identified GLT draws}

In addition to estimating the number of factors as in  Subsection~\ref{estr},  further Bayesian inference can be performed for the  $M_V$ variance identified draws without resolving trivial rotation.
Evidently,  %(regardless whether the rotation issue has been addressed)
 posterior inference is possible for all idiosyncratic variances  $\idiov_1, \ldots, \idiov_\dimy$ in $\Vare$.
 Functionals of $\Vare$, such as the trace  of $\Vare$ and $\Vare^{-1}$ as well as  the (log)
determinant  of $\Vare$ are useful means of assessing convergence of the MCMC sampler. Furthermore, for each variable  $y_{it}$ inference with respect to  the proportion of the variance
 explained by the common factors (also known as communalities $R^2_i$) is possible:
  \begin{eqnarray}
R^2_i =   \sum_{j=1}^{\nfac} R_{ij}^2,  %\load_{ij}^2 }{\sum_{j=1}^{\nfac} \load_{ij}^2+\idiov_i}.
\qquad  R_{ij}^2 = \frac{\loadtrue_{ij}^2 }{\sum_{l=1}^{\nfactrue} \loadtrue_{il}^2+\idiov_i} .
 \label{faccum}
\end{eqnarray}
 %\paragraph*{Irrelevant variables and overall sparsity.}
In addition, due to Lemma~\ref{theirr}, irrelevant variables can be identified through the position of
 zero rows. This allows to estimate the (marginal) posterior probability $\Prob{q_i=0|\ym}$ for all variables $y_{it}$ by counting the frequency of the event $q_i=\sum_{j=1}^\nfac \delta_{ij}= 0$ during MCMC
  sampling for each row $i=1,\ldots,\dimy$.
Finally,  overall sparsity  in terms of the number $d$ of nonzero elements in $\deltav$,
   \begin{eqnarray} \label{modeddd}
d = \sum_{j=1}^\nfac  \sum_{i=1}^\dimy \delta_{ij},
 \end{eqnarray}
 can be evaluated. % for each (variance identified)  draw  $\deltav$.
 Posterior draws  of $d$ are particularly  useful to check convergence and assessing efficiency of the MCMC sampler, as $d$ captures the ability of the sampler to move across (variance identified) factor models of different dimensions.

\subsubsection{Resolving trivial rotation issues}  \label{subGL}

  For  all unordered GLT draws $\betatilde$ that are variance identified,
 the  factor loading matrix $\facloadtrue$  and the corresponding indicator matrix $\deltavlam$
 % including the leading indices $\lm=(l_1, \ldots , l_\nfactrue)$
are uniquely identified  from the $\nfactrue$ nonzero columns  $\betatilde$ and  $\deltavtilde$ of  $ \facload$ and the corresponding indicator matrix
$\deltav$ by Theorem~\ref{theGLT}.
Since the MCMC draws   $\betatilde  $ and  $\deltavtilde $
% $\betatilde  = \facloadtrue   \bP _{\pm}'  \bP _{\rho} ' $  and  $\deltavtilde  = \deltavlam \bP _{\rho}'$
 are trivial rotations
of  $\facloadtrue$ and $\deltavlam$, column and  sign switching are  easily resolved. % in a post-processing manner
    %to derive  $\facloadtrue$  and of  from  $\betatilde$ and$\deltavtilde$.
First, % column  switching  is resolved  by reordering
  the columns of $\deltavtilde$ are ordered such that the leading indices $\lm=(l_1, \ldots , l_\nfactrue)$
  obey $l_1 < \ldots < l_\nfactrue$; i.e. $\deltavlam =\deltavtilde \bP _{\rho}$.
   Then,   the sign of the entire column $j$ of $\betatilde \bP _{\rho}$ is switched if the leading element
   is negative; i.e.   $\facloadtrue= \betatilde    \bP _{\rho} \bP _{\pm}  $.
  In addition,  the  factors $\tilde{\facm}_t$  corresponding to the nonzero columns of  $\deltav$ are reordered
  through $ \trans{\bP _{\pm}} \trans{\bP _{\rho}} \tilde{\facm}_t$ for  $t=1,\ldots,T$.
  Finally, $\bP _{\rho}$ is also used to reorder the draws of  the hyperparameter $\hypv$ of the prior $p(\deltav)$.

  % through  $\bP _{\rho} (\tau_1, \ldots, \tau_\nfactrue)' $.
  %%

  The draws of  $(\facloadtrue, \deltavlam)$  are exploited in various ways.
  Their leading indices $l_1, \ldots , l_\nfactrue$  are draws  from  the marginal
  posterior distribution $p(l_1, \ldots , l_\nfactrue|\ym)$  allowing
 posterior inference w.r.t to $\lm$.   In particular,  the identifiability constraint
    $\lm ^\star=(l_1^\star, \ldots, l_{\rstar}^\star)$ visited most often is determined
     together with its frequency $p_{L}$ which reflects posterior uncertainty with respect to
     choosing the leading indices.  The number  $\rstar$ of elements in $ \lm ^\star$ provide yet another estimator of the number of factors.
     Furthermore, the
    highest probability model (HPM),~i.e. the indicator matrix  $\deltavlam_H$ visited most often,
    its frequency $p_{H}$  (an estimator of the posterior probability of the HPM), its model size $d_H$,  and its leading indices $\lm _H$  are  of interest, and whether  $\lm _H$  coincides with $\lm ^\star$.

      Bayesian inference  with respect to the loading matrix  $\facloadtrue$
     is performed conditional on $\lm^\star$, to avoid switches between
  different  leading indices.
  % and is based on all      (variance identified) posterior draws where  the leading indices  of $\facloadtrue$ coincide with $\lm ^\star$.
    %
 Averaging over  the corresponding $M_V p_{L} $
 MCMC draws  provides an estimate  of  $\facloadtrue$
%the idiosyncratic variances and the communalities are obtained by averaging over all (variance identified) posterior  draws where the leading indices of $\deltavlam$  coincides with $\lm ^\star$.
and the marginal inclusion probabilities
$\Prob{\delta^\Lambda_{ij}=1|\ym, \lm ^\star}$ for all elements of the corresponding indicator matrix.  Also,
     the median probability model (MPM) $\deltavlam_M$,
      obtained by setting each indicator to one
   %   $\hat{\delta}^\Lambda_{ij} = 1$
whenever $\Prob{\delta_{ij}^\Lambda=1|\ym, \lm ^\star}\geq 0.5$,
%\footnote{Note that the leading indices of $\deltavlam_M$ are equal to $\lm ^\star$ for all draws.}
  and its model size $d_M$ are of interest.

  \section{Applications}  \label{secalpp}

All computations are based on the designer MCMC algorithm introduced in Algorithm~\ref{Algo3},  with boosting in Step~(A)
being based on ASIS with choosing $\sqrt{\Psi_j}$ as the largest loading (in absolute values) in each nonzero column
 (see  Appendix~\ref{accelerate_App}),   choosing  $U^2 \sim \Betadis{3, 1.5}$  as proposal $g(u)$  in Step~(R)  % (which implies a mode at $ \pm 0.9$) as well as $p_0=p_s=0.5$
 (see Appendix~\ref{RJdetails})
and  choosing $\pshift = p_{\mbox{\rm \footnotesize switch}}=1/3, p_a=0.5$ in Step~(L)  (see Appendix~\ref{updatelead}).

  \subsection{Sparse factor analysis for exchange rate data}  \label{applicEx22}

To analyze exchange rates with respect to the Euro,  data was obtained from
the European Central Bank’s Statistical Data Warehouse and ranges from January 3, 2000
to December 3, 2007. It contains $m = 22$  exchange rates listed in Table~\ref{abbrev}  from which we derived $T=96$ monthly returns, based on the first trading day in a month. The data are demeaned and standardized.\footnote{A similar set of  exchange rates (however with daily returns) was studied in \citet{kas-etal:eff}.}

\begin{Tabelle}{Currency abbreviations.}{abbrev}
{  \small \begin{tabular}{cccc}
\begin{tabular}{rllrlrl}
  \hline
1 &   AUD & Australia dollar  \\
2 &   CAD & Canada dollar \\
3&  CHF & Switzerland franc \\
4 &  CZK & Czech R.\ koruna \\
5  &DKK & Denmark krone \\
 6 & GBP & UK pound \\
7  & HKD & Hong Kong dollar \\
8 &IDR & Indonesia rupiah \\
9 &JPY & Japan yen \\
10 &KRW & South Korea won \\
11  & MXN& Mexican Peso\\  \hline
\end{tabular} &&
\begin{tabular}{rrlrlrl} \hline
12  &MYR & Malaysia ringgit \\
13  &NOK & Norway krone \\
 14 &NZD & New Zealand dollar \\
 15 &PHP & Philippines peso \\
 16 &PLN & Poland zloty \\
 17 &RON & Romania fourth leu \\
 18  &RUB & Russian ruble \\
 19 &SEK & Sweden krona \\
 20 &SGD & Singapore dollar \\
 21 &THB & Thailand baht \\
22  &USD & US dollar \\  \hline
\end{tabular} \end{tabular}
}
\end{Tabelle}

% The data is displayed in Figure~\ref{dataexchange} in Appendix~\ref{data}.

Since the number of factors is unknown, an overfitting factor model is applied with  maximum degree of overfitting $S=3$ and
 the maximum number of  factors $\nfac=9$ obeying  inequality (\ref{kbound_extend}).
  The  hyperparameter $b_0$ of the prior  (\ref{prigen}) for the indicators   is  chosen as  $b_0=0.6$, while $a_0=0.1714$  is chosen such that
a prior simplicity of $ \Ew{q_i}=2$  is achieved. This implies   $\alpha=2.57$  in the parameterization
 (\ref{prialt}). This  prior  introduces column sparsity, see  the corresponding prior distributions $p(\nfacr)$ for the
number of nonzero columns reported in Table~\ref{ken_tab1},  with most of  the prior mass being considerably
smaller than $\nfac=9$.\footnote{This prior distributions was determined by simulating $\dimmat{\dimy}{\nfac}$ indicator matrices $\deltav$ from the prior
(\ref{prigen}), restricted to GLT structures, and rejecting all draws that did not fulfill  condition \AR\ for the $\nfactrue=\nfacr$ nonzero columns.}

%\begin{Tabelle}{Prior distribution $p(\nfacr)$ of the number on nonzero columns $\nfacr$ under prior (\ref{prialt}) with $k=10, a_0=0.15, b_0=0.6$ obtained by simulation; draws that did not satisfy the condition \AR\ for variance identification with  $\nfactrue=\nfacr$ were rejected. The fraction of identified draws is equal to $p_V=0.25$ apriori.}{priorr}
%{\begin{tabular} {ccccccccccc}\hline
% $ \nfacr$  &        0 &  1          &      2         & 3      & 4    & 5       & 6      & 7 & 8 - 10 & $p_V$ (in percent) \\ \hline
%$p(\nfacr)$  &  $<$0.01  &  0.04 &    0.12  &   0.22 &    0.25  &  0.21 &    0.11  &   0.04  & $<$0.01 %&$<$0.01& $<$0.01 &   25\% \\
% \hline \end{tabular}} \end{Tabelle}

The prior   (\ref{priorsiid})  on the idiosyncratic variances  is selected % to avoid a Heywood problem,
with  $c_0=2.5$ and  $\widehat{\Vary^{-1}}$ being estimated  from (\ref{Varyhat}) with $\nu_o =3 $ and   $\Sm_o=\identy{\dimy}$.
To  study sensitivity to further prior choices,  we consider   fractional priors  (\ref{priorfrac})
with   %$b=10^{-5} < b_R < b=10^{-4} < b_N < b=10^{-3}$.
$b=10^{-5}, b_R, 10^{-4}, b_N, 10^{-3}$. %, where  $b_N=4.735\cdot10^{-4} $ and $b_R =3.265 \cdot 10^{-5} $
Since $d(\nfac,\dimy)=175 << N=2112$, choosing $b_N $ is the recommended choice.  % Obviously, $b_R $ implies a much stronger penalty than $b_N$, since  $d(\nfac,\dimy) > \sqrt{T\dimy}=46$.
   In addition,  the standard prior  (\ref{prior1})  is considered with
    $\bV_{i0}^{\deltav}=\identm$, $c_0=1.1$ and   $C_{i0} \equiv  0.055$ \citep{lop-wes:bay}. %  and $c_0=1$ and  $C_{i0}=0.2$  \citep[GD]{gho-dun:def}.

\begin{Tabelle}{Exchange rate data;  Bayesian inference for an overfitting factors model with  $k=9$.
The first row shows the prior distribution $p(\nfacr)$ of the number on nonzero columns $\nfacr$ under prior (\ref{prialt}) with $\Ew{q_i}=2$ and  $b_0=0.6$. % obtained by simulation. %; draws that did not satisfy the condition \AR\ for variance identification with  $\nfactrue=\nfacr$ were rejected.
 The upper part shows the posterior distribution $p(\nfacr|\ym)$ of
 $\nfacr$  (bold number corresponds to the posterior mode $\tilde{r}$) for various  fractional  priors with different fractions $b$ ($b_N=4.735\cdot10^{-4} $, $b_R =3.265 \cdot 10^{-5} $) and the prior of \citet{lop-wes:bay} (LW) using only  draws  satisfying \AR\ ($p_V= M_V/M$ is the corresponding fraction). The lower part shows  the posterior distribution $p(\nfacr|\ym)$ of
 $\nfacr$ without imposing
 variance identification.  Probabilities  smaller
than $<$$10^{-2}$ are indicated by $\approx 0$.}{ken_tab1}
{ \small \begin{tabular}{lccccccccc} \hline
 &  \multicolumn{8}{c}{$\nfacr$} &   \\ \cline{2-9}
&   0-1& 2  & 3 & 4 & 5 & 6 & 7 & 8-  9&  $100 \cdot p_V$  \\  \hline
$p(\nfacr)$ &       0.0434   % 0.0058    0.0376
                           &    0.112 &    0.231 &   0.2642 &   0.1996  &  0.1106 &   0.0336  & 0.0054   %  0.0047 &   0.0007
                            &  27.4 \\ \hline
  $p(\nfacr|\ym)$     &&  & & & & &  & &     \\ \cline{1-1}
  $b=10^{-5}$                           &      0&  0 &   \textbf{0.96}  &   0.04        &  0&  0&  0&  0        & 58.2 \\
   $b=b_R$                          &     0     &    0   & 0.36   &  \textbf{0.63}     &  $\approx 0$   &     0 &  0 &0           & 74.1 \\
   $b=10^{-4}$                       &      0      &   0 &   0.04 &   \textbf{0.95}  &   $\approx 0$ &  0  & 0&  0 &  80.6  \\
       $b= b_N$                         &     0    &     0    & $\approx 0$ &   \textbf{0.88}  &  0.11   & $\approx 0$   &0   &0   & 58.9  \\
      $b=10^{-3}$                       &      0     &    0   &      0   & \textbf{0.63}&    0.34  &  0.02 &  $\approx 0$   &    0     &   42.9  \\
      LW    &     0 &  0 &0     &  $\approx 0$  &   0.19 &  \textbf{ 0.47} & 0.29   &0.05       &22.1 \\
\hline
 no varide     &&  & & & & &  & &     \\ \cline{1-1}
 $b=10^{-5}$  &      0&  0 &   0.89  &  0.10  &  $\approx 0$  &  0&  0&  0  &  \\
  $b=b_R$ &  0    &     0  &  0.40   & 0.54  &  0.06  &  $\approx 0$ &    0& 0 &      \\
 $b=10^{-4}$  &     0    &    0  &  0.04   & 0.80   & 0.15 &  $\approx 0$ & $\approx 0$ & 0  &    \\
    $b= b_N$  &   0   &  0 &   $\approx 0$ &   0.54  &  0.38  &  0.08  & $\approx 0$   &$\approx 0$ &    \\
  $b=10^{-3}$  &     0      &   0    &     0  &  0.28   & 0.44   & 0.23   & 0.05    & $\approx 0$     &   \\
LW &    0 &  0 &0     &    $\approx 0$   & 0.05 &   0.27   & 0.43   &0.24  &   \\
\hline
\end{tabular}}
\end{Tabelle}

%The  designer MCMC scheme outlined in
Algorithm~\ref{Algo3} is run for $M=100,000$ draws
after a burn-in  of $M_0=50,000$ draws. %\footcomment{Currently, $M=50.000$ draws after a burn-in phase of $M_0=50.000$. Rerun.}
To verify convergence, independent MCMC chains were started respectively with $\nfacr^{(0)}=2$ and
$\nfacr^{(0)}=9$  nonzero columns.
As discussed in Subsection~\ref{mcmc}, this sampler navigates in
the space of all unordered GLT structures with an unknown number of nonzero columns and unknown leading indices, without
forcing variance identification. Apart from $\Vary$ no further parameters are identifiable from the unrestricted draws, and
as outlined in Subsection~\ref{exbaykrandom},
 we screen for variance identified draws during post-processing. The  fraction $p_V$ %$p_V=M_V/M$
 of  variance identified  draws is reasonably high, as reported in
Table~\ref{ken_tab1} for each prior.

We  use only variance identified draws for further inference. Most importantly, for these draws
 the number $\nfacr$ of nonzero columns of $\deltav$ may be regarded as draws of the number $\nfactrue$ of factors.
 Table~\ref{ken_tab1} reports the posterior distribution $p(\nfacr|\ym)$  for all  priors under investigation and the left-hand side of Figure~\ref{fig_ken_0} shows  posterior draws of  $\nfacr$  for the fractional prior  $b=b_N$ for  illustration.
  All fractional priors based on $b=10^{-3}, 10^{-4},b_R, b_N$ point at a four factor
solution.   The fractional prior with $b=10^{-5}$ introduces too strong shrinkage leading to a three factor model, whereas
  the standard prior of  \citet{lop-wes:bay}  leads to an overfitting model with six factors.

%  \comment{Adress no varide?}

 Our designer  MCMC scheme shows good mixing across   models of different dimension, as illustrated by  Figure~\ref{fig_ken_0} showing  posterior
 draws of $\nfacr$   and the model size $d$ for the fractional  prior $b=b_N$, with an  inefficiency factor of  roughly 8 for $d$.
 This good behaviour  is particularly due to the RJMCMC Step~(R) in Algorithm~\ref{Algo3}, which has an acceptance rate of  18.9\% for a split and 30.8\% for a merge move.

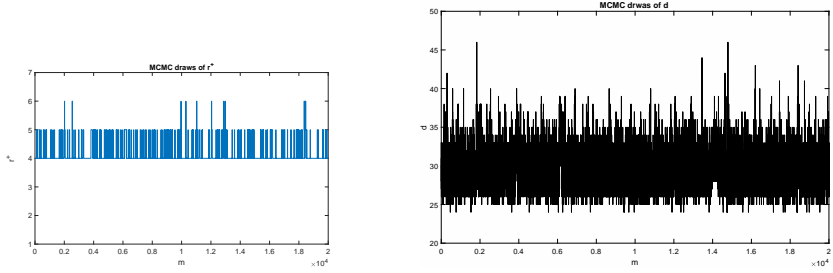
\begin{Figure2}{Exchange rate data;  fractional prior with   $b=b_N$. Posterior draws  of the number $\nfacr$ of nonzero columns (left-hand side) and  model size $d$ (right-hand side). The figure shows the last 20,000  among all % ($M_V=58,900$ )
variance identified draws.}{fig_ken_0}{mcmc_fac}{ex22_mcmc_m}{0.2} \end{Figure2}

% \begin{Figure2}{Exchange rate data;  fractional prior with   $b=b_N$. Posterior draws  of the number $\nfacr$ of nonzero columns (left hand side) and  model size $d$ (right hand side); the figure shows the last 20,000, among all $M_v=88,339$ variance identified draws.}{fig_ken_0}{mcmc_fac}{ex22_mcmc_m}{0.2}  \end{Figure2}

As outlined in Subsection~\ref{exbaykrandom}, the variance identified draws can be post-processed further.
For instance,  it is possible to investigate, if some  measurements are uncorrelated with the remaining measurements. This is investigated in Table~\ref{uncorex}
 through the posterior probability  $\Prob{q_i=0|\ym}$, where  $q_i$ is the row sum of  $\deltav$.
Various currencies  appear to be uncorrelated with the rest, namely
 Swiss franc (CHF), Czech koruna (CZK),    the Mexican peso (MXN),  the New Zealand dollar (NZD ), the Romania fourth leu (RON), and the  Russian ruble (RUB).

   \begin{Tabelle}{Exchange rate data;  posterior probability of the event $\Prob{q_i=0|\ym}$, where  $q_i$ is the row sum of  $\deltav$  for various exchange rates.}{uncorex}
{\small \begin{tabular}{lccccccc} \hline
% & & \multicolumn{7}{c}{$i$} \\
%&  & \cline{1-7} \\
          & \multicolumn{7}{c}{$\Prob{q_i=0|\ym}$}\\ \cline{2-8}
  % $i$ & 3  & 4 & 11& 14 &  17& 18 & remaining  \\
   Currency & CHF  & CZK & MXN& NZD & RON& RUB & remaining  \\
  \hline %1-2, 5-10, 12-13, 15-16, 19-22  \\  \hline
%  & &      \\ \hline
 $b=10^{-5}$  &  0.98  &  0.95  &  0.98  &  0.85  &   0.97 &   0.97        &  0   \\
 $b=b_R$ &    0.96  &  0.89   & 0.94    & 0.75  &   0.90   & 0.91          &  0   \\
$b=10^{-4}$  &    0.93   & 0.82   & 0.91   & 0.68    & 0.78   &  0.81     &  0    \\
 $b= b_N$  &       0.83   & 0.59    & 0.78   &  0.44  &  0.56   & 0.51         &  0  \\
  $b=10^{-3}$  &   0.28 &   0.64  &   0.35   &  0.76  & 0.59 &   0.73     &  0    \\
   \hline
 LW &   0.14   &  0.01  &   0.11  &  0.01  & 0.02  &  $\approx 0$   & 0  \\
\hline
\end{tabular}}
\end{Tabelle}

   \begin{Tabelle}{Bayesian inference  under the GLT structures with unknown
number of factors and unknown leading indices  (posterior draws of  $\lm=(l_1,\ldots,\l_r)$  ordered by size), based on the $M_V$
variance identified draws. Posterior mode estimator $\tilde{r}$ of
the number of factors; posterior expectation
  $\hat{d}=\Ew{d|\ym}$ of the model size $d$;  total number of visited models $N_v$;
 frequency  $p_H$ (in percent),  leading indices $\lm _H$ and model size $d_H$ of the HPM; %  $\deltavlam_H$ (model visited most often);
 leading indices $\lm ^\star$ visited most often,  corresponding frequency  $p_L$ (in percent) and correspding number of factors $\rstar$; model size $d_M$ of the MPM . % $\deltavlam_M$ (median probability model).
 }{ken_tab3}
{\small \begin{tabular}{lcccccccccc}
 \hline Prior              &  $\tilde{r}$ &$\hat{d}$ & $N_v$   &   $100p_H$  & $\lm_H$ & $d_H$  & $\lm^{\star}$ &  $100p_L$  & $\rstar$ & $d_M$  \\  \hline
$b=10^{-5}$                & 3  & 21  & 2709  & 42.8 &  (1,2,5) &  20& (1,2,5) &  88.5 &3 & 20 \\
$b=b_R $    & 4  & 24  & 10809  &  10.6 & (1,2,5,7) & 20 & (1,2,5,7) & 49.7 & 4  & 20 \\
$b=10^{-4}$    & 4 &  27  &19198  & 11.9 & (1,2,5,7) & 26& (1,2,5,7)& 85.5 & 4 & 26  \\
 $b=b_N$  &4  &29  & 42906 & 2.9 & (1,2,5,7) & 26 & (1,2,5,7) &  65.3 & 4  & 26 \\
 $b=10^{-3}$      &  4& 32 &  50920 & 0.5 & (1,2,5,7) & 26  & (1,2,5,7) & 37.3 & 4 &   27 \\
\hline
  LW & 6 &  59 & 32921  & 0.01 & (1,2,3,4,5,6) & 56 & (1,2,3,4,5,6)  & 11.2& 6  & 52\\
\hline
\end{tabular}}
\end{Tabelle}

Further Bayesian inference   is reported in Table~\ref{ken_tab3}, including
the  posterior mode  estimator $\tilde{r}$, the posterior mean $\hat{d}$ of the model size $d$ defined in (\ref{modeddd}),
 the total number $N_v$ of visited GLT structures,   the identifiability constraint
    $\lm ^\star=(l_1^\star, \ldots, l_{\rstar}^\star)$ visited most often  together with its frequency $p_{L}$ (in percent),
 as well as  the  frequency  $p_H$ (in percent),  the leading indices $\lm _H$ and model size $d_H$ of the highest probability model (HPM)  $\deltavlam_H$.  % (model visited most often).
  For all priors, $\lm ^\star$ coincides with   $\lm _H$.
       For all 4-factor models, the GLT constraint $\lm ^\star=(1,2,5,7)$ turns out to be the
most likely constraint, whereas for the 3-factor models the GLT constraints
$\lm ^\star=(1,2,5)$  is preferred.
Once more we find that   a standard prior as in   \citet{lop-wes:bay}  leads to an overfitting model both in terms of the  factors as well in terms of the model size. Too many models are visited, leading to a very small posterior probability $p_H$ for the HPM.

As a  final step, the factor loadings $\facloadtrue$ and the MPM are identified for a 4-factor model.
This inference  is based on all posterior draws where the
 leading indices of $\deltav$  (after reordering) coincide with the GLT constraint
  $\lm ^\star=(1, 2,5,7)$.   From these draws,  the marginal inclusion probabilities
$\Prob{\delta_{ij}=1|\ym, \lm ^\star}$ and the corresponding median probability model (MPM)  are derived.   Its  model size $d_M$  is   reported in Table~\ref{ken_tab3} for all priors.

For most fractional priors,  the HPM and  the MPM coincide.  Table~\ref{ken_mpm_tab2} reports   the  marginal inclusion probabilities
$\Prob{\delta_{ij}=1|\ym, \lm ^\star}$ for the  fractional prior $b=b_N$ and
Figure~\ref{fig_deltaplot} displays  both models  for illustration. The resulting model indicates considerable sparsity, with  many factor loadings being shrunk toward zero.  Factor~2 is a  common factor  among  the correlated currencies, while the remaining factors are three group specific, for the most part dedicated  factors.

%\begin{Tabelle}{Indicator matrix $\deltav$ visited most often among the  $M_V= XXXX$ variance identified draws  for the fractional prior  $b=b_N$  (number of factors unknown).}{ken_mpm}
% \begin{tabular}{cccccccccccccccccccccc}
  %   1   &  1   &  \bf{0}  &   \bf{0}  &  0 &    0    & 0 &    0  &   0   &  0   &   \bf{0}  &   1   &  0    &  \bf{0}   &  0  &   0  &   \bf{0}   &  \bf{0}  &    0 &0&     0   &  0 \\
  %   0   &  1   &   \bf{0}  &    \bf{0}  &   1  &   1   &  1  &   1  &   1  &   1   &   \bf{0}  &   0  &   1  &    \bf{0}  &   1   &  1  &   \bf{0}   &   \bf{0}   &   1& 1     &     1 &    1\\
   %  0   &  0   &   \bf{0}  &    \bf{0}&    1   &  0  &   1  &   0 &    0 &    0    &  \bf{0}  &   0   &  0 &     \bf{0}  &   1  &   0  &   \bf{0}   &  \bf{0}   &   0 &0 &    0  &   1\\
   %  0   &  0  &    \bf{0}  &   \bf{0} &    0    & 0 &    1  &   1   &  0   &  0    &  \bf{0}  &   0  &   0   &   \bf{0}  &   0  &   0  &   \bf{0}   &   \bf{0}   &   1& 1&     0  &   0  \\
   %  \end{tabular}}\end{Tabelle}

\begin{Tabelle}{Inclusion probabilities for the indicator matrix $\deltav$ for the fractional prior  $b=b_N$  averaged over  the  variance identified draws
with  $\lm ^\star= (1,2,5,7) $ (leading indices $\lm=(l_1,l_2,l_3,l_4)$ ordered by size).}{ken_mpm_tab2}
{  \small \begin{tabular}{ccccc}  \hline
 Currency & Factor~1 & Factor~2 & Factor~3 & Factor~4   \\ \hline
   AUD &       1 &      0 &       0 &       0 \\
 CAD   &       1 &      1 &       0 &       0 \\
     CHF &  0.01  &   0.12 &       0 &       0 \\
    CZK &  0.01 &   0.21 &       0 &       0 \\
    DKK &   0.02 &      1 &       1 &       0 \\
    GBP &   0.07 &      1 &    0.05 &       0 \\
    HKD  &   0.01 &      1 &    0.97 &       1 \\
   IDR  &   0.04 &      1 &   0.03 &       1 \\
   JPY &    0.13 &      1 &   0.01 &   0.02 \\
    KRW  &  0.01 &      1 &   0.06  &   0.02 \\
    MXN &  0.01 &   0.16 &   0.01 &   0.01 \\
    MYR  &       1 &  0.06 &  0.01 &  0.01 \\
    NOK  &  0.01 &      1 &   0.01 &   0.02 \\
    NZD  &   0.09 &   0.42 &   0.04 &   0.01 \\
    PHP &  0.01 &      1 &    0.95 &   0.04 \\
    PLN &  0.01 &      1 &   0.02 &    0.73 \\
    RON &    0.14 &  0.06  &    0.24 &  0.01 \\
    RUB &    0.27 &   0.11 &    0.09 &    0.16 \\
    SEK &  0.01 &      1 &   0.01  &    0.99 \\
   SGD  &   0.03 &      1 &   0.03 &    0.99 \\
   THB  &   0.01 &      1 &  0.01 &   0.02 \\
    USD &   0.02 &      1 &       1 &   0.01 \\
     \hline
     \end{tabular}
}
\end{Tabelle}

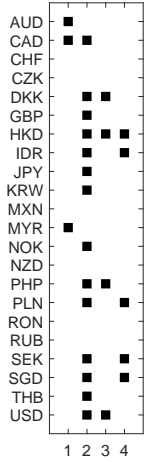
\begin{Figure}{Exchange rate data;   indicator matrix  $\deltav$ corresponding both to the HPM and the MPM for a fractional prior with   $b=b_N$. The number of estimated factors is equal to 4.}{fig_deltaplot}{fig_delta}{0.5} \end{Figure}

Finally, Table~\ref{ken_fac4} shows the posterior mean of the factor loading matrix,
the idiosyncratic variances and the communalities, obtained by averaging over all draws
where the leading indices of $\deltav$  coincide with $\lm ^\star$.
Sign switching in the posterior draws of $\facloadtrue$ is resolved  through the constraint $\loadtrue_{11} >0$, $\loadtrue_{22} >0$, $\loadtrue_{53} >0$,   and  $\loadtrue_{74} >0$.  As expected,  nonzero factors loading have
       relatively high  communalities for the different currencies, whereas for zero rows the communalities are practically equal to zero.

\begin{Tabelle}{Exchange rate data; posterior mean of the factor loadings  $\loadtrue_{ij}$,  the communalities  $R^2_{ij}$ (in percent)
and the idiosyncratic variances  $\sigma_i^2$   (fractional prior  $b=b_N$)
 for a 4-factor model with the GLT constraint
$\lm ^\star=%(l_1,l_2,l_3,l_4)=
(1,3,5,7)$. Entries %for  $\loadtrue_{ij}$
 with  %  absolute value smaller than
$|\loadtrue_{ij}|  <0.01$  and entries %for  $R^2_{ij}$
with % value smaller than
$ R^2_{ij} < 0.1$  are  indicated by $\approx 0$.}{ken_fac4}
{\small \begin{tabular}{lccccccccc} \hline
 &   \multicolumn{4}{c}{Factor loadings} &  \multicolumn{4}{c}{Communalities}&  \\ %\cline{2-4} \cline{6-8}%\hline
  Currency & $\loadtrue_{i1}$ & $\loadtrue_{i2}$ & $\loadtrue_{i3}$  & $\loadtrue_{i4}$  & $R^2_{i1} $ & $R^2_{i2}$ & $R^2_{i3}$  & $R^2_{i4}$ % $R_i^2$
  & $\sigma_i^2$ \\ \hline
  AUD &    0.96 &         0 &         0 &         0 &      88 &      0 &      0 &        0 &   0.12 \\
  CAD  &    0.39 &       0.6 &         0 &         0 &      17 &     39 &      0 &        0 &   0.42 \\
 CHF &  $\approx 0$  &    -0.02 &         0 &         0 &   $\approx 0$ &   0.36 &      0 &        0 &   0.98 \\
  CZK &  $\approx 0$ &     0.04 &         0 &         0 &  $\approx 0$ &   0.96 &      0 &        0 &   0.98 \\
 DKK & $\approx 0$ &       1.1 &      0.22 &         0 &  $\approx 0$ &     95 &    4.2 &        0 &  0.01 \\
 GBP  &  0.01  &      0.57 &    -0.01 &         0 &    0.39 &     32 &   0.27 &        0 &    0.70 \\
 HKD & $\approx 0$ &       0.5 &      0.39 &      0.76 &  $\approx 0$ &     22 &     14 &       49 &   0.17 \\
 IDR   &  0.01 &       0.8 &   -0.01 &      0.42 &   $\approx 0$ &     58 &  $\approx 0$ &       16 &   0.29 \\
   JPY &  0.02 &      0.93 &    $\approx 0$ &    $\approx 0$ &    0.35 &     76 &  $\approx 0$ &   $\approx 0$ &   0.27 \\
 KRW  & $\approx 0$ &       1.1 &    0.01 &   $\approx 0$ &   $\approx 0$ &     96 &  $\approx 0$ &    $\approx 0$  &0.01 \\
MXN &    $\approx 0$ &     0.03 &   $\approx 0$ &    $\approx 0$ &  $\approx 0$ &   0.65 &  $\approx 0$ &   $\approx 0$  &   0.98 \\
MYR &    0.79 &   $\approx 0$ &   $\approx 0$ &  $\approx 0$ &      61 &  $\approx 0$ & $\approx 0$ &   $\approx 0$ &    0.40 \\
NOK &  $\approx 0$ &      0.89 &  $\approx 0$ &     $\approx 0$ &   $\approx 0$ &     70 &  $\approx 0$  &  $\approx 0$ &   0.33 \\
NZD  &   0.025 &      0.11 &     -0.01 &     $\approx 0$ &    0.75 &    3.2 &   0.29 &   $\approx 0$  &   0.95 \\
 PHP  &  $\approx 0$ &      0.55 &     -0.42 &    0.01 &   $\approx 0$  &     29 &     18 &     0.14 &   0.56 \\
PLN  &  $\approx 0$ &         1 &     $\approx 0$ &      0.12 &   $\approx 0$ &     86 &  $\approx 0$  &      1.9 &   0.14 \\
 RON &    0.04 &    $\approx 0$ &    -0.08 &    $\approx 0$  &     1.3 &   0.11 &      3 &    $\approx 0$ &   0.95 \\
RUB &   -0.09 &     0.02 &     0.03 &     0.05 &     3.2 &   0.35 &   0.84 &      1.6 &   0.94 \\
SEK & $\approx 0$  &      0.98 &   $\approx 0$ &      0.31 &   $\approx 0$ &     82 &  $\approx 0$ &      8.5 &   0.11 \\
SGD &  $\approx 0$ &      0.75 &    $\approx 0$ &      0.39 &   $\approx 0$  &     51 &  $\approx 0$  &       14 &   0.37 \\
 THB   & $\approx 0$ &      0.59 &   $\approx 0$ &     $\approx 0$ &  $\approx 0$  &     33 &  $\approx 0$ &    $\approx 0$  &    0.7 \\
USD & $\approx 0$ &       1.1 &      0.22 &    $\approx 0$ &   $\approx 0$  &     95 &    4.2 &   $\approx 0$ &  0.01 \\
\hline
\end{tabular}}
\end{Tabelle}

%\clearpage

 \subsection{Sparse factor analysis for  NYSE100 returns}

To show that our approach also scales to higher dimensions, we consider  monthly log returns from $m=73$ firms from NYSE100
observed for $T=240$ months from January 1992  to  December  2011. Again, the data are standardized.
Since the number of factors is unknown, an overfitting factor model is applied with the maximum degree of overfitting $S=4$ and
$\nfac=20$ being considerably smaller than the upper bound given by  (\ref{kbound_extend}).
The  hyperparameters of the prior  (\ref{prigen}) for the indicators are chosen as
$ a_0=0.05$ and   $b_0=0.1$, implying  a prior simplicity of $ \Ew{q_i}=6.\dot{6}$   and   $\alpha=10$  in  parameterization (\ref{prialt}).
The prior  (\ref{priorsiid})   is chosen for $\sigma^2_i$  with  $c_0=2.5$ and  $\widehat{\Vary^{-1}}$ being  estimated as in (\ref{Varyhat}), with $\nu_o =3 $ and   $\Sm_o=\identy{\dimy}$. Since $d(\nfac,\dimy)=1,270 << N=17,520$, we consider   fractional priors with $ b= 10^{-5}, b_N, 10^{-4}$, where $b_N =5.71\cdot 10^{-5}$.
Further tuning is exactly as in  Subsection~\ref{applicEx22}.

The  designer MCMC scheme outlined in Algorithm~\ref{Algo3} is used to obtain  $M=100,000$ draws
after a burn-in of $M_0=50,000$ draws starting, respectively, with $\nfacr^{(0)}=7$ and
$\nfacr^{(0)}=20$. Functionals of the posterior draws  were used to monitor MCMC convergence.
The  fraction $p_V$  of  MCMC draws satisfying \AR\ is smaller than in the previous subsection but, being in the order of  8 to 11\%, still acceptable.
Although the prior  $p(\nfacr)$ is fairly wide-spread, the posterior distribution $p(\nfacr|\ym)$  derived from all variance identified draws turns out to be strongly centered on $\tilde{\nfactrue}=12$ for all three fractional priors, see Table~\ref{NASDAQ_tab1}. % in Appendix~\ref{appApp:NasDaq}.

\begin{Tabelle}{NYSE100 return data;  Bayesian inference for an unknown number of factors (maximum number of factors $k=20$)
% Prior distribution $p(\nfacr)$ of the number on nonzero columns $\nfacr$   obtained by simulation.
under prior (\ref{prigen}) with $ a_0=0.05$ and   $b_0=0.1$.  %; draws that did not satisfy the condition \AR\ for variance identification with  $\nfactrue=\nfacr$ were rejected.
 $p_V$ is the fraction of  draws satisfying \AR . Posterior distribution $p(\nfacr|\ym)$ of the number
 $\nfacr$ of nonzero columns (bold number corresponding to the posterior mode $\tilde{r}$) for various fractional priors  on $\facload_{i\cdot}^{\deltav}$ with $ b= 10^{-5}, b=b_N =5.71\cdot 10^{-5}, b= 10^{-4}$. Upper part: variance identified draws; lower part: all posterior draws.}{NASDAQ_tab1}
{\small \begin{tabular}{lccccccc} \hline
  & \multicolumn{6}{c}{$\nfacr$} &   \\ \cline{2-8}
  & $\leq 11$ & 12  & 13 & 14 & 15 & $\geq 16$  &  $100 p_V$  \\  \hline
% $p(\nfacr)$ &  0 &   CHANGE &    0.231 &   0.2642 &   0.1996  &  0.1106       &  27.4 \\ \hline
  $p(\nfacr|\ym)$ &&&&&&&\\ \cline{1-1}
  $b=10^{-5}$  &      0& \textbf{ 0.98}   &   0.02    &  0&  0&  0        & 7.6 \\
  $b= b_N$ &     0    &  \textbf{ 0.70} &    0.27 &    0.02    &0   &0   & 10.7   \\
 $b=10^{-4}$    &      0      &  \textbf{0.56} &  0.35 &    0.09 & 0 &  0 & 9.4  \\ \hline
 no varide &&&&&&&\\ \cline{1-1}
 $b=10^{-5}$  &      0&      0.88 &    0.12 &   0.01   &  0 & 0 &   \\
   $b= b_N$  &   0     &  0.30 &    0.45 &    0.22 &    0.03  &  0  &    \\
  $b=10^{-4}$  &     0    &    0.21 &   0.52 &    0.23 &  0.04   & 0  &    \\
  \hline
\end{tabular}}
\end{Tabelle}

 The  MCMC scheme shows good mixing, despite the high dimensionality, as illustrated by  Figure~\ref{fig_nyse_0} showing    draws from the posterior distributions
  $p(\nfacr|\ym)$ and  $p(d|\ym)$  for  $b=b_N$.  The RJMCMC Step~(R) in Algorithm~\ref{Algo3} has an acceptance rate of  8.6\% for a split and 14.7\% for a merge move and the inefficiency factor  for $d$ is equal to 8.

\begin{Tabelle}{NYSE100 return data;  sequence of leading indices  $ \lm ^\star$ visited most often together with its frequency $100 p_{L}$ (in percent)
  for various fractional priors.}{nyse_listar}
{ \small   \begin{tabular}{llcr}
     \hline
     &   $\lm ^\star $ &  $\rstar$ & $100 p_{L}$\\
     \hline
   $b= b_N$  & (1,2,3,4,5,6,7,8,9,14,15,26) & 12 & 10.3 \\
             &  (1,2,3,4,5,6,7,8,9,14,15,26) &  12 &  9.9 \\
  $b= 10^{-4} $  & (1,2,3,4,5,6,7,8,9,14,15,26)    &  12 &  9.8 \\
                          & (1,2,3,4,5,6,7,8,9,14,15,26)&  12 &  10.8\\
    $b= 10^{-5}$ &  (1,2,3,4,5,6,7,14,15,19,25,26)  &  12 &  19.0 \\
              &  (1,2,3,4,5,6,7,9,14,15,25,26)  &  12 &  25.2\\
     \hline
   \end{tabular}
}
\end{Tabelle}

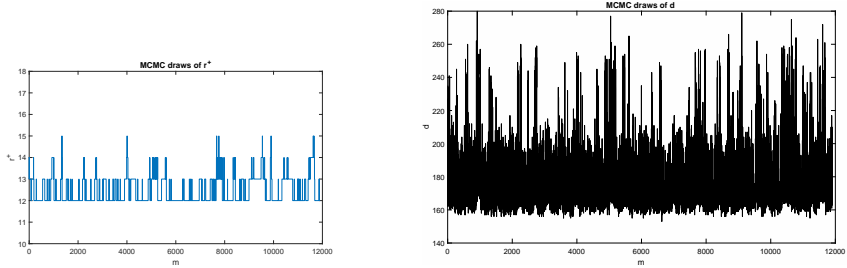
  \begin{Figure2}{NYSE100 return data;  fractional prior with   $b=b_N$. All (11906 variance identified) posterior draws  of the number $\nfacr$ of factors (left-hand side) and
 model size $d$ (right-hand side).}{fig_nyse_0}{nyse_mcmc_nfac}{nyse_mcmc_m}{0.2}
 \end{Figure2}

 In Table~\ref{nyse_listar}, the identifiability constraint
    $\lm ^\star=(l_1^\star, \ldots, l_{\rstar}^\star)$ visited most often is reported together with its frequency $p_{L}$ for all three priors for both runs.
   Also  $\lm ^\star$ points at  a 12-factor model for all priors and coincides for both runs for  $b= b_N$ and $b= 10^{-4} $.
   Further inference with respect to  $\facloadtrue$ and $\deltav$ is based on all posterior draws where the
 leading indices of $\deltav$  (after reordering) are equal to  $\lm ^\star$.
 The corresponding median probability model (MPM)
 is shown  for  $b= b_N$ in Figure~\ref{fig_deltanyse} and is extremely sparse with only $d_M=156$ nonzero loadings.
 The MPM clearly indicates that all   returns are correlated\footnote{This confirmed by the posterior probabilities $\Prob{q_i=0|\ym}$ which are equal to 1 for all firms.}  and  one main factor is present which loads on all  returns. The remaining factors  are for the most part dedicated  factors that capture cross-sectional correlations between specific firms.

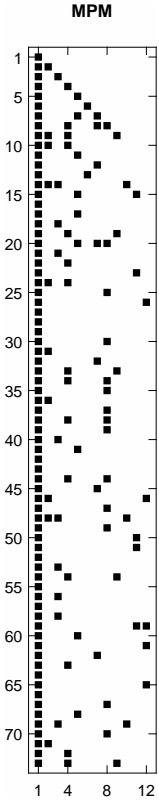
  \begin{Figure}{NYSE100 return data;  $\deltav$ corresponding to the MPM with $\lm ^\star = (1,2,3,4,5,6,7,8,9,14,15,26)$ for a fractional prior with   $b=b_N$.}{fig_deltanyse}{nyse_delta}{0.5}
 \end{Figure}

%\newpage

\section{Concluding remarks} \label{secconcluse}

We have characterised, identified and estimated (from a Bayesian viewpoint) a fairly important and highly implemented class of sparse factor models when the number of common factors is unknown.  More specifically, we have explicitly and rigorously addressed identifiability issues that arise in this class of models by going well beyond and much deeper than simply applying rotation for identification and seeking instead uniqueness of the variance decomposition.

In addition, our framework leads to a natural, efficient and simultaneous coupling of model estimation and selection on one hand and model identification and reduction as well as rank estimation (number of factors) on the other hand.  More precisely, by combining point-mass mixture priors with overfitting sparse factor modelling, in a generalised lower triangular loadings representation, we obtain posterior summaries regarding factor loadings, common factors as well as the number of common factors via postprocessing our highly efficient and customised MCMC scheme.  Two applications, one with $m=22$ variables and $T=96$ observations and one with $m=73$ and $T=240$, illustrates in detail  many of the existing and new aspects of estimating a parsimonious and sparse factor model when the number of factors is unknown.

The new framework is readily available for some straightforward extensions.  Theorem~\ref{Lemma2}, for example,
is not confined to GLT structures and is applicable to any (sparse) loading matrix  which arises
in statistics  and machine learning  (see e.g. the web appendix of \citet{roc-geo:fas} where the factor model fitted to the applicants data obviously is not identified) or  to spatial
factor models with 0-1 neighbouring structures (see \citet{lop-etal:spa} and \citet{sch-lop:dyn}, and their references),
but also in economics  and  genetics \citep{car-etal:hig}.

 Other relatively immediate extensions are
           (i) idiosyncratic errors following Student's $t$-distributions or more general Gaussian mixtures
            and (ii) dynamic sparse factor models with stationary common factors; both extensions commonly found in econometrics applications, see e.g. the recent papers    by \citet{pia-pap:bay} and     \citet{kau-sch:bay}.
Finally, extending our   approach,  in particular Theorem~\ref{theoverGLT},  to correlated factors
could prove useful towards generalizing the work of \citet{con-etal:bay} to simple structures with more than one nonzero loading per factor.

\bibliographystyle{chicago}

%\bibliography{sylvia_hedi}
%\bibliography{sylvia_kyoto}

\begin{thebibliography}{}

\bibitem[\protect\citeauthoryear{Akaike}{Akaike}{1987}]{aka:fac}
Akaike, H. (1987).
\newblock Factor analysis and {AIC}.
\newblock {\em Psychometrika\/}~{\em 52}, 317--332.

\bibitem[\protect\citeauthoryear{Anderson}{Anderson}{2003}]{and:int}
Anderson, T.~W. (2003).
\newblock {\em An Introduction to Multivariate Statistical Analysis\/} (3 ed.).
\newblock Chichester: Wiley.

\bibitem[\protect\citeauthoryear{Anderson and Rubin}{Anderson and
  Rubin}{1956}]{and-rub:sta}
Anderson, T.~W. and H.~Rubin (1956).
\newblock Statistical inference in factor analysis.
\newblock In {\em Proceedings of the Third Berkeley Symposium on Mathematical
  Statistics and Probability}, Volume~V, pp.\  111--150.

\bibitem[\protect\citeauthoryear{A{\ss}mann, Boysen-Hogrefe, and
  Pape}{A{\ss}mann et~al.}{2016}]{ass-etal:bay}
A{\ss}mann, C., J.~Boysen-Hogrefe, and M.~Pape (2016).
\newblock {B}ayesian analysis of static and dynamic factor models: {An} ex-post
  approach toward the rotation problem.
\newblock {\em Journal of Econometrics\/}~{\em 192}, 190--206.

\bibitem[\protect\citeauthoryear{Bai and Ng}{Bai and Ng}{2002}]{bai-ng:det2002}
Bai, J. and S.~Ng (2002).
\newblock Determining the number of factors in approximate factor models.
\newblock {\em Econometrica\/}~{\em 70}, 191--221.

\bibitem[\protect\citeauthoryear{Bartholomew}{Bartholomew}{1987}]{bar:lat}
Bartholomew, D.~J. (1987).
\newblock {\em Latent Variable Models and Factor Analysis}.
\newblock London: Charles Griffin.

\bibitem[\protect\citeauthoryear{Bhattacharya and Dunson}{Bhattacharya and
  Dunson}{2011}]{bha-dun:spa}
Bhattacharya, A. and D.~Dunson (2011).
\newblock Sparse {B}ayesian infinite factor models.
\newblock {\em Biometrika\/}~{\em 98}, 291--306.

\bibitem[\protect\citeauthoryear{Bitto and Fr\"{u}hwirth-Schnatter}{Bitto and
  Fr\"{u}hwirth-Schnatter}{2016}]{bit-fru:ach}
Bitto, A. and S.~Fr\"{u}hwirth-Schnatter (2016).
\newblock Achieving shrinkage in a time-varying parameter model framework.
\newblock {\em submitted\/}~({\tt arXiv:1611.01310}).

\bibitem[\protect\citeauthoryear{Boivin and Ng}{Boivin and
  Ng}{2006}]{boi-ng:are}
Boivin, J. and S.~Ng (2006).
\newblock Are more data always better for factor analysis?
\newblock {\em Journal of Econometrics\/}~{\em 132}, 169--194.

\bibitem[\protect\citeauthoryear{Carvalho, Chang, Lucas, Nevins, Wang, and
  West}{Carvalho et~al.}{2008}]{car-etal:hig}
Carvalho, C.~M., J.~Chang, J.~E. Lucas, J.~Nevins, Q.~Wang, and M.~West (2008).
\newblock High-dimensional sparse factor modeling: Applications in gene
  expression genomics.
\newblock {\em Journal of the American Statistical Association\/}~{\em 103},
  1438--1456.

\bibitem[\protect\citeauthoryear{Chan, Leon-{G}onzalez, and Strachan}{Chan
  et~al.}{2018}]{cha-etal:inv}
Chan, J. C.~C., R.~Leon-{G}onzalez, and R.~W. Strachan (2018).
\newblock Invariant inference and efficient computation in the static factor
  model.
\newblock {\em Journal of the American Statistical Association\/}, forthcoming.

\bibitem[\protect\citeauthoryear{Conti, Fr\"uhwirth-Schnatter, Heckman, and
  Piatek}{Conti et~al.}{2014}]{con-etal:bay}
Conti, G., S.~Fr\"uhwirth-Schnatter, J.~J. Heckman, and R.~Piatek (2014).
\newblock Bayesian exploratory factor analysis.
\newblock {\em Journal of Econometrics\/}~{\em 183}, 31--57.

\bibitem[\protect\citeauthoryear{Fan, Fan, and Lv}{Fan
  et~al.}{2008}]{fan-etal:hig_je}
Fan, J., Y.~Fan, and J.~Lv (2008).
\newblock High dimensional covariance matrix estimation using a factor model.
\newblock {\em Journal of Econometrics\/}~{\em 147}, 186--197.

\bibitem[\protect\citeauthoryear{Forni, Giannone, Lippi, and Reichlin}{Forni
  et~al.}{2009}]{for-etal:ope}
Forni, M., D.~Giannone, M.~Lippi, and L.~Reichlin (2009).
\newblock Opening the black box: {S}tructural factor models with large cross
  sections.
\newblock {\em Econometric Theory\/}~{\em 25}, 1319--1347.

\bibitem[\protect\citeauthoryear{Foster and George}{Foster and
  George}{1994}]{fos-geo:ris}
Foster, D.~P. and E.~I. George (1994).
\newblock The risk inflation criterion for multiple regression.
\newblock {\em The Annals of Statistics\/}~{\em 22}, 1947--1975.

\bibitem[\protect\citeauthoryear{Fr{\"u}hwirth-Schnatter and
  Lopes}{Fr{\"u}hwirth-Schnatter and Lopes}{2010}]{fru-lop:par}
Fr{\"u}hwirth-Schnatter, S. and H.~Lopes (2010).
\newblock {Parsimonious {B}ayesian Factor Analysis when the Number of Factors
  is Unknown}.
\newblock Research report, Booth School of Business, University of Chicago.

\bibitem[\protect\citeauthoryear{Fr{\"u}hwirth-Schnatter and
  T{\"u}chler}{Fr{\"u}hwirth-Schnatter and T{\"u}chler}{2008}]{fru-tue:bay}
Fr{\"u}hwirth-Schnatter, S. and R.~T{\"u}chler (2008).
\newblock Bayesian parsimonious covariance estimation for hierarchical linear
  mixed models.
\newblock {\em Statistics and Computing\/}~{\em 18}, 1--13.

\bibitem[\protect\citeauthoryear{Fr{\"u}hwirth-Schnatter and
  Wagner}{Fr{\"u}hwirth-Schnatter and Wagner}{2010}]{fru-wag:sto}
Fr{\"u}hwirth-Schnatter, S. and H.~Wagner (2010).
\newblock Stochastic model specification search for {G}aussian and partially
  non-{G}aussian state space models.
\newblock {\em Journal of Econometrics\/}~{\em 154}, 85--100.

\bibitem[\protect\citeauthoryear{Geweke and Singleton}{Geweke and
  Singleton}{1980}]{gew-sin:int}
Geweke, J.~F. and K.~J. Singleton (1980).
\newblock Interpreting the likelihood ratio statistic in factor models when
  sample size is small.
\newblock {\em Journal of the American Statistical Association\/}~{\em 75},
  133--137.

\bibitem[\protect\citeauthoryear{Geweke and Zhou}{Geweke and
  Zhou}{1996}]{gew-zho:mea}
Geweke, J.~F. and G.~Zhou (1996).
\newblock Measuring the pricing error of the arbitrage pricing theory.
\newblock {\em Review of Financial Studies\/}~{\em 9}, 557--587.

\bibitem[\protect\citeauthoryear{Ghahramani, Griffiths, and Sollich}{Ghahramani
  et~al.}{2007}]{gha-etal:bay}
Ghahramani, Z., T.~L. Griffiths, and P.~Sollich (2007).
\newblock Bayesian nonparametric latent feature models (with discussion and
  rejoinder).
\newblock In J.~M. Bernardo, M.~J. Bayarri, J.~O. Berger, A.~P. Dawid,
  D.~Heckerman, A.~F.~M. Smith, and M.~West (Eds.), {\em Bayesian Statistics
  8}, pp.\  ADD--ADD. Oxford: Oxford University Press.

\bibitem[\protect\citeauthoryear{Ghosh and Dunson}{Ghosh and
  Dunson}{2009}]{gho-dun:def}
Ghosh, J. and D.~B. Dunson (2009).
\newblock Default prior distributions and efficient posterior computation in
  {B}ayesian factor analysis.
\newblock {\em Journal of Computational and Graphical Statistics\/}~{\em 18},
  306--320.

\bibitem[\protect\citeauthoryear{J\"{o}reskog}{J\"{o}reskog}{1967}]{joe:som}
J\"{o}reskog, K.~G. (1967).
\newblock Some contributions to maximum likelihood factor analysis.
\newblock {\em Psychometrika\/}~{\em 32}, 443--482.

\bibitem[\protect\citeauthoryear{Kastner}{Kastner}{2018}]{kas:spa}
Kastner, G. (2018).
\newblock Sparse {B}ayesian time-varying covariance estimation in many
  dimensions.
\newblock {\em Journal of Econometrics\/}, forthcoming.

\bibitem[\protect\citeauthoryear{Kastner and Fr\"{u}hwirth-Schnatter}{Kastner
  and Fr\"{u}hwirth-Schnatter}{2014}]{kas-fru:anc}
Kastner, G. and S.~Fr\"{u}hwirth-Schnatter (2014).
\newblock Ancillarity-sufficiency interweaving strategy {(ASIS)} for boosting
  {MCMC} estimation of stochastic volatility models.
\newblock {\em Computational Statistics and Data Analysis\/}~{\em 76}, 408–423.

\bibitem[\protect\citeauthoryear{Kastner, Fr\"{u}hwirth-Schnatter, and
  Lopes}{Kastner et~al.}{2017}]{kas-etal:eff}
Kastner, G., S.~Fr\"{u}hwirth-Schnatter, and H.~F. Lopes (2017).
\newblock Efficient {B}ayesian inference for multivariate factor stochastic
  volatility models.
\newblock {\em Journal of Computational and Graphical Statistics\/}~{\em 26},
  905--917.

\bibitem[\protect\citeauthoryear{Kaufmann and Schuhmacher}{Kaufmann and
  Schuhmacher}{2017}]{kau-sch:ide}
Kaufmann, S. and C.~Schuhmacher (2017).
\newblock Identifying relevant and irrelevant variables in sparse factor
  models.
\newblock {\em Journal of Applied Econometrics\/}~{\em 32}, 1123--1144.

\bibitem[\protect\citeauthoryear{Kaufmann and Schuhmacher}{Kaufmann and
  Schuhmacher}{2018}]{kau-sch:bay}
Kaufmann, S. and C.~Schuhmacher (2018).
\newblock Bayesian estimation of sparse dynamic factor models with
  order-independent and ex-post identification.
\newblock {\em Journal of Econometrics\/}, forthcoming.

\bibitem[\protect\citeauthoryear{Ledermann}{Ledermann}{1937}]{led:ran}
Ledermann, W. (1937).
\newblock On the rank of the reduced correlational matrix in multiple-factor
  analysis.
\newblock {\em Psychometrika\/}~{\em 2}, 85--93.

\bibitem[\protect\citeauthoryear{Lee and Song}{Lee and
  Song}{2002}]{lee-son:bay}
Lee, S.~Y. and X.~Y. Song (2002).
\newblock {B}ayesian selection on the number of factors in a factor analysis
  model.
\newblock {\em Behaviormetrika\/}~{\em 29}, 23--39.

\bibitem[\protect\citeauthoryear{Lopes, Salazar, and Gamerman}{Lopes
  et~al.}{2008}]{lop-etal:spa}
Lopes, H.~F., E.~Salazar, and D.~Gamerman (2008).
\newblock Spatial dynamic factor analysis.
\newblock {\em Bayesian Analysis\/}~{\em 3}, 759--792.

\bibitem[\protect\citeauthoryear{Lopes and West}{Lopes and
  West}{2004}]{lop-wes:bay}
Lopes, H.~F. and M.~West (2004).
\newblock Bayesian model assessment in factor analysis.
\newblock {\em Statistica Sinica\/}~{\em 14}, 41--67.

\bibitem[\protect\citeauthoryear{Lucas, Carvalho, Wang, Bild, Nevins, and
  West}{Lucas et~al.}{2006}]{luc-etal:spa}
Lucas, J., C.~Carvalho, Q.~Wang, A.~Bild, J.~R. Nevins, and M.~West (2006).
\newblock Sparse statistical modelling in gene expression genomics.
\newblock In K.~Do, P.~M\"uller, and M.~Vannucci (Eds.), {\em Bayesian
  Inference for Gene Expression and Proteomics}, pp.\  155--176. Cambridge, UK:
  Cambridge University Press.

\bibitem[\protect\citeauthoryear{Martin and {McD}onald}{Martin and
  {McD}onald}{1975}]{mar-mcd:bay}
Martin, J.~K. and R.~P. {McD}onald (1975).
\newblock Bayesian estimation in unrestricted factor analysis: a treatment for
  {H}eywood cases.
\newblock {\em Psychometrika\/}~{\em 40}, 505--517.

\bibitem[\protect\citeauthoryear{{O'H}agan}{{O'H}agan}{1995}]{oha:fra}
{O'H}agan, A. (1995).
\newblock Fractional {B}ayes factors for model comparison.
\newblock {\em Journal of the Royal Statistical Society, Ser. B\/}~{\em 57},
  99--138.

\bibitem[\protect\citeauthoryear{{Pati}, {Bhattacharya}, {Pillai}, and
  {Dunson}}{{Pati} et~al.}{2014}]{pat-etal:pos}
{Pati}, D., A.~{Bhattacharya}, N.~S. {Pillai}, and D.~B. {Dunson} (2014).
\newblock {Posterior contraction in sparse Bayesian factor models for massive
  covariance matrices}.
\newblock {\em Annals of Statistics\/}~{\em 42}, 1102--1130.

\bibitem[\protect\citeauthoryear{Piatek and Papaspiliopoulos}{Piatek and
  Papaspiliopoulos}{2018}]{pia-pap:bay}
Piatek, R. and O.~Papaspiliopoulos (2018).
\newblock A bayesian nonparametric approach to factor analysis.
\newblock {\em Submitted\/}.

\bibitem[\protect\citeauthoryear{Ro\v{c}kov\'{a} and George}{Ro\v{c}kov\'{a}
  and George}{2017}]{roc-geo:fas}
Ro\v{c}kov\'{a}, V. and E.~I. George (2017).
\newblock Fast {B}ayesian factor analysis via automatic rotation to sparsity.
\newblock {\em Journal of the American Statistical Association\/}~{\em 111},
  1608--1622.

\bibitem[\protect\citeauthoryear{Rue and Held}{Rue and
  Held}{2005}]{rue-hel:gau}
Rue, H. and L.~Held (2005).
\newblock {\em {G}aussian {M}arkov Random Fields: {T}heory and Applications},
  Volume 104 of {\em Monographs on Statistics and Applied Probability}.
\newblock London: Chapman {\rm \&} Hall/CRC.

\bibitem[\protect\citeauthoryear{Sato}{Sato}{1992}]{sat:stu}
Sato, M. (1992).
\newblock A study of an identification problem and substitute use of principal
  component analysis in factor analysis.
\newblock {\em Hiroshima Mathematical Journal\/}~{\em 22}, 479--524.

\bibitem[\protect\citeauthoryear{Schmidt and Lopes}{Schmidt and
  Lopes}{2018}]{sch-lop:dyn}
Schmidt, A.~M. and H.~F. Lopes (2018).
\newblock Dynamic models.
\newblock In A.~Gelfand, M.~Fuentes, J.~Hoeting, and R.~Smith (Eds.), {\em
  Handbook of Econometrics}. Chapman \& Hall.

\bibitem[\protect\citeauthoryear{Smith and Kohn}{Smith and
  Kohn}{2002}]{smi-koh:par}
Smith, M. and R.~Kohn (2002).
\newblock Parsimonious covariance matrix estimation for longitudinal data.
\newblock {\em Journal of the American Statistical Association\/}~{\em 97},
  1141--1153.

\bibitem[\protect\citeauthoryear{Stock and Watson}{Stock and
  Watson}{2002}]{sto-wat:mac}
Stock, J.~H. and M.~W. Watson (2002).
\newblock Macroeconomic forecasting using diffusion indexes.
\newblock {\em Journal of Business {\rm \&} Economic Statistics\/}~{\em 20},
  147--162.

\bibitem[\protect\citeauthoryear{T{\"u}chler}{T{\"u}chler}{2008}]{tue:bay}
T{\"u}chler, R. (2008).
\newblock Bayesian variable selection for logistic models using auxiliary
  mixture sampling.
\newblock {\em Journal of Computational and Graphical Statistics\/}~{\em 17},
  76--94.

\bibitem[\protect\citeauthoryear{Tumura and Sato}{Tumura and
  Sato}{1980}]{tum-sat:ide}
Tumura, Y. and M.~Sato (1980).
\newblock On the identification in factor analysis.
\newblock {\em TRU Mathematics\/}~{\em 16}, 121--131.

\bibitem[\protect\citeauthoryear{van {D}yk and Meng}{van {D}yk and
  Meng}{2001}]{van-men:art}
van {D}yk, D. and X.-L. Meng (2001).
\newblock The art of data augmentation.
\newblock {\em Journal of Computational and Graphical Statistics\/}~{\em 10},
  1--50.

\bibitem[\protect\citeauthoryear{West}{West}{2003}]{wes:bay_fac}
West, M. (2003).
\newblock Bayesian factor regression models in the \lq\lq large p, small
  n\rq\rq\ paradigm.
\newblock In J.~M. Bernardo, M.~J. Bayarri, J.~O. Berger, A.~P. Dawid,
  D.~Heckerman, A.~F.~M. Smith, and M.~West (Eds.), {\em Bayesian Statistics
  7}, pp.\  733--742. Oxford: Oxford University Press.

\bibitem[\protect\citeauthoryear{Yu and Meng}{Yu and Meng}{2011}]{yu-men:cen}
Yu, Y. and X.-L. Meng (2011).
\newblock To center or not to center: that is not the question - {a}n
  ancillarity-suffiency interweaving strategy {(ASIS)} for boosting {MCMC}
  efficiency.
\newblock {\em Journal of Computational and Graphical Statistics\/}~{\em 20},
  531--615.

\end{thebibliography}

\newpage

\appendix

\begin{center}
{\LARGE Parsimonious Bayesian Factor Analysis when the Number of Factors is Unknown} \\[5mm]
{\LARGE  Webappendix} \\[5mm]

{\Large
Sylvia Fr\"uhwirth-Schnatter\footnote{Department of Finance, Accounting, and Statistics, WU Vienna University of Economics and Business, Austria. Email: {\tt sfruehwi@wu.ac.at}}
 and
 Hedibert Freitas Lopes\footnote{Insper Institute of Education and Research, S\~ao Paulo, Brazil. Email: {\tt hedibertfl@insper.edu.br}} }\\[5mm]

%{\large Institute for Statistics and Mathematics}\\
% {\large  Department of Finance, Accounting and Statistics}\\
%{\large WU Vienna University of Economics and Business, Vienna, Austria}\\
%{\large Email: {\tt  angela.bitto@wu.ac.at, sfruehwi@wu.ac.at}}\\

\end{center}

% Change equation numbering
\setcounter{equation}{0}
\setcounter{figure}{0}
\setcounter{table}{0}
\setcounter{page}{1}

\renewcommand{\thesection}{\Alph{section}}
\renewcommand{\thetable}{\Alph{section}.\arabic{table}}
\renewcommand{\thefigure}{\Alph{section}.\arabic{figure}}
\renewcommand{\theequation}{\Alph{section}.\arabic{equation}}

\section{Proofs and further details on identification}

\subsection{Proofs}   \label{app:proof}

\paragraph*{Proof of Theorem~\ref{theGLT}.}

Assume that two pairs $(\facloadtrue ,\Varetrue)$
and $(\facloadtilde ,\Varetilde)$ satisfy  (\ref{fac4}), where both $\facloadtrue $ and
$\facloadtilde$ are GLT matrices with, respectively, leading indices $l_1 < \ldots <  l_\nfactrue$
and $\tilde{l}_1 < \ldots <  \tilde{l}_\nfactrue$. Uniqueness of  the variance decomposition (\ref{fac4})
implies
 \begin{eqnarray} \label{conglt}
  \facloadtrue  \trans{\facloadtrue }  = \facloadtilde \trans{\facloadtilde} .
\end{eqnarray}
We need to prove that all columns of $\facloadtrue $ and $\facloadtilde$ are identical.%  up to sign switching.

First, we prove that $l_1=\tilde{l}_1$ by contradiction. Assume $\tilde{l}_1 \neq l_1 $. Exploiting the GLT  structure of both matrices,
we obtain from (\ref{conglt}):
\begin{eqnarray} \label{conglt:eq1}
 && \loadtrue_{l_1,1}^2= \sum_{j=1}^\nfactrue \loadtruetilde_{l_1,j}^2 \neq 0,\\
 && \sum_{j=1}^\nfactrue \loadtrue_{\tilde{l}_1,j}^2=  \loadtruetilde_{\tilde{l}_1,1}^2 \neq 0. \label{conglt:eq2}
\end{eqnarray}
Assuming $\tilde{l}_1>l_1 $  implies $\loadtruetilde_{l_1,j}=0$ for $j=1, \ldots,\nfactrue$, which contradicts (\ref{conglt:eq1}),
assuming $l_1>\tilde{l}_1$ implies $\loadtrue_{\tilde{l}_1,j}=0$ for $j=1, \ldots,\nfactrue$, which contradicts (\ref{conglt:eq2});
hence $l_1=\tilde{l}_1$. By definiton, $\loadtruetilde_{l_1,j}=0$ for $j=2, \ldots,\nfactrue$,
and (\ref{conglt:eq1}) implies:
\begin{eqnarray*} \label{conglt:eq3}
   \loadtruetilde_{l_1,1}^2 = \loadtrue_{l_1,1}^2 \Rightarrow \loadtruetilde_{l_1,1} = \loadtrue_{l_1,1}.
\end{eqnarray*}
For all $i>l_1$ we obtain from (\ref{conglt}):
 \begin{eqnarray*} \label{conglt:eq4}
  \Cov{y_{l_1,t}, y_{i t}}= \loadtrue_{l_1,1}  \loadtrue_{i 1} = \loadtruetilde_{l_1,1}  \loadtruetilde_{i 1} =  \loadtrue_{l_1,1} \loadtruetilde_{i 1}.
\end{eqnarray*}
Therefore $\loadtruetilde_{i 1}=  \loadtrue_{i 1}$ %$\loadtruetilde_{i,1}= (-1)^{s_1} \loadtrue_{i,1}$
for all $i=l_1, \ldots,\dimy$, hence the first
columns of $\facloadtrue $ and $\facloadtilde$ are identical.

We show identity of  the remaining columns by induction. Assume that the first $q-1$
columns of $\facloadtrue $ and $\facloadtilde$ are identical. %  up to sign switching.
 Similarly as above, we prove  $l_q=\tilde{l}_q$ by contradiction. Exploiting the GLT
structure of both matrices,
we obtain from (\ref{conglt}):
\begin{eqnarray*} %\label{conglt:eq1}
 && \sum_{j=1}^{q-1} \loadtrue_{l_q,j}^2 + \loadtrue_{l_q,q}^2 =
  \sum_{j=1}^{q-1} \loadtruetilde_{l_q,j}^2 + \sum_{j=q}^{\nfactrue}  \loadtruetilde_{l_q,q}^2 =
   \sum_{j=1}^{q-1} \loadtrue_{l_q,j}^2 + \sum_{j=q}^{\nfactrue}  \loadtruetilde_{l_q,q}^2 ,\\
&& \sum_{j=1}^{q-1} \loadtrue_{\tilde{l}_q,j}^2 + \sum_{j=q}^{\nfactrue} \loadtrue_{\tilde{l}_q,j}^2 =
  \sum_{j=1}^{q-1} \loadtruetilde_{\tilde{l}_q,j}^2 + \loadtruetilde_{\tilde{l}_q,q}^2 =
   \sum_{j=1}^{q-1} \loadtrue_{\tilde{l}_q,j}^2 + \loadtruetilde_{\tilde{l}_q,q}^2 . \label{conglt:eq2A}
\end{eqnarray*}
Therefore:
\begin{eqnarray} \label{conglt:eq5}
 && \loadtrue_{l_q,q}^2 = \sum_{j=q}^{\nfactrue}  \loadtruetilde_{l_q,q}^2  \neq 0,\\
&&  \sum_{j=q}^{\nfactrue} \loadtrue_{\tilde{l}_q,j}^2 =  \loadtruetilde_{\tilde{l}_q,q}^2 \neq 0 . \label{conglt:eq6}
\end{eqnarray}
Assuming $\tilde{l}_q>l_q $  implies $\loadtruetilde_{l_q,j}=0$ for $j=1, \ldots,\nfactrue$,
which contradicts (\ref{conglt:eq5}).
Assuming $l_q>\tilde{l}_q$ implies $\loadtrue_{\tilde{l}_q,j}=0$ for $j=1, \ldots,\nfactrue$, which
contradicts (\ref{conglt:eq6});
hence $l_q=\tilde{l}_q$. By definition, $\loadtruetilde_{l_q,j}=0$ for $j=q+1, \ldots,\nfactrue$,
and (\ref{conglt:eq5}) implies:
\begin{eqnarray} \label{conglt:eq7}
   \loadtruetilde_{l_q,q}^2 = \loadtrue_{l_q,q}^2 \Rightarrow   \loadtruetilde_{l_q,q} = \loadtrue_{l_q,q} .
  % \loadtruetilde_{l_q,q} = \loadtrue_{l_q,q} (-1)^{s_q},
\end{eqnarray}
% for $s_q \in \{0,1\}$.
For all $i>l_q$ we obtain from (\ref{conglt}):
 \begin{eqnarray*} %\label{conglt:eq8}
 % \Cov{y_{l_q,t}, y_{i,t}}=
  \sum_{j=1}^{q-1} \loadtrue_{l_q,j} \loadtrue_{i j}
  + \loadtrue_{l_q,q}  \loadtrue_{i q} = \sum_{j=1}^{q-1} \loadtruetilde_{l_q,j} \loadtruetilde_{i j}
  +  \loadtruetilde_{l_q,q}  \loadtruetilde_{i q} =
 \sum_{j=1}^{q-1} \loadtrue_{l_q,j} \loadtrue_{i j} +  \loadtrue_{l_q,q}  \loadtruetilde_{i q}   .
 %(-1)^{s_q}  \loadtrue_{l_q,q}  \loadtruetilde_{i,q}   .
\end{eqnarray*}
Therefore $\loadtruetilde_{i q}=  \loadtrue_{i q}$  % $\loadtruetilde_{i,q}= (-1)^{s_q} \loadtrue_{i,q}$
for all $i=l_q, \ldots,\dimy$, hence also the $q$th
column of $\facloadtrue $ and $\facloadtilde$ is identical. We repeat this procedure
till $q=\nfactrue$.

Finally, since
$\loadtruetilde_{i j}= 0 \Leftrightarrow \loadtrue_{i j}= 0 $,  also the indicators  $\delta_{ij}$
are uniquely identified for all $i,j$. This completes the proof.

 \paragraph{Proof of Theorem~\ref{rule357}.}

    \citet[Theorem~3.3]{sat:stu} shows that  condition  \NC\ is necessary for \AR\ for every  nonsingular $\Gm$, hence also for  $\Gm=\bP _{\pm} \bP _{\rho}$.
 As  conditions \AR\ and  \NC\  are invariant to trivial rotations of $\facloadtrue$,  it is sufficient to verify
 for a single unordered GLT structure $\facloadtilde =  \facloadtrue \bP _{\pm} \bP _{\rho}$ that   \NC\  implies  \AR .
\citet{and-rub:sta} prove that   \NC\ is sufficient for \AR\ for $\nfactrue=1$. The proof that   \NC\ is sufficient for \AR\ also for $\nfactrue>1$ follows by induction.

Assume that   \NC\ is sufficient for \AR\ for some $\nfacind > 1$. Consider an arbitrary unordered GLT structure $\facloadtilde$ with $\nfactrue=\nfacind +1$ nonzero columns and assume that   \NC\ holds for $\facloadtilde$.  Reorder the rows and the columns of $\facloadtilde$ such that the following block structure is obtained:
\begin{eqnarray*}
 \facloadsc = \Pim_r \facloadtilde  \Pim_c  =  \left( \begin{array}{cc}
   \cm  & \bfz \\
   \bm  & \Am \\
  \end{array} \right),
\end{eqnarray*}
where $\Am$ is a $\dimmat{m_n}{\nfacind}$ matrix with $m_n$ nonzero rows  and
$\cm$ and   $\bm $ are column vectors  of dimension $\dimmat{(m-m_n)}{1}$ and $\dimmat{m_n}{1}$, respectively.
Since $\facloadtilde$ is an  (unordered) GLT structure,  $\cm$ contains at least one nonzero element.  Since   \NC\ holds for $\facloadtilde$,  it holds for  $\Am$, implying that \AR\  with $\nfactrue=\nfacind$ holds for   $\Am$.
 Evidently, \AR\  holds for $\facloadtilde$, if it holds
for $ \facloadsc $.  To prove  that \AR\  holds for $ \facloadsc $,  the following cases are distinguished.

(a) If $\cm$ contains at least two nonzero elements $c_{i_1}$ and $c_{i_2}$, then deleting any row below  $\cm$  yields two sub matrices $\Am_1$ and  $\Am_2$ of $\Am$ of rank $\nfacind $. The two sub matrices $ \Bm_1$ and $ \Bm_2$ defined by
\begin{eqnarray}
 \Bm_1=   \left( \begin{array}{cc}
  c_{i_1} & \bfz \\
   \bm_1  & \Am_1 \\
  \end{array}
\right), \quad    \Bm_2 =   \left( \begin{array}{cc}
  c_{i_2} & \bfz \\
   \bm_2  & \Am_2 \\
  \end{array}
\right),  \label{subproof1}
\end{eqnarray}
  obviously have rank  $\nfactrue=\nfacind +1$.

 (b)  If $\cm$ contains at least three nonzero elements, then whenever a row in $\cm$ is deleted, two elements $c_{i_1}$ and $c_{i_2}$ remain  to construct matrices as in (\ref{subproof1}). %, even without deleting any row below $\cm$.
 Together with (a),  this implies that \AR\ holds for this case.

 (c) If $\cm$ contains exactly two nonzero elements, then whenever one of these elements is deleted, another nonzero elements $c_{i_1}$ remains.
  \NC\ for  $\facloadtilde$  implies that   $\bm$ contains at least one nonzero element $b_{i_2}$.   Deleting the corresponding row  $i_2$ from $\Am$ yields two sub matrices $\Am_1$ and  $\Am_2$  of rank $\nfacind $. The two sub matrices  $ \Bm_1$ and $ \Bm_2$ defined by
   \begin{eqnarray}
\Bm_1 =   \left( \begin{array}{cc}
    c_{i_1} & \bfz \\
   \bm_1  & \Am_1 \\
  \end{array}
\right), \quad    \Bm_2=   \left( \begin{array}{cc}
b_{i_2} & \times   \\
   \bm_2  & \Am_2 \\
  \end{array}
\right),  \label{subproof2}
\end{eqnarray}
 have rank  $\nfactrue=\nfacind +1$,  except for a set of Lebesgue measure 0 concerning $ \Bm_2$.   Together with (a),  this implies that \AR\ holds for this case, except for a set of Lebesgue measure 0.

  (d) If $\cm$ contains exactly one nonzero element $c_{i_1}$, then
  \NC\ for  $\facloadtilde$  implies that   $\bm$ contains at least two nonzero element.
 Deleting any row below  $\cm$  yields two sub matrices $\Am_1$ and  $\Am_2$ of $\Am$ of rank $\nfacind $ and
 leaves  at least one  nonzero element $b_{i_2}$ in  $\bm$.  The two sub matrices  $ \Bm_1$ and $ \Bm_2$ are then defined as in
 (\ref{subproof2}).

  (e) Finally, to prove \AR\ if the only nonzero element $c_{i_1}$ in $\cm$  is deleted, we use the fact that  \NC\ for  $\facloadtilde$  implies that
 the  matrix  $\facloadtilde_{- i_1} $ satisfies a similar counting rule:  for  each $q =1,\ldots,\nfacind+1$ and for each submatrix consisting of $q$ column of $\facloadtilde _{- i_1}$, the  number  of nonzero rows in this sub-matrix is at least equal to $2q$.  The columns of   $\facloadtilde _{- i_1}$  are reordered such that the first columns contains at least 2, the first two columns contain at least 4, or more generally, the first  $q$ columns contain at least $2q$  nonzero elements and the resulting matrix is denoted by $ \facloadtilde ^\star$. Evidently,
      the first column of $\facloadtilde ^\star$ has at least  two nonzero elements $\loadtruetilde_{l_1,1}$
      and $\loadtruetilde_{u_1,1}$ in rows $l_1$ and $u_1$, the second column
      of $\facloadtilde ^\star$ has at least  two nonzero elements $\loadtruetilde_{l_2,2}$
      and $\loadtruetilde_{u_2,2}$ in rows $l_2$ and $u_2$, different from $\{l_1,u_1\}$.
      In general, the $q$th column
      of $\facloadtilde ^\star$ has at least  two nonzero elements $\loadtruetilde_{l_q,q}$
      and $\loadtruetilde_{u_q,q}$ in rows $l_q$ and $u_q$, different from
      $\{l_1,\ldots, l_{q-1},u_1,\ldots, u_{q-1}\}$.  Proceeding in this way till $q=\nfacind+1$
       yields   two disjunct sub matrices   $\Bm_1$ and $\Bm_2$
        defined elementwise
     as $B_{1,qj}=\loadtruetilde_{l_q,j}$ and $ B_{2,qj} = \loadtruetilde_{u_q,j}$
      for each $q,j=1,\ldots, \nfactrue$, where all diagonal elements of are nonzero.
      Hence,  $\Bm_1$ and $\Bm_2$  are of   rank $\nfacind+1$, except for a set of Lebesgue measure 0.
     Together with (d),  this implies that \AR\ holds for this case, except for a set of Lebesgue measure 0.

\paragraph*{Proof of Theorem~\ref{Lemma2}.}

%Using induction, it is straightforward to show that the sequential procedure introduced in this lemma leads to following representation of the matrix   $\Bsub{1}$ containing the nonzero rows  of  $\facload$:
%\begin{eqnarray*} \label{blockl}
%\Pim_r  \Bsub{1} \Pim_c =
%\left( \begin{array}{llll}
 %  \Asub{1} & \bfzmat &\bfzmat &\bfzmat \\
 %  \times & \ddots & \bfzmat &\bfzmat \\
 % \times & \times &  \Asub{Q-1} & \bfzmat \\
  % \times & \times & \times &  \Asub{Q} \\
 % \end{array}\right),
 %\end{eqnarray*}
% where $\Asub{Q}=\Bsub{Q}$  and $\Pim_r$ and  $\Pim_c$ are suitable permutation matrices.

Since the row deletion property is invariant to % removing zero rows from $\facload$ and
 reordering the rows and the columns, %  of  remaining matrix,
\AR\ holds for  $\facloadsc$  iff \AR\ holds for  $\facloadstar = \Pim_r  \facloadsc \Pim_c$.

 If each  of the matrices   $\Asub{1}, \ldots, \Asub{Q}$ satisfies   \AR\ with  $\nfactrue=r_q$, then \AR\ with
  $\nfactrue=\nfacr$ is easy to proof for $\facloadstar$.
Whatever row $i$ is deleted from  $\facloadstar$,  a specific row  is deleted from a corresponding  submatrix $\Asub{\tilde{q}}$.
Since $\Asub{\tilde{q}}$  satisfies  \AR\ with  $r=r_{\tilde{q}}$, two disjoint submatrices  $\Asub{\tilde{q}}_1$
 and   $\Asub{\tilde{q}}_2$, each of rank $r_{\tilde{q}}$,  remain. In a similar manner,  disjoint  submatrices $\Asub{q}_1$ and   $\Asub{q}_2$, each of  rank $r_q$,  can be obtained for all other submatrices  $\Asub{q}$, with $q \neq \tilde{q}$.
  From the resulting sequence of   submatrices $\Asub{q}_1$ and   $\Asub{q}_2$, $q=1, \ldots, Q$, following disjoint submatrices
of  $\facloadstar$ can be constructed:
 \begin{eqnarray} \label{bloclam}
\left(
  \begin{array}{lll}
   \Asub{1}_1 & \bfzmat &\bfzmat \\
   \times & \ddots & \bfzmat  \\
    \times & \times &    \Asub{Q}_1 \\
  \end{array}
\right),  \qquad    \left(
  \begin{array}{lll}
   \Asub{1}_2 & \bfzmat &\bfzmat \\
   \times & \ddots & \bfzmat  \\
    \times & \times &    \Asub{Q}_2 \\
  \end{array}
\right).
 \end{eqnarray}
Due to the  block-diagonal  structure appearing in (\ref{bloclam}), the rank of both matrices is equal to $\sum_{q=1}^Q r_q=\nfacr$. Hence,  the
 row deletion property \AR\  with $\nfactrue=\nfacr$ is satisfied for $\facloadstar$.  This proves part~(a).

 Since  $\Asub{Q}$ is a submatrix of  $\facloadstar$  with $r_Q$ columns, a condition necessary for the row deletion property for  $\facloadstar$
is that \NC\  holds for  $\Asub{Q}$ with $r=r_Q$.   Since this condition is violated, \AR\ cannot hold. This proves part~(b).

 \paragraph*{Proof  of Lemma~\ref{theirr}.}  For any row $i$ of $\facloadtrue$ and $\facloadtilde$
 uniqueness of the variance decomposition  implies
 %\begin{eqnarray}
  $\facloadtrue_{i\cdot}  \trans{\facloadtrue_{i\cdot}} = \facloadtilde_{i\cdot} \trans{\facloadtilde_{i\cdot}}$.
%\end{eqnarray}
Hence, if $\facloadtrue_{i\cdot}=\bfz$ is a zero row,
 then $\facloadtilde_{i\cdot} \trans{\facloadtilde_{i\cdot}}=\| \facloadtilde_{i\cdot} \| _2 ^2 =0$, therefore
 $\facloadtilde_{i\cdot}=\bfz$.
 On the other hand, if $\facloadtilde_{i\cdot}=\bfz$ is a zero row,
 then $\facloadtrue_{i\cdot} \trans{\facloadtrue_{i\cdot}}=\| \facloadtrue_{i\cdot} \| _2 ^2 =0$ and
 $\facloadtrue_{i\cdot}=\bfz$ is also a zero row. % $\diamondsuit$ \\[1mm]

\paragraph*{Proof of Theorem~\ref{theoverGLT}.}  Let $\betatilde ^\star$, $\Vare^\star$, $\Mm ^\star$, $\facloadtrue ^\star$, and $\Varetrue^\star $,
 be the matrices that result from deleting  the $s$ spurious rows $n_1, \ldots, n_s$ from the  matrices $\betatilde$, $\Vare$, $\Mm $, $\facloadtrue$,  and $\Varetrue$.  Condition \TS\ for $\facloadtrue$  implies that $\facloadtrue ^\star$ satisfies condition \AR\ and the  variance decomposition
   % \begin{eqnarray}  \label{facuniqr}
  $\Vary ^\star=    \facloadtrue ^\star \trans{(\facloadtrue ^\star) } + \Varetrue^\star $
   % \end{eqnarray}
 is unique. Hence, for  any
GLT matrix  $\facload^\star_r $ of rank $\nfactrue$
%with leading indices $\tilde{l}_{1}, \ldots, \tilde{l}_{\nfactrue}$
that satisfies
$ \facload^\star_r \trans{( \facload^\star_r)}= \facloadtrue ^\star \trans{(\facloadtrue ^\star)}$, we can apply Theorem~\ref{theGLT} to show that
$ \facload_r ^\star  =  \facloadtrue ^\star $ and the leading indices of both matrices are identical.
This strategy
is applied to a submatrix of $\betatilde^\star$.
%, hence the  leading $\tilde{l}_{1}, \ldots, \tilde{l}_{\nfactrue}$  of $\betatilde_r $ are identical to the leading indices $l_{1} , \ldots , l_{\nfactrue}$ of $\facloadtrue $.
%
 Since $\Mm ^\star =\bfzmat$,   we obtain from (\ref{decover})  that
\begin{eqnarray} \label{eqeq}
 \betatilde^\star  \Tm =  \left(\begin{array}{cc}
                 \facloadtrue ^\star & \bfzmat
               \end{array} \right) ,    \qquad   \Vare ^\star =   \Varetrue ^\star,
 \end{eqnarray}
 hence $\betatilde^\star \trans{(\betatilde^\star)} = \facloadtrue ^\star \trans{(\facloadtrue ^\star)}$
 and  $\betatilde^\star$ has reduced rank   $\rank{\betatilde^\star }= \rank{\facloadtrue ^\star} = \nfactrue$.
 %$\facloadtrue ^\star$ has full rank due to condition \TS\ and (\ref{eqeq}) implies that $\rank{\facloadtrue ^\star} =\rank{\betatilde^\star  \Pm}= \rank{\betatilde^\star}=\nfactrue$.
 Since $\betatilde^\star$  is obtained by deleting the $s$ rows ${n_1}, \ldots, {n_s}$
  from an unordered GLT matrix $\betatilde$ of rank $\nfactrue+s$, it follows that $\rank{\betatilde^\star}=\nfactrue$, iff
$\betatilde^\star$ contains exactly $s$ zero columns. This implies  that
 $s$  leading indices of $\betatilde$   are equal to the deleted spurious row indices  $n_1< \ldots < n_s$,
 while the remaining $\nfactrue$  elements  lead the  nonzero columns of $\betatilde^\star$.\footnote{A trivial result about an unordered GLT matrix $\facloadtilde$  is the following: let $\facloadtilde_1$ be a submatrix of $\facloadtilde$ with $q< \nfactrue$ columns and let the submatrix $\facloadtilde_2$ contain  the  $\nfactrue-q$ remaining columns. Then $\facloadtilde_1$ and $\facloadtilde_2$ are unordered GLT matrices
where the leading indices lie in different rows.}
 Hence, a   trivial permutation $\bP _{\pm} \bP _{\rho}$ exists which
yields following representation of $\betatilde^\star$:
\begin{eqnarray} \label{eqeq2}
 \betatilde^\star  \bP _{\pm} \bP _{\rho} =  \left(\begin{array}{cc}
                 \facload^\star_r  & \bfzmat \end{array} \right),
 \end{eqnarray}
 where $\facload^\star_r$ is a GLT  matrix of rank $\nfactrue$ with leading indices being equal to the leading indices $\tilde{l}_{1}< \ldots < \tilde{l}_{\nfactrue}$ of $\betatilde \bP _{\pm} \bP _{\rho}$.
 Application of Theorem~\ref{theGLT} to $\facload^\star_r  $,
  which satisfies $ \facload^\star_r \trans{( \facload^\star_r)} =\betatilde^\star \trans{(\betatilde^\star)} =   \facloadtrue ^\star \trans{(\facloadtrue ^\star)}$,  yields $ \facload_r ^\star  =  \facloadtrue ^\star $. Comparing  representations (\ref{eqeq}) and (\ref{eqeq2}) yields  $\Tm = \bP _{\pm} \bP _{\rho}$ and proves (a).

  To prove (b),  we first show that  $\Mm \trans{\Mm} = \Dm$ is equal to a % being a $\dimy$-dimensional $\Dm$
diagonal matrix  of rank $s$, with  $s$ nonzero entries $d_{n_1}, \ldots, d_{n_s}$  in rows $n_1, \ldots, n_\nspu$.   From $ \rank{\betatilde  \Tm}=\min(\rank{\betatilde} , \rank{\Tm})= \nfactrue+s$, we obtain  that
 $\Mm$ must have full column rank, i.e.~$\rank{\Mm}=s$. Therefore
    $ \rank{\Dm} =\rank{\Mm}=s$  and  only  $\nspu$ diagonal elements $d_{n_1}, \ldots, d_{n_s}$  in  rows $n_1, \ldots, n_\nspu$ are different from 0 in $\Dm$.

 It is straightforward to show that the  matrix $\Mm $ has exactly the same $s$ nonzero rows $n_1, \ldots, n_\nspu$ as $\Dm$: using  %$\Mm \trans{\Mm} = \Dm$, we obtain
for each row $\Mm_{i,\cdot}$ of  $\Mm $ that  $\Mm_{i,\cdot} \trans{\Mm_{i,\cdot}}=\|\Mm_{i,\cdot}\|_2^2=d_i$, %, with $d_i$ being the entry in the $i$th row of $\Dm$.
 it follows for any  $i \neq \{n_1, \ldots, n_\nspu\}$  that $\|\Mm_{i,\cdot}\|_2^2=0$ and, therefore, $\Mm_{i,\cdot}=\bfz$, whereas the remaining rows with $i \in \{n_1, \ldots, n_\nspu\}$ are nonzero since $\|\Mm_{i,\cdot}\|_2^2 >0$.
 The  submatrix $\Mm _0$ of nonzero rows in $\Mm $ satisfies $ \Mm _0 \trans{\Mm _0}= \Dm_0^2$ with $\Dm_0 ^2=\Diag{d_{n_1}, \ldots, d_{n_s} }$ being a diagonal matrix of rank $s$.
 It follows that  $\Dm_0^{-1} \Mm  \trans{\Dm_0^{-1}  \Mm}   = \identm$,
 hence  $ \Dm_0^{-1} \Mm  = \Qm $
 for any arbitrary rotation matrix $\Qm $ of rank $s$. Therefore:
 \begin{eqnarray} \label{decmmm}
  \Mm _0 =  \Dm_0 \Qm , \qquad  \Dm_0= \Diag{d_{n_1}, \ldots, d_{n_s} }^{1/2},
  \end{eqnarray}
for any arbitrary rotation matrix $\Qm $ of rank $s$.

  Since $\Tm = \bP _{\pm} \bP _{\rho}$ is a trivial rotation and  $\betatilde^\star$ is an unordered GLT matrix, we obtain from (\ref{decover}) that also $\Mm$ is a   GLT matrix
  with leading indices  $n_1< \ldots < n_s$. Therefore, the only possible
  rotation $\Qm$ of $\Mm _0 =  \Dm_0 \Qm$  in (\ref{decmmm}) is  equal to sign switching and
  $\Mm$ is a  spurious GLT matrix.  This proves (b).

  Finally, the identity $ \facload_r ^\star  =  \facloadtrue ^\star $  implies in particular that the leading indices  $\tilde{l}_{1}, \ldots, \tilde{l}_{\nfactrue}$  of $\facload^\star_r  $ are identical to the leading indices $l_{1} , \ldots , l_{\nfactrue}$ of $\facloadtrue $.  This proves (c), since the remaining leading indices $\tilde{l}_{\nfactrue+1} < \ldots < \tilde{l}_{\nfactrue+ s}$ are equal to the spurious rows $n_1< \ldots < n_s$, as shown above. $\diamondsuit$\\[1mm]

\subsection{Verifying the row deletion property in practice} \label{mcmcpart}

For $q=1,2$ as well as for $q=\nfactrue -1 , \nfactrue$   the \CountAR\
counting rule,  introduced in  Subsection~\ref{varidesp}  to verify variance identification, can be directly verified for the indicator   matrix
$\deltav$. These  simple counting rules are  outlined in  Corollary~\ref{Lemma1} in  Subsection~\ref{simcount}.
Using Corollary~\ref{Lemma1},  it is easy to  verify, if \NC\ (and hence \AR ) holds for a factor model with up to $\nfactrue \leq 4$ factors.
For   $\nfactrue > 4$, Corollary~\ref{Lemma1} provides  necessary  conditions for \AR\  and helps  to quickly  identify
indicator matrices  $\deltav$ where  \NC\ (and hence \AR ) is violated. If the conditions of  Corollary~\ref{Lemma1} hold,
 then  \NC\   can be verified  by Algorithm~\ref{algARIDE}, outlined in Subsection~\ref{verpartbig}.
 This procedure is summarized in Algorithm~\ref{algVer}.\\

\begin{alg}{{\bf Verifying the row deletion property.}}\label{algVer}
 \begin{itemize}\itemsep 1mm
   \item[(V-1)]  Check the simple counting rules outlined in  Corollary~\ref{Lemma1}. If any of these conditions is violated, then \AR\ does not holds.
  \item[(V-2)] If all conditions are satisfied and $\nfactrue \leq 4$, then \AR\ holds.
 \item[(V-3)] If all conditions are satisfied and $\nfactrue > 4$, then  apply Algorithm~\ref{algARIDE}  to verify \AR .
\end{itemize}
\end{alg}

%the  \CountARfour\ rule can be verified easily by  simple counting rules for $\deltav$,  outlined in  Corollary~\ref{Lemma1}.

% Hence, for $\nfactrue \leq 4$ it is easy to  verify, if \NC\ (and hence \AR ) holds;\footcomment{MOVE APPENDIX?}  see  Subsection~\ref{app:proof} for further comments on Corollary~\ref{Lemma1}.

\subsubsection{Simple counting rules} \label{simcount}

All conditions of the following  corollary are special cases of  \NC . Corollary~\ref{Lemma1} is sufficient for  \NC\  for   $\nfactrue \leq 4$, and necessary, otherwise.\\[1mm]

  \begin{cor}[{\bf Simple counting rules.}]\label{Lemma1}
The following conditions on the indicator matrix $\deltav$ % corresponding to the factor loading matrix $ \facload$
  are  necessary for the row deletion property \AR\ to hold:
%\begin{itemize}
%   \item[(a)]  The total number of  nonzero rows of $\deltav$ is at least equal to   $2\nfactrue +1$, or equivalently,
 % For  $\nfactrue\geq 2$,
 each column contains at least  3 and each pair of columns contains at least 5 nonzero rows, the total number of  nonzero rows  is at least equal to   $2\nfactrue +1$, and  each submatrix of $\nfactrue-1$ columns has at least  $2\nfactrue -1$ nonzero rows,
  or equivalently:
  \begin{eqnarray}
&\ones_{\nfactrue \times \dimy} \cdot   \deltav +   \trans{\deltav} ( \ones_{ \dimy \times  \nfactrue} - \deltav )  \geq   5- 2 \identy{\nfactrue} , &  \label{checkNC2} \\
&\ones_{1 \times \dimy} \cdot  \indic{ \deltav^\star >0} \geq  2\nfactrue +1, \quad     \deltav^\star=  \deltav \cdot \ones_{\nfactrue \times 1}, &
\label{checkNC1} \\
 &  \ones_{1 \times \dimy} \cdot   \indic{\deltav^\star >0}  \geq  2\nfactrue -1 , \quad     \deltav^\star= \deltav ( \ones_{ \dimy \times \dimy} -  \identy{\dimy}),
 & \label{checkNC3}
\end{eqnarray}
where the   indicator function is applied element-wise to  $\deltav^\star$ to define the $\dimy \times \dimy$ matrix $ \indic{\deltav^\star >0}$
and $\ones_{ n  \times k}$ denotes a $ n  \times k$ matrix of ones.
For  $\nfactrue \leq 4$ these conditions are sufficient for  \AR .
\end{cor}

% \paragraph*{Comments on Corollary~\ref{Lemma1}.}

\noindent
 (\ref{checkNC2}) simultaneously checks $q=1$ and $q=2$.
     The matrix on the right hand side of (\ref{checkNC2}) has diagonal elements equal to 3 and off-diagonal elements equal to 5. The  elements of  the matrix on the left hand side are given  by
%\begin{eqnarray*}
$ d_j + \sum_{i=1}^\dimy \delta _{il} (1-\delta _{ij} )$,  where $d_j=\sum_{i=1}^{m} \delta _{ij}$ is the column size.
%\end{eqnarray*}
For $j \neq l$, they  count the number of nonzero rows in column $j$ and $l$
and  they  are equal to  $d_j$ for $j=l$, since $\delta _{ij} (1-\delta _{ij} )=0$ for all rows.

 (\ref{checkNC1})  verifies (for  $q=\nfactrue$)   that the total number of  nonzero rows of $\deltav$ is at least equal to   $2\nfactrue +1$.
Note that the column vector $\deltav^\star$  in (\ref{checkNC1}) is equal to the  row sum.

Finally, (\ref{checkNC3}) correspond to  $q=\nfactrue -1$ and verifies that each submatrix of $\nfactrue-1$ columns has at least  $2\nfactrue -1$ nonzero rows.  The $j$th column of the matrix $\deltav^\star$ appearing in (\ref{checkNC3}) contains the row sums of the submatrix $\deltav_{-j}$ not containing the $j$th column of  $\deltav$.  The  matrix $ \indic{\deltav^\star >0}$  indicates nonzero rows  in $\deltav_{-j}$.
Hence,  the $j$th  element  of  the row vector  $\ones_{1 \times \dimy} \cdot    \indic{\deltav^\star >0}$ counts the number of  nonzero rows  in $\deltav_{-j}$.

 For  $\nfactrue < 4$, some of these conditions overlap, e.g. for $\nfactrue =3$ condition  (\ref{checkNC3}) is covered by  (\ref{checkNC2}).

\subsubsection{Factor models  with more than four factors}  \label{verpartbig}

 The following Algorithm~\ref{algARIDE}  is used to  verify  the row deletion property for factor models with more than four factors.

\begin{alg}{{\bf Verifying the row deletion property for models with more than four factors.}}\label{algARIDE}
Starting with the matrix $\Bsub{1}$ containing  the $\Bsubr{1}= \nfacr$ nonzero  columns and % the $m_n$
 all nonzero rows of  $ \facload$,  we proceed sequentially for $q=1, 2, \ldots$:
  \begin{enumerate} \itemsep 1mm
   \item[(a)]   The submatrix $\Asub{q}$ is constructed
  from  columns of $\Bsub{q}$  where sufficiently many
  measurements  are dedicated  to fewer than $\Bsubr{q}$ factors.
   % to guarantee  that the row deletion property \AR\ holds for  $\Asub{q}$.
   Assume, for instance, that column  $j_1$ of $\Bsub{q}$  contains $m_q\geq 3$  measurements  % in row $\{i_1,\ldots ,i_{d_q}\}$
     that are dedicated to factor $j_1$, i.e.  for all   rows $ i \in \{i_1,\ldots ,i_{m_q}\}$:
  $\delta_{i, j_1}=1$ and $\delta_{i j}=0$ for $j\neq j_1$ (note that the remaining measurements $ i \notin \{i_1,\ldots ,i_{m_q}\}$ need not be dedicated).
    If we define % $\Asub{q}$  and  $\Bsub{q+1}$ are constructed  in the following way.
    $\Asub{q}$ as the  vector   containing all  dedicated measurements
    in column $j_1$  in rows $\{i_1,\ldots ,i_{m_q}\}$, then  evidently,  $\Asub{q}$ satisfies \NC\ with $r_q=1$.
    % and therefore \AR\ holds for    $ \facload$.

     \item[(a*)]  If $\Bsub{q}$ does not contain such a column $j_1$,   assume that a pair of  columns  $(j_1,j_2)$ of $\Bsub{q}$  contains  at least $m_q\geq 5$ measurements
% in row $\{i_1,\ldots ,i_{d_q}\}$
 that are dedicated to factor $j_1$ or $j_2$ or both,  i.e.  for all $ i \in \{i_1,\ldots ,i_{m_q}\}$,
 $\delta_{i j}=0$ for $j\neq (j_1,j_2)$ (again the remaining measurements $ i \notin \{i_1,\ldots ,i_{m_q}\}$ need not be dedicated).
%Then the rows and the columns of $\Bsub{q}$ are permuted such that the upper part contains only  measurements that are dedicated to the first two factor  and  $\Asub{q}$  and  $\Bsub{q+1}$ are constructed  in the following way.  $\Asub{q}$  is  equal to the two columns  containing the   factor loadings of the dedicated measurements, whereas the submatrix  $\Bsub{q+1}$ is obtained from the remaining measurements by  deleting  the first two columns  and then deleting all zero rows in the remaining loading matrix.
In this case,  % $\Asub{q}$  and  $\Bsub{q+1}$ are constructed  in the following way.
$\Asub{q}$ with  $r_q=2$   is  constructed from  all   measurements %  in row $\{i_1,\ldots ,i_{m_q}\}$
    dedicated to  column $j_1$ and $j_2$.  If no such columns exist, we  search for least $m_q\geq 7$ measurements % in row $\{i_1,\ldots ,i_{m_q}\}$
 that are dedicated to three factor  (leading to  $\Asub{q}$ with $r_q=3$) or at least $m_q\geq 9$ measurements % in row $\{i_1,\ldots ,i_{m_q}\}$
 that are dedicated to four factors (leading to $\Asub{q}$ with $r_q=4$) . In any of these cases,  Corollary~\ref{Lemma1} implies
that  $\Asub{q}$ satisfies \NC\ with $r=r_q$.

   \item[(b)] If  a suitable matrix $\Asub{q}$ has been identified,  then  a submatrix  $\Bsub{q+1}$  with  $\Bsubr{q+1}=\Bsubr{q}-r_q$ columns  is determined   from  $\Bsub{q}$  by  first removing  the  columns   corresponding to $\Asub{q}$  and then deleting  all zero rows in the resulting matrix.
   Then, we apply  Theorem~\ref{Lemma2} with $Q=q+1$,   $\Asub{Q}=\Bsub{q+1}$, and $r_Q=\Bsubr{q+1}$:
 %If  no suitable submatrix $\Asub{q}$ has been identified,  we apply  Theorem~\ref{Lemma2} with $Q=q$,   $\Asub{Q}=\Bsub{q}$, and $r_Q=\Bsubr{q}$:
 \begin{enumerate}
\item[(b1)]   Check, if   $\Asub{Q}$ satisfies the simple counting rules in  Corollary~\ref{Lemma1} with $r=r_Q$.
   \item[(b2)]  If the simple counting rules are not fulfilled for $\Asub{Q}$, then Theorem~\ref{Lemma2},  part~(b)  implies that  the  factor loading matrix $ \facload$  does not satisfy \AR\ and the procedure is terminated.

   \item[(b3)] If the simple counting rules are fulfilled for $\Asub{Q}$ and $r_Q\leq 4$, then  Theorem~\ref{Lemma2}, part~(a)  implies that  the  factor loading matrix $ \facload$   satisfies \AR\ and the procedure is terminated.

   \item[(b4)] If the simple counting rules are fulfilled for $\Asub{Q}$ and $r_Q> 4$, then two options exist:
   \begin{itemize}
     \item[(b4-A)]  The search procedure is continued by  increasing $q$ by 1 and searching for a suitable submatrix $\Asub{q+1}$ in Step~(a) (or~(a*)).
    \item[(b4-B)]   The remaining counting rules of \NC\ are verified  for $\Asub{Q}$ and the procedure is terminated. Depending on the outcome,    Theorem~\ref{Lemma2}  implies  that the  factor loading matrix $ \facload$   either satisfies \AR\ or not.
   \end{itemize}

 \end{enumerate}

    \item[(c)] If  no suitable submatrix $\Asub{q}$ has been identified in Step~(a) (or~(a*)),  then we apply  Theorem~\ref{Lemma2} with $Q=q$,   $\Asub{Q}=\Bsub{q}$,
 and $r_Q=\Bsubr{q}$ and proceed as in step (b1)-(b4) above, with step (b4-B)  being the only option in step~(b4).

 \end{enumerate}
\end{alg}

%If $\rank{\Bsub{q+1}}>2$  (in which case  it is not easy to verify if the submatrix  $\Bsub{q+1}$  satisfies \AR ), then the search procedure is continued by identifying a suitable submatrix $\Asub{q+1}$ of  $\Bsub{q+1}$ as described above, until for some  $q=Q-1$ a submatrix  $\Bsub{q+1}=\Bsub{Q}$ is obtained that allows a clear distinction between the two cases described above.

   %Then the rows and the columns of $\Bsub{q}$ are permuted such that the upper part contains only  measurements that are dedicated to the first factor  and  $\Asub{q}$  and  $\Bsub{q+1}$ are constructed  in the following way.  $\Asub{q}$  is  equal to the subvector containing the  corresponding nonzero factor loading, whereas the submatrix  $\Bsub{q+1}$ is obtained from the remaining measurements by deleting  the first column  and then deleting all zero rows in the remaining loading matrix.

  %  The submatrix  $\Bsub{q+1}$ is obtained from the remaining measurements in matrix   $\Bsub{q}$ by  first removing   columns $j_1$ and  $j_2$   and  then deleting all zero rows in the resulting loading matrix.

\newpage

\section{Details on MCMC estimation}   \label{app:mcmc}

The designer MCMC  scheme introduced in Algorithm~\ref{Algo3}  in Subsection~\ref{mcmc} on  one hand includes  standard steps of MCMC estimation for the basic factor model such as Step~(F) and Step~(P). In Subsection~\ref{mcmcgen}, these standard steps  are revisited and modifications and improvements are discussed in the light of sparsity.  On the other hand,  Algorithm~\ref{Algo3}  % as well as postprocessing the MCMC as discussed in Subsection~\ref{exbaykrandom}
includes  a number of MCMC  steps that are specifically designed  to achieve identification for a GLT structure with an unknown number of factors, such as Step~(R), Step~(L)  and Step~(D).  Full details for these steps are provided in Subsection~\ref{mcmcide}.
 Finally, Subsection~\ref{accelerate_App}  provides further details for  the boosting Step~(A) of Algorithm~\ref{Algo3}.

\subsection{Revisiting MCMC for factor models in the light of sparsity}   \label{mcmcgen}

%\subsection{MCMC and data augmentation}
%

In this subsection, various standard steps of MCMC estimation for the basic factor model are revisited and modifications and improvements are suggested  in the light of sparsity.

\subsubsection{Sampling the latent factors}    \label{SectionF_fac}

Step~(F) in Algorithm~\ref{Algo3} is a standard step in Bayesian factor analysis, see \citet{gew-sin:int} and \citet{lop-wes:bay}, among many others.  Given prior independence of the factors,  the joint posterior $ p(\facm_1,\ldots,\facm_T|\facload,\idiov_1,\ldots,\idiov_{\dimy},\ym) $ factors into $T$ independent normal distributions  given by:
 \begin{eqnarray}  \label{filtPXsim}
{\facm}_t
|\ym_t,  \facload, \Vare \sim \Normult{\nfac}{(\identy{\nfac} + \trans{\facload} \Vare ^{-1} \facload) ^{-1} \trans{\facload} \Vare ^{-1} \ym_t , (\identy{\nfac} + \trans{\facload}  \Vare ^{-1}  \facload) ^{-1} }.
\end{eqnarray}
A simplification is possible for sparse Bayesian factor models, as usually $k-\nfacr$ columns  and $\dimy_0$ rows  of the coefficient matrix $ \facload$ are equal to zero.  Let  $k_1,\ldots,k_{k-\nfacr}$ and $j_1, \ldots, j_{\nfacr}$ denote,  respectively, the column indices of the zero and the nonzero columns.
Evidently, the posterior of the  latent factors
$ \fac_{k_1,t}, \ldots, \fac_{k_{k-\nfacr},t} $  of the  zero columns  is equal to the prior for $t=1,\ldots, T$.
% Since the covariance matrix in (\ref{facfiltKF}) is the same for all $t$,
 An  extremely efficient sampling step is available for jointly sampling  the
factors $ \tilde{\facm}_{t}=(\fac_{j_1,t}, \ldots, \fac_{j_{\nfacr},t}) $ of the nonzero columns
simultaneously % for all  $\nfacr$ nonzero columns and
for all observations $t=1, \ldots,T$. Step~(F) for sparse factor models is summarized in  Algorithm~\ref{algofact}.

%\begin[]{alg} \label{algofact}
\begin{alg}{\textbf{Sampling the latent factors for  a sparse Bayesian factor model}} \label{algofact}
\begin{itemize}
  \item[(F-a)] Sample $\fac_{jt} \sim \Normal{0,1}$ for  all zero columns  $j \in \{ k_1,\ldots, k_{k-\nfacr}\}$, for $ t=1,\ldots, T$.
  \item[(F-b)] Let $\zm$ be a $\dimmat{\nfacr}{T}$  array of \iid\ random variables $z_{jt} \sim \Normal{0,1}$; %, generated from the standard normal distribution.
      let $\betatilde$ be the $\dimmat{(\dimy-\dimy_0)}{\nfacr}$ matrix containing the nonzero columns  and the nonzero rows of $\facload$,
      let  $\Sigmatilde$ be the covariance matrix of the corresponding  idiosyncratic errors, and
      let $\tilde{\ym}$  be the  $\dimmat{(\dimy-\dimy_0)}{T}$  array, where the $t$th column $\tilde{\ym}_t$ corresponds to  $\ym_t$, with the uncorrelated measurements  corresponding to the zero rows of  $\facload$ being removed.
      The factors  $ \tilde{\facm}_{t}=(\fac_{j_1,t}, \ldots, \fac_{j_{\nfacr},t}) $ then are given as the $t$th column of the $\dimmat{\nfacr}{T}$ array  $\tilde{\facm}$, generated in the following way:
 \begin{eqnarray} \label{facfiltKF}
  && \Pm=(\identy{\nfacr} + \trans{\betatilde} \Sigmatilde ^{-1} \betatilde) ^{-1},  \\
&&  \Km=  \Pm \trans{\betatilde} \Sigmatilde ^{-1},  \quad   \Cm  \trans{\Cm}= \Pm,   \nonumber \\
&& \tilde{\facm}= \Km \tilde{\ym} + \Cm \zm.  \nonumber
%&\tilde{\facm}= \Km \tilde{\ym} + \Cm \zm,   & \\
%&   \quad   \Km=  \Pm \trans{\betatilde} \Vare ^{-1},  \qquad
 %\Pm=(\identy{\nfacr} + \trans{\betatilde} \Vare ^{-1} \betatilde) ^{-1}, \qquad \Cm  \trans{\Cm}= \Pm,  \nonumber
\end{eqnarray}
\end{itemize}
\end{alg}

\vspace*{2mm} \noindent
%\noindent {\em Proof.}
 (\ref{facfiltKF}) is easily derived  from (\ref{filtPXsim}):
\begin{eqnarray} \label{facsubKF}
\tilde{\facm}_t
|\ym_t,  \facload, \Sigmatilde \sim \Normult{\nfacr}{(\identy{\nfacr} + \trans{\betatilde} \Sigmatilde ^{-1} \betatilde) ^{-1} \trans{\betatilde} \Sigmatilde ^{-1} \tilde{\ym}_t , (\identy{\nfacr} + \trans{\betatilde} \Sigmatilde ^{-1}  \betatilde) ^{-1} },
\end{eqnarray}
where $\betatilde$, $\Sigmatilde$ and $\tilde{\ym}_t$ are defined as above.  Since the covariance matrix in (\ref{facsubKF}) is the same for all $t$,  a single Cholesky decomposition of the covariance matrix $\Pm$ is required to sample the factors $\tilde{\facm}_t$ for all $t$.  The simulation step in (\ref{facfiltKF}) is a vectorized version of this sampling step, which does not require any loop over $t$. % $\diamondsuit$ \\[1mm]

 %A problem of special interest for sparse factor models, that  will be relevant in Subsection~\ref{DimChange}, is sampling the latent factors for a spurious column. Assume that  the only nonzero element in that column, say $j$, lies in row $l_j$, i.e. $\delta_{l_j,j}=1$. ADDSP

\subsubsection{Posterior distributions in a confirmatory sparse factor model} \label{postdisfac}

Step~(P) of Algorithm~\ref{Algo3}  updates the parameters in a  confirmatory sparse factor model, where  the  indicator matrix $\deltav$ imposes
a certain zero structure on the loading matrix.
The joint posterior  distribution $p(\facload_{i\cdot}^{\deltav}, \idiov_i| \ym,\facm ,\deltav)$
of the nonzero factor loadings $\facload_{i\cdot}^{\deltav}$ and the idiosyncratic variance  $\idiov_i$ is derived for each row $i$ ($i=1, \ldots, \dimy$)
conditional on the factors  $\facm$ and the  indicator matrix $\deltav$
from  the  following  regression model:
\begin{eqnarray}
\tilde{\ym}_i= \Xb_i ^{\deltav} \facload_{i\cdot}^{\deltav} + \tilde{\errorm}_i,
 \label{regnonpAPP}
\end{eqnarray}
where  $\tilde{\ym}_i=\trans{(y_{i1} \cdots y_{iT})}$ and
$\tilde{\errorm}_i=\trans{(\error_{i1} \cdots \error_{iT})} \sim \Normult{T}{0,  \idiov_i \identm}$.   $\Xb_i ^{\deltav}$ is a regressor matrix  for
 $\facload_{i\cdot}^{\deltav}$  constructed from the $\dimmat{T}{\nfac}$ dimensional latent factor   matrix $\Fm=\trans{(\facm_1 \cdots \facm_T)}$ in the following way.
 If no element  in row $i$ of $\facload$ is restricted to 0, then  $\Xb_i ^{\deltav}=\Fm$.
 If some elements are restricted to 0, then $ \Xb_i ^{\deltav}$ is obtained from  $\Fm$  by deleting all columns $j$ where $\delta_{ij}=0$, i.e. $ \Xb_i ^{\deltav}=   \Fm \Pim_i ^{\deltav} $, where $\Pim_i ^{\deltav}$ is a $\dimmat{\nfac}{\sum_{j=1}^{\nfac} \delta_{ij} }$ selection matrix, selecting those columns $j$ of $\Fm$ where $\delta_{ij}\neq 0$.

Concerning $\idiov_i$, the inverted Gamma prior (\ref{priorsiidg}) with prior moments $ c_0 $ and  $ C_{i0}$ is considered. The precise form of  $p(\facload_{i\cdot}^{\deltav}, \idiov_i| \ym,\facm ,\deltav)$   depends
 the prior chosen for  $\facload_{i\cdot}^{\deltav}|  \idiov_i$, and has different posterior moments for
 the  standard prior  (\ref{prior1}) and for  the  fractional prior (\ref{priorfrac}).
 In a sparse factor model,  the dimension of this posterior depends
on the number of nonzero elements in the $i$th row of  $\facload$,  i.e.  $q_i=\sum_{j=1}^{\nfac} \delta_{ij}$.
There are basically three types of rows, when  it comes to updating the parameters: zero rows, dedicated rows and
rows with multiple loadings.

 \paragraph{Zero rows.} For zero rows (i.e.~$q_i=0$),  (\ref{regnonpAPP}) reduces to a \lq\lq null\rq\rq\ model without regressors $\Xb_i ^{\deltav}$,
 that is  $\tilde{\ym}_i=  \tilde{\errorm}_i$. Hence,   the posterior of $ \idiov_i$
is  simply given by
 \begin{eqnarray} \label{idiozero} %&&
\idiov_i|\tilde{\ym}_i,\facm, \deltav  \sim \Gammainv{c_{T} ^{\nullmod},C_{iT}^{\nullmod}}, %\\&&
\qquad c_{T}^{\nullmod}= c_0 +\frac{T}{2}, \qquad  C_{iT}^{\nullmod}=C_{i0} +\frac{1}{2}\sum_{t=1}^T  y_{it}^2 .
\end{eqnarray}

\paragraph{Non zero  rows.}
For all nonzero rows (i.e.~$q_i>0)$, the posterior    $(\facload_{i\cdot}^{\deltav}, \idiov_i)$ for  a specific row $i$ is given by:
 \begin{eqnarray}  \label{idiopos}   \label{rowpos}
 \idiov_i|  \tilde{\ym}_i,\facm, \deltav  \sim \Gammainv{c_{T},C_{iT}^{\deltav}},  \qquad %\\&& \label{rowpos}
\facload_{i\cdot}^{\deltav} | \idiov_i, \tilde{\ym}_i, \facm, \deltav
 \sim \Normult{q_i}{\bV_{iT} ^{\deltav} \cv_{iT} ^{\deltav}, \bV_{iT} ^{\deltav}\idiov_i}.
\end{eqnarray}
For the standard prior (\ref{prior1}),  the moments are given by:
\begin{eqnarray}  \label{postmomunit}
& ( \bV_{iT} ^{\deltav} ) ^{-1}  =  (\bV_{i0} ^{\deltav} ) ^{-1} + \trans{(\Xb_i^{\deltav})} \Xb_i^{\deltav} , \qquad
 \cv_{iT} ^{\deltav} =   \trans{(\Xb_i ^{\deltav})}\tilde{\ym}_i, &\\
 & c_T = c_0 +\frac{T}{2}, \qquad  % \\&&
  C_{iT}^{\deltav} =C_{i0} +  \frac {1}{2} \SSR_i , \quad
   \SSR_i= \ \trans{\tilde{\ym}_i} \tilde{\ym}_i - \trans{(\cv_{iT} ^{\deltav})}
  \bV_{iT}^{\deltav}  \cv_{iT} ^{\deltav}. & \nonumber %  \label{SSRI2} .
\end{eqnarray}
 For the fractional prior (\ref{priorfrac}),  the moments are given by:
 \begin{eqnarray}%&&
&  (\bV_{iT} ^{\deltav}) ^{-1}   = \label{postmomfrac}
 \trans{(\Xb_i ^{\deltav})} \Xb_i ^{\deltav}  , \quad  \cv_{iT} ^{\deltav} =   \trans{(\Xb_i ^{\deltav})}\tilde{\ym}_i, &\\[1mm]
 &  c_T  = c_0 +\frac{(1-b)T}{2},  \quad%\\&&
C_{iT}^{\deltav} = C_{i0}+ \frac {(1-b)}{2}  \SSR_i ,  \nonumber % \label{SSRI}
%\quad   \SSR_i= \ \trans{\tilde{\ym}_i} \tilde{\ym}_i - \trans{(\cv_{iT} ^{\deltav})}  \bV_{iT}^{\deltav}  \cv_{iT} ^{\deltav}. &
\end{eqnarray}
where $\SSR_i$ is defined the same way as in  (\ref{postmomunit}).
It is easy to show that for the fractional prior, $\SSR_i$ is identical to  the residual sum of squares errors.\footnote{If  the residual $\errorm_i = \tilde{\ym}_i -   \Xb_i^{\deltav}   \bV_{iT} ^{\deltav} \cv_{iT} ^{\deltav}$ is defined in the usual way, then:
\begin{eqnarray*} %&&
 \trans{\errorm}_i\errorm _i &=&  \trans{\tilde{\ym}_i} \tilde{\ym}_i -    \trans{(\cv_{iT} ^{\deltav})}   \bV_{iT} ^{\deltav} \trans{(\Xb_i^{\deltav} )}  \tilde{\ym}_i-
\trans{ \tilde{\ym}_i }  \Xb_i^{\deltav}  \bV_{iT} ^{\deltav}   \cv_{iT} ^{\deltav}
+   \trans{(\cv_{iT} ^{\deltav})}    \bV_{iT} ^{\deltav}    \trans{(\Xb_i^{\deltav})}  \Xb_i^{\deltav}   \bV_{iT} ^{\deltav} \cv_{iT} ^{\deltav}\\
&=&   \trans{\tilde{\ym}_i} \tilde{\ym}_i -
 \trans{(\cv_{iT} ^{\deltav})}   \bV_{iT} ^{\deltav}  \cv_{iT} ^{\deltav}
-   \trans{(\cv_{iT} ^{\deltav})}  \bV_{iT} ^{\deltav}   \cv_{iT} ^{\deltav}
+   \trans{(\cv_{iT} ^{\deltav})}    \bV_{iT} ^{\deltav}    \cv_{iT} ^{\deltav} = \SSR_i.
\end{eqnarray*}}
 % with $\errorm_i = \tilde{\ym}_i -  \Xb_i ^{\deltav} \bm_{iT}$.

\paragraph{Dedicated rows.}
For dedicated rows (i.e.~$q_i=1$)  only a single nonzero factor loading $\load_{i,\ji}$ is present in a particular column $\ji$ and the posterior
given in (\ref{idiopos}) simplifies considerably:
%  $p(\load_{i,\ji}, \idiov_i| \ym,\facm,\deltav)$
  \begin{eqnarray}  \label{rowdedi}
 \idiov_i|  \tilde{\ym}_i,\facm,\deltav  \sim \Gammainv{c_{T},C_{iT}}, \qquad %\\&& \label{rowpos}
\load_{i,\ji} | \idiov_i, \tilde{\ym}_i, \facm ,\deltav
 \sim \Normal{b_{iT} ,  B_{iT}  \idiov_i}.
\end{eqnarray}
%  with moments that do not depend  does not depend on any other indicator in $\deltav_{-(i,j_i)}$.
For a fractional prior,
 the posterior moments   are given  by:
  \begin{eqnarray}  \label{singloadfr}
&  B_{iT} = 1/ \left(\sum_{t=1}^T \fac_{\ji,t}^2\right),  \quad  b_{iT} =   B_{iT}  \left(\sum_{t=1}^T \fac_{\ji,t} y_{it}\right), &\\
&   c_T  = c_0 +\frac{(1-b)T}{2},  \quad C_{iT}= C_{i0}+ \frac {(1-b)}{2} \sum_{t=1}^T  (y_{it}- \fac_{\ji,t} b_{iT})^2. & \nonumber
\end{eqnarray}
  and  for  the standard prior by
\begin{eqnarray}
&   B_{iT}  = 1  /(B_{i0,(\ji,\ji)}^{-1} +  \sum_{t=1}^T \fac_{\ji,t}^2),
 \quad  b_{iT} = B_{iT} \left(\sum_{t=1}^T \fac_{\ji, t} y_{it} \right),  & \label{singloadfi}\\
& c_T = c_0 +\frac{T}{2}, \qquad  C_{iT}=C_{i0} +\frac{1}{2}\sum_{t=1}^T  (y_{it}- \fac_{\ji, t} b_{iT})^2
% \frac{1}{2B_{i0,jj}}(b_{iT}-b_{i0,j})^2. % \label{c2t}    \\ % b_{i0,j} different from 0
-b_{iT}^2 /(2B_{i0,(\ji,\ji)}) , & \nonumber
\end{eqnarray}
where
  $B_{i0,(\ji,\ji)}$ is $\ji$th diagonal element  of the prior covariance  matrix $\bV_{i0}$.

\subsubsection{Block sampling of idiosyncratic variances and factor loadings} \label{jointfac}

Step~(P)  in Algorithm~\ref{Algo3} could be implemented as in \citet{lop-wes:bay},  by sampling $\facload_{i\cdot}^{\deltav}$ and $ \idiov_i$ from  the posterior  distribution $p(\facload_{i\cdot}^{\deltav}, \idiov_i| \ym,\facm ,\deltav)$ derived in Subsection~\ref{postdisfac} row by row.
However,   an important improvement  is  feasible
through block sampling of all idiosyncratic variances and all nonzero factor loadings, summarized in Algorithm~\ref{algoP}.\footnote{This algorithm has been implemented for the first time in the unpublished research report by \citet{fru-lop:par}.}
The use of  the Cholesky decomposition of the information matrix (instead of the covariance) to sample  from a high-dimensional
density  is  fashioned  after  \citet[Theorem~2.5 and Algorithm~2.5]{rue-hel:gau} who consider  Gaussian random fields.

% This algorithm requires that the parameters $(\facload_{i\cdot}^{\deltav}, \idiov_i)$ are independent {\em a posteriori} across rows conditional in the factors $\facm$  and conditional on the hyper parameters in the definition of the priors  of  $\facload_{i\cdot}^{\deltav}$ and  $\idiov_i$.

\begin{alg}{\textbf{Sampling parameters for a sparse Bayesian factor model }} \label{algoP}
\begin{itemize}
  \item[(P-a)] For all zero rows, sample  $ \idiov_i$ from (\ref{idiozero}), which can  be trivially vectorized.

    \item[(P-b)] If the remaining rows are all dedicated  with  a single nonzero loading in column $j_i$ (which can be different  for different rows), then sampling from  (\ref{rowdedi}) is easily vectorized, since all posterior  moments
      are  univariate.% \footcomment{Is not implemented so far in MATLAB. Implement?}

   \item[(P-c)] Even if some of the nonzero  rows are not dedicated,  joint sampling of all idiosyncratic variances and all factor loadings  is  feasible for all nonzero rows. Let $i_1, \ldots,i_n$ be the indices of  all $n=\dimy - m_0$ nonzero rows of $\facload$, i.e.  $q_{i_l}>0$ for  $l=1,\ldots,n $. Let $\facload^{\deltav}=(\facload_{i_1\cdot}^{\deltav}, \ldots, \facload_{i_n\cdot}^{\deltav})$ be a vector obtained by stacking row by row all nonzero elements in each  row. Let $d=\sum_i q_i$ be the total number of nonzero elements in $\facload^{\deltav}$.
 To sample the idiosyncratic variances $\idiov_{i_1}, \ldots,\idiov_{i_n}$ and the nonzero factor loadings $\facload^{\deltav}$ jointly, proceed in the following way:
\begin{enumerate}
  \item[(P-c1)] Construct  the information matrix $\Omegav$ and the covector $\cv$ of the joint posterior $$\facload^{\deltav}| \idiov_{i_1}, \ldots,\idiov_{i_n} ,\facm,\ym  \sim \Normult{d}{ \Omegav^{-1} \cv,   \Omegav^{-1} \mathbf{D} }.$$
  The matrix   $\mathbf{D} = \Diag{ \idiov_{i_1} \unit{1 \times q_{i_1}} \cdots \idiov_{i_n} \unit{1 \times q_{i_n}}}$,
with $\unit{1 \times l}$  being a $\dimmat{1}{l}$ row vector of ones, is a $\dimmat{d}{d}$ diagonal matrix containing the idiosyncratic variances, while  the  $\dimmat{d}{d}$  matrix  $\Omegav$ and  the $\dimmat{d}{1}$  vector  $\cv$ are given by:
\begin{eqnarray*}
&&\Omegav = \left(
              \begin{array}{cccc}
                (\bV_{i_1,T} ^{\deltav}) ^{-1}  & \bfzmat & \cdots & \bfzmat \\
                \bfzmat &   (\bV_{i_2,T}^{\deltav}) ^{-1} & \ddots & \vdots \\
                   \vdots     &     \ddots               & \ddots & \bfzmat\\
                \bfzmat &  \cdots & \bfzmat &  (\bV_{i_n,T} ^{\deltav}) ^{-1} \\
              \end{array}
            \right),
 \qquad \cv= \left(
              \begin{array}{c}
                \cv_{i_1,T} ^{\deltav} \\
                \vdots  \\
                \cv_{i_n,T} ^{\deltav} \\
              \end{array}
            \right) ,
\end{eqnarray*}
where $(\bV_{i_l,T} ^{\deltav}) ^{-1}$ and $\cv_{i_l,T} ^{\deltav}$ are
 the information matrix and the covector  appearing in
the    posterior (\ref{rowpos}) of the nonzero elements in row $i_l$. $\Omegav$ is a sparse  band matrix with maximal band width equal to $\max q_{i_l}$.

  \item[(P-c2)]     Compute the Cholesky decomposition
  $\Omegav=\Lv \trans{\Lv}$, where $\Lv$ is lower triangular, using a special algorithm developed for
   band matrices.   Next, solve $\Lv \xm = \cv$ for $\xm$ using  an algorithm  specially designed for triangular matrices.
  Evidently,  $\xm$ is a $\dimmat{d}{1}$  vector.

   \item[(P-c3)] Sample  $\idiov_{i_1},\ldots,  \idiov_{i_n}$  jointly from  (\ref{idiopos}).
  The squared sum $\trans{\xm} \xm$  can be used to vectorize the computation of  $C_{i_l,T}^{\deltav}$  for  each $l=1, \ldots, n$, since
    \begin{eqnarray} \label{citfrac}
   \trans{\xm_{i_l}} \xm_{i_l} =   \trans{(\cv_{i_l,T} ^{\deltav})} \bV_{i_l,T}^{\deltav } \cv_{i_l,T} ^{\deltav},
    \end{eqnarray}
     where  $ \xm_{i_l}$  is the $q_{i_l}$-dimensional sub vector  of $\xm$ corresponding to  $\facload_{i_l,\cdot}^{\deltav}$.

 \item[(P-c4)]   Finally, define the   diagonal matrix $\mathbf{D}$  from $ \idiov_{i_1}, \ldots,
\idiov_{i_n}$  as described above and draw $\zm \sim \Normult{d}{\bfz, \mathbf{D}}$. Solving the system
 \begin{eqnarray} \label{pposbb}
 \trans{\Lv} \facload^{\deltav}= \xm +   \zm
%\qquad  \zm \sim \Normult{d}{\bfz, \mathbf{D}},
\end{eqnarray}
for $\facload^{\deltav}$ leads to a  draw  from  the joint posterior
$ \facload^{\deltav}| \idiov_{i_1}, \ldots, \idiov_{i_n},\ym,\facm$.
\end{enumerate}
 \end{itemize}
\end{alg}

\vspace*{1mm}  \noindent
 To derive  (\ref{citfrac}),  let $\Lv_{i_l}$ be the $\dimmat{q_{i_l}}{q_{i_l}}$
 submatrix of $\Lv$ corresponding to  $\facload_{i_l,\cdot}^{\deltav}$.  Evidently, $\Lv_{i_l}$  is
 equal to the Cholesky decomposition of the individual
 information matrix $(\bV_{i_l,T}^{\deltav}) ^{-1}$. Furthermore,
 the $q_{i_l}$-dimensional sub vector
   $\xm_{i_l}$ corresponding to  $\facload_{i_l,\cdot}^{\deltav}$ satisfies $\Lv_{i_l}\xm_{i_l}=\cv_{i_l,T} ^{\deltav}$.
   Therefore:
   \begin{eqnarray*}
&&  \trans{\xm_{i_l}} \xm_{i_l} =   \trans{(\cv_{i_l,T} ^{\deltav})}
  (\trans{\Lv_{i_l}}) ^{-1} \Lv_{i_l} ^{-1}  \cv_{i_l,T} ^{\deltav}=
 \trans{(\cv_{i_l,T} ^{\deltav})} (\Lv_{i_l} \trans{\Lv_{i_l}} ) ^{-1}  \cv_{i_l,T} ^{\deltav} =
 \trans{(\cv_{i_l,T} ^{\deltav})} \bV_{i_l,T}^{\deltav } \cv_{i_l,T} ^{\deltav}.
\end{eqnarray*}
 It is easy to  prove that the solution $\facload^{\deltav}$ of (\ref{pposbb}) is a draw from the posterior
 $p(\facload^{\deltav}|\idiov_{i_1}, \ldots, \idiov_{i_n},\ym,\facm)$.  Note that
%\begin{eqnarray*}&&
$\Lv \trans{\Lv} \facload^{\deltav}= \Lv \xm + \Lv  \zm  = \cv + \Lv  \zm $.
%\end{eqnarray*}
Therefore
\begin{eqnarray*}
&&   \facload^{\deltav}= ( \Lv \trans{\Lv})^{-1} \cv + ( \Lv \trans{\Lv})^{-1} \Lv  \zm=  \Omegav ^{-1} \cv + (\trans{\Lv})^{-1}  \zm.
\end{eqnarray*}
Evidently,  $\Ew{\facload^{\deltav}}= \Omegav ^{-1} \cv$.   Since
   for each $l=1, \ldots,n$,  $  \Lv_{i_l} \idiov_{i_l}  = \idiov_{i_l}\Lv_{i_l} $, it holds that $  \Lv \mathbf{D} =  \mathbf{D} \Lv $
   and therefore $\mathbf{D}  \Lv  ^{-1}=   \Lv ^{-1}\mathbf{D}$. Since $\V{\facload^{\deltav}}=  (\trans{\Lv})^{-1}  \mathbf{D}  \Lv  ^{-1}=  (\trans{\Lv})^{-1}   \Lv  ^{-1}\mathbf{D} = \Omegav ^{-1}\mathbf{D}$,
it follows  that
   $ \facload^{\deltav} \sim \Normult{d}{\Omegav ^{-1} \cv,\Omegav ^{-1}\mathbf{D}}$.

\subsubsection{Marginal likelihoods when the factors are known} \label{marfac}

Although we work throughout  this paper with a factor model where the factors $\facm_t$ are latent,
Step~(L) and Step~(D) of Algorithm~\ref{Algo3} perform model selection with respect to $\deltav$  conditional
on the most recent draw of  the factors $\facm=(\facm_1, \ldots, \facm_T)$. Hence, to sample new indicators $\deltav_{i \cdot}$ in row $i$,
 the marginal likelihood $p(\tilde{\ym}_i|  \facm, \deltav_{i \cdot})$
of  regression model  (\ref{regnonpAPP}) is  needed.

If   $\deltav_{i \cdot}$ is a zero row (i.e $q_i=0$), then the marginal likelihood  simplifies to
\begin{eqnarray}
 p(\tilde{\ym}_i|  \facm, \deltav_{i \cdot})  =  p(\tilde{\ym}_i) =
   \frac{\Gamfun{c_{T}^{\nullmod} } (C_{i0})^{c_0} }
{(2\pi)^{T/2}\Gamfun{c_0} (C_{iT}^{\nullmod})^{c_{T}^{\nullmod}}} ,  \label{marnull}
\end{eqnarray}
where $c_{T}^{\nullmod}$ and $C_{iT}^{\nullmod}$ are the posterior moments of $\idiov_i$ under
 the \lq\lq null\rq\rq\ model given by (\ref{idiozero}).
If at least one element of $\deltav_{i \cdot}$ is different from zero,
then the marginal likelihood  computation differs between the standard prior  (\ref{prior1})  and the fractional prior (\ref{priorfrac}).

  \paragraph{Marginal likelihoods for the standard prior.}

  For the standard prior,  a  well-known  exercise in Bayesian regression analysis yields:
\begin{eqnarray}
 p(\tilde{\ym}_i|  \deltav_{i \cdot}, \facm )  =
  \frac{1}{(2\pi)^{T/2} }   \frac{|\bV_{iT}^{\deltav}|^{1/2}}{|\bV_{i0}^{\deltav } |^{1/2}}
   \frac{\Gamfun{c_{T}} (C_{i0})^{c_0} }
{\Gamfun{c_0} (C_{iT}^{\deltav})^{c_{T}}} ,  \label{ADDGAD}
\end{eqnarray}
where $\bV_{iT} ^{\deltav}$, $c_{T}$ and $C_{iT}^{\deltav}$ are the posterior moments of
 $p(\facload^{\deltav}_{i\cdot},\idiov_i|\deltav_{i\cdot},\tilde{\ym}_i, \facm)$
 given by (\ref{postmomunit}).

  \paragraph{Marginal likelihoods for a fractional prior.}

 For  a fractional prior, the derivation of the marginal likelihood  is less standard  and can be
   obtained in a  similar way as in \citet{fru-wag:sto}.
   A fraction $b$ of the full conditional likelihood  of regression model  (\ref{regnonpAPP})
   is used to  define  the fractional prior $p(  \facload_{i\cdot}^{\deltav} | \idiov_i \addb ,\facm )$: % (\ref{priorfrac}):
    \begin{eqnarray*}
          p(\tilde{\ym}_i|  \facm, \facload_{i\cdot}^{\deltav} ,\idiov_i)=   p(\tilde{\ym}_i|  \facm,\facload_{i\cdot}^{\deltav} ,\idiov_i)^{1-b}
           p(\tilde{\ym}_i|  \facm, \facload_{i\cdot}^{\deltav} ,\idiov_i)^b \propto  p(\tilde{\ym}_i|  \facm, \facload_{i\cdot}^{\deltav} ,\idiov_i)^{1-b}
              p(  \facload_{i\cdot}^{\deltav} | \idiov_i \addb ,\facm ) .
    \end{eqnarray*}
    The remaining part of  the likelihood, that is $ p(\tilde{\ym}_i|  \facm,\facload_{i\cdot}^{\deltav} ,\idiov_i)^{1-b} $, is used for model selection and is combined with the   prior  $p(\idiov_i)$  defined  in  (\ref{priorsiidg})  and the  {\em normalized} fractional prior  $ p(  \facload_{i\cdot}^{\deltav} | \idiov_i \addb ,\facm )$, given by:
  \begin{eqnarray*}
  p(  \facload_{i\cdot}^{\deltav} | \idiov_i \addb ,\facm )  =\frac{ p(\tilde{\ym}_i|  \facm, \facload_{i\cdot}^{\deltav} ,\idiov_i)^b}{c_i (\idiov_i, \facm ,b)}.
    \end{eqnarray*}
 The normalising constant $c_i (\idiov_i, \facm ,b)$ is   given by:
\begin{eqnarray} \label{normfrac}
  c_i (\idiov_i, \facm ,b)=  \int p(\tilde{\ym}_i| \facm, \facload_{i\cdot}^{\deltav} ,\idiov_i)^b  \, d \,  \facload_{i\cdot}^{\deltav} =
%\left(  \frac {1}{2\pi \idiov_i}\right)  }
( 2\pi \idiov_i)^{\frac{ q_i-Tb}{2}}
b^{-\frac{ q_i}{2}}  |\bV_{iT}^{\deltav}|^{1/2} \exp\left(  -  \frac {b}{2  \idiov_i} \SSR_i \right),
% \\ &   \SSR_i = \|    \tilde{\ym}_i -  \Xb_i ^{\deltav} \cv_{iT} ^{\deltav}  \|_2^2. &
 \end{eqnarray}
 where $\bV_{iT} ^{\deltav}$  and $\SSR_i $  are the posterior moments of
 $p(\facload^{\deltav}_{i\cdot},\idiov_i|  \facm, \deltav_{i\cdot},\tilde{\ym}_i )$
 given by (\ref{postmomfrac}). % and (\ref{SSRI}).
   Integrating the  fractional posterior
  \begin{eqnarray*}
 p(\tilde{\ym}_i|  \facm,\facload_{i\cdot}^{\deltav} ,\idiov_i)^{1-b}   p(  \facload_{i\cdot}^{\deltav} | \idiov_i \addb ,\facm ) p(\idiov_i)
  \end{eqnarray*}
   over $\facload_{i\cdot}^{\deltav}$, yields the fractional likelihood  $p(\tilde{\ym}_i|  \facm,\idiov_i,b)$:
 \begin{eqnarray*}
 \displaystyle p(\tilde{\ym}_i|  \facm,\idiov_i,b) &=&  \int p(\tilde{\ym}_i|  \facm,\facload_{i\cdot}^{\deltav} ,\idiov_i)^{1-b}   p(  \facload_{i\cdot}^{\deltav} |  \facm, \idiov_i \addb ,\facm )  \, d \,  \facload_{i\cdot}^{\deltav} = \frac{ 1}{ c_i (\idiov_i, \facm ,b) } \int p(\tilde{\ym}_i|  \facm,\facload_{i\cdot}^{\deltav} ,\idiov_i)   \, d \,  \facload_{i\cdot}^{\deltav} \\
 &=& \displaystyle  \left(  \frac {1}{2\pi \idiov_i}\right) ^{\frac{(T - q_i) -(Tb - q_i)}{2} }  b^{ \frac{q_i}{2}}  \frac{ |\bV_{iT}^{\deltav}|^{1/2}} { |\bV_{iT}^{\deltav}|^{1/2}}  \exp \left(  -  \frac {(1-b)}{2  \idiov_i} \SSR_i \right)  \\
&=& \displaystyle  \left(  \frac {1}{2\pi \idiov_i}\right) ^{\frac{T(1 - b)}{2} }  b^{\frac{q_i}{2}}  \exp\left(  -  \frac{(1-b)}{2  \idiov_i} \SSR_i \right).
 \end{eqnarray*}
   When we combine  $p(\tilde{\ym}_i|  \facm,\idiov_i,b)$  with  the prior $p(\idiov_i)$, then  we obtain:
\begin{eqnarray*}
% p(\idiov_i|  \facm,\tilde{\ym}_i ,b) \propto
p(\tilde{\ym}_i|  \facm, \idiov_i,b)  p(\idiov_i) =   \frac{C_{i0}^{c_0} }{\Gamfun{c_0}}
 \left( \frac {1}{2\pi } \right) ^{\frac {T(1 - b)}{2} }  b^{\frac{q_i}{2}} \left(  \frac {1}{\idiov_i}\right) ^{ \frac{c_0 + T(1 - b)}{2} }
 \exp\left(  -  \frac{ C_{i0} +   \SSR_i (1-b)/2 }{\idiov_i} \right),
 \end{eqnarray*}
which  is the kernel of  the  inverted Gamma distribution in (\ref{idiopos}). Integrating the right hand side with respect to $\idiov_i$ yields
the marginal likelihood  under the fractional prior: %  (\ref{frac_marlik}).
 \begin{eqnarray}
p(\tilde{\ym}_i|\deltav_{i,\cdot}, \facm)  =
 \frac {b^{q_i /2}\Gamma(c_T)(C_{i0})^{c_0}}{(2\pi)^{T(1-b)/2}\Gamma(c_0) (C_{iT}^{\deltav})^{c_T}}.
 \label{frac_marlik}
\end{eqnarray}

\subsubsection{Multimove sampling of a set of  indicators in a column}   \label{mcmcsmodi}

Another important building block of MCMC inference for sparse Bayesian factor models is sampling all
indicators $\delta_{ij}$ in column $\deltacol{j}$   for a set of  rows $i \in I_j  \subseteq  \{1, \ldots, m\}$, % =\{i_1, \ldots, i_{d_j}\}$,
conditional on  the factors $\facm=(\facm_1,\ldots,\facm_T)$, the remaining columns $\deltacol{-j}$  and
the hyperparameter $\tau_j$,  without conditioning on the model parameters $  \facload$ and $\idiov_1,\ldots,\idiov_{\dimy}$, see Step~(D) of Algorithm~\ref{Algo3}.

 According to the prior   (\ref{prigen}),  these indicators $\delta_{ij}$  are independent apriori
  conditional on the hyperparameter $\tau_j$, with the log prior odds $\oddpr_{ij}$ of  $\delta_{ij}=1$ versus
$\delta_{ij}=0$ being given by:
\begin{eqnarray} \label{samdelpr}
 \oddpr_{ij}= \log  \frac{\Prob{\delta_{ij}=1| \tau_j}}{\Prob{\delta_{ij}=0| \tau_j}} =  \log \frac{\tau_j}{1- \tau_j}.
\end{eqnarray}
Let $ \deltav_{i,-j}$ be all indicators in row $i$, except  $\delta_{ij}$.
    To sample $\delta_{ij}$ conditional $ \deltav_{i,-j}$ and $\facm$, without conditioning on   $  \facload$ and $(\idiov_1,\ldots,\idiov_{\dimy})$,
 the   log posterior odds $\oddpost_{ij}$, given by
    \begin{eqnarray}
\oddpost_{ij} &= &\log  \frac{\Prob{\delta_{ij}=1| \deltav_{i,-j}, \tau_j,  \tilde{\ym}_i, \facm}}
{\Prob{\delta_{ij}=0| \deltav_{i,-j},\tau_j, \tilde{\ym}_i, \facm}}  = \log  \frac{p(\tilde{\ym}_i|\delta_{ij}=1, \deltav_{i,-j} ,  \facm)}
{p(\tilde{\ym}_i|\delta_{ij}=0, \deltav_{i,-j},  \facm)} +
 \log  \frac{\Prob{\delta_{ij}=1| \tau_j}}{\Prob{\delta_{ij}=0| \tau_j}} \nonumber
 \\ &=&   \odd_{ij} + \oddpr_{ij},  \label{samdelmult}
\end{eqnarray}
  is required which combines  the log prior odds  $\oddpr_{ij}$ given in (\ref{samdelpr}) with the  log likelihood ratio $ \odd_{ij}$, given by:
     \begin{eqnarray}  \label{samplpost}
  \odd_{ij} = \log  \frac{p(\tilde{\ym}_i|\delta_{ij}=1, \deltav_{i,-j} ,  \facm)}
{p(\tilde{\ym}_i|\delta_{ij}=0, \deltav_{i,-j},  \facm)}.
\end{eqnarray}
 The  likelihood ratio  $ \odd_{ij}$  is easily  computed from  the
 marginal likelihoods $p(\tilde{\ym}_i|\delta_{ij}, \deltav_{i,-j},  \facm)$ where,  %  and $p(\tilde{\ym}_i|\delta_{ij}=0, \deltav_{i,-j},  \facm)$
respectively,  $\delta_{ij}=1$  and $\delta_{ij}=0$.  As discussed in  Subsection~\ref{marfac}, these marginal likelihoods are available in closed form  both for  the fractional as well as the standard prior and
marginal likelihood computation  can be done  individually for each row $i \in I_j$,  separately for  $\delta_{ij}=0$ and  $\delta_{ij}=1$.  However, this procedure is likely to be  inefficient, in particular, if  the set $ I_j$  is large.
To achieve greater efficiency,   Algorithm~\ref{AlgoInd} outlined below provides a technique to
  compute  directly the log likelihood  ratio $\odd_{ij}$ % in (\ref{samplpost})
   (rather than the individual marginal likelihoods) {\em simultaneously}  for all rows  $i \in I_j$.
  This allows  joint sampling of all indicators $\delta_{ij}$ in column $j$ for all  rows  $i \in I_j$.

The precise form of    the log  likelihood ratio $\odd_{ij}$ of  $\delta_{ij}=1$ versus  $\delta_{ij}=0$  defined  in (\ref{samplpost})
  depends on the remaining indicators  $\deltav_{i,-j}$ in row $i$.
  The computation of
 $\odd_{ij}$   is easily vectorized for all rows $i \in I_j$   where all elements of $\deltav_{i,-j}$ are zero. In this case,   a model  where observation $y_{it}$ is
 dedicated to factor $j$ ($\delta_{ij}=1$) is compared to a  model  where  $y_{it}$ is
 uncorrelated   with all remaining observations ($\delta_{ij}=0$). In this case,   $\odd_{ij}$ is easily
obtained from the marginal likelihood  of   a  dedicated model with $\ji=j$
 and the  \lq\lq null\rq\rq\ model. % given, respectively, by (\ref{mardedic}) and  (\ref{marnull}).
 As shown in Algorithm~\ref{AlgoInd}, it is   possible (but less straightforward)  to vectorize   the computation of the  log  likelihood ratio
  also for the remaining rows $i \in I_j$
  where at least one element of $\deltav_{i,-j}$ is different from
  zero.  \\  %Algorithm~\ref{Algoratio} describes the underlying  technique.\\

\begin{alg}{\textbf{Multimove sampling indicators in a column.}} \label{AlgoInd}
Sample all
indicators $\delta_{ij}$ in column $\deltacol{j}$  jointly for all rows $i \in I_j  \subseteq  \{1, \ldots, m\}$ % =\{i_1, \ldots, i_{d_j}\}$,
conditional on  the factors $\facm=(\facm_1,\ldots,\facm_T)$, the remaining indicators $\deltav_{i,-j}$  and the hyperparameter $\tau_j$,  without conditioning on the model parameters $  \facload$ and $\idiov_1,\ldots,\idiov_{\dimy}$ using the following steps:
\begin{itemize}
  \item[(I-a)]   Compute the log likelihood  ratio for  all rows $i \in I_j$  where  all elements of $\deltav_{i,-j}$ are zero  as
  \begin{eqnarray}  \label{saliklmult}
&& \odd_{ij} = \log  \frac{p(\tilde{\ym}_i|\delta_{ij}=1, \deltav_{i,-j},   \facm)}
{p(\tilde{\ym}_i|\delta_{ij}=0, \deltav_{i,-j},  \facm)}  =
\log  \frac{\Gamma(c_T) ( C_{iT} ^{\nullmod} )^{c_T ^{\nullmod}}}
{\Gamma(c_T^{\nullmod}) (C_{iT})^{c_T }}
  +  \Bodd_{ij}.
\end{eqnarray}
$c_T ^{\nullmod}$ and $C_{iT} ^{\nullmod}$ are the posterior moments of the null model  given in (\ref{idiozero}).
$c_T$ and $ C_{iT}$  are the  posterior moments  of $\sigma_i^2$ for a dedicated measurement with $\ji=j$, given in  (\ref{singloadfr})
and   (\ref{singloadfi}), respectively  for a fractional prior  and   the standard prior.
 For  a fractional prior, $  \Bodd_{ij} = 0.5 \log(b(2\pi)^{bT})$.
 For the standard prior,  $   \Bodd_{ij}  = 0.5  \log(B_{iT}/B_{i0,jj})$, where
  $B_{i0,jj}$ is $j$th diagonal element
 of the prior variance  $\bV_{i0}$ and  $B_{iT}$  is the posterior scale factor  for a dedicated measurement with $\ji=j$, given in  (\ref{singloadfi}).
This step is  trivial  to vectorize.

   \item[(I-b)] For all  rows   $i \in \{i_1,\ldots,i_n\} \subseteq I_j$  where  $\deltav_{i,-j}$ is not  zero,   compute
\begin{eqnarray}  \label{saliklmult2}
\odd_{ij}=  \log \frac{p(\tilde{\ym}_i |\delta_{ij}=1, \deltav_{i,-j},   \facm)}
{p(\tilde{\ym}_i |\delta_{ij}=0, \deltav_{i,-j},  \facm)}   =
c_T   \log  \frac{C_{iT} ^{0}} { C_{iT} ^1}   +  \Bodd_{ij},
\end{eqnarray}
where  $c_T$ and $C_{iT} ^{\delta_{ij}}$  are the posterior moments of $\sigma^2_i|\delta_{ij},\cdot  $
given in   (\ref{idiopos}) and % and (\ref{postmomfrac}), respectively, for a standard and a fractional prior.
$C_{iT}^0$ refers to  a model with $\delta_{ij}=0$,  while $C_{iT}^1$ refer to  a model
 with  $\delta_{ij}=1$.
  For a fractional prior,
   \begin{eqnarray}
    \Bodd_{ij} \equiv 0.5 \log b .
    \end{eqnarray}
  For a standard  prior,
  \begin{eqnarray}
   \Bodd_{ij} = 0.5  \log(|\bV_{iT}^1 |/|\bV_{iT}^0|)- 0.5\log ( |\bV_{i0} ^1|/|\bV_{i0}^0| ), \label{B38A}
   \end{eqnarray}
  where  $\bV_{i0}^{\delta_{ij}}$ and $\bV_{iT}^{\delta_{ij}}$ refer to the prior and posterior moments
  of  $ \facload_{i\cdot}^{\deltav} |  \delta_{ij}, \cdot $  given in  (\ref{postmomunit}).
  $\bV_{i0}^1$ and $\bV_{iT}^1$ refer to the prior and posterior moments for  a model
where   $\delta_{ij}=1$,
  while $\bV_{i0}^0$ and $\bV_{iT}^0$   refer to the prior and posterior moments for  a model
where $\delta_{ij}=0$.

Use Algorithm~\ref{Algoratio} to determine
 $C_{iT}^1$, $C_{iT}^0$,  as well as $\Bodd_{ij}$   for the standard prior,   simultaneously for all rows  $i \in \{i_1,\ldots,i_n\} \subseteq I_j$.

   \item[(I-c)]   Determine the vector  of the log posterior odds $\oddpost_{ij}=\odd_{ij} + \oddpr_{ij} $ for all rows $i \in I_j$.
   %\item[(I-d)]
    Joint sampling of   $\delta_{ij}| \tau_j,\cdot $  is easily vectorized:
       \begin{itemize}
         \item[(I-c1)] Propose  $\delta_{ij}\new=1-\delta_{ij}$  for   $i \in I_j$.
         \item[(I-c2)] Draw a   vector  of  $| I_j|$   random variables $U_i \sim \Uniform{0,1}$, indexed by   $i \in I_j$.
         \item[(I-c3)] For all rows $i \in I_j$, where $\delta_{ij}=0$, accept the proposal $\delta_{ij}\new=1$, iff  $\log  U_i \leq \oddpost_{ij}$;
         \item[(I-c4)] For all rows $i \in I_j$,  where $\delta_{ij}=1$, accept the proposal $\delta_{ij}\new=0$, iff  $\log  U_i  \leq - \oddpost_{ij}$.
         % otherwise stay with $\delta_{ij}\old$.
       \end{itemize}
       \end{itemize}
\end{alg}

\vspace*{1mm}  \noindent
 Using, respectively,  (\ref{ADDGAD}) and   (\ref{frac_marlik}),  the expression for $ \odd_{ij}$ in (\ref{saliklmult2}) is easily derived.
  Since the indicators   in column $j$ are  independent  given $\tau_j$, Step~(I-c)  is based  on $| I_j|$
 independent  Metropolis-Hastings (MH) steps  each of which proposes to update $\delta_{ij}$ by flipping  the   indicator,
 i.e. $\delta_{ij}\new=1-\delta_{ij}$.\footnote{Alternatively,  a Gibbs step  may be used, i.e.
 set $\delta_{ij} \new=1$, iff $\log (U_i/(1-U_i)) \leq \oddpost_{ij}$, otherwise  $\delta_{ij} \new=0$. However,  simulation experiments  indicate that the MH step is more efficient.}  It easy to verify
 that the acceptance rules formulated in Step~(I-c3) and  (I-c4)  are equivalent to the more convential
rule  to accept  $\delta_{ij}\new$ with probability
$$
 \min\left\{1, \frac{\Prob{\delta_{ij}\new| \deltav_{i,-j}, \tau_{j},  \tilde{\ym}_i , \facm}}
{\Prob{\delta_{ij} | \deltav_{i,-j},\tau_{j}, \tilde{\ym}_i , \facm}} \right\} = \min\left\{1,  \exp( \oddpost_{ij} )  \right\}.$$

\begin{alg} % {\textbf{Computing likelihood ratio of 1 versus 0.}}
 \label{Algoratio}
 To compute  all relevant posterior moments in (\ref{saliklmult2}) simultaneously for all rows $\{i_1,\ldots,i_n\}$,  proceed as follows:
\begin{itemize}
     \item[(a)] Set  the indicator $\delta_{i_l,j}=1$ in each row $i_l \in \{i_1,\ldots,i_n\}$. Reorder the columns of the factor loading matrix
 in such a way, that the $j$th column appears
last.  This is simply done by permuting the column of $\Fm$ appropriately before defining
 $\Xb_{i_l} ^{\deltav}$.\footnote{While the fractional prior is not affected by this, it might be necessary to reorder
 the prior mean and the prior  covariance matrix for the standard prior.}
\item[(b)]   Set up the information matrix $\Omegav$ and the covector $\cv$ of the corresponding joint
posterior of all nonzero factor loadings in the rows  $i_1,\ldots,i_n$ as described in Algorithm~\ref{algoP}.
Compute the Cholesky decomposition $\Lv$  of $\Omegav$   and the corresponding vector $\xm$  solving
$ \Lv  \xm= \cv$.

    \item[(c)] Knowing $ \Lv$ and  $\xm$,  a   vectorized computation of  the
log  likelihood ratio   (\ref{saliklmult})   for  all rows $i_l \in \{i_1,\ldots,i_n\}$  is possible.
    The posterior moments $C_{i_l,T} ^{1}$ are  directly available from the appropriate
sub vectors $\xm_{i_l}$ of $\xm$, defined in  (\ref{citfrac}).  When we switch from  $\delta_{i_l,j}=1$   to a model where $\delta_{i_l,j}=0$, then  for the fractional prior
  \begin{eqnarray} \label{Citposta}
  C_{i_l,T}^{0} =  C_{i_l,T} ^{1}  + \frac{1-b}{2 }  (x^{\star}_{i_l})^2,
\end{eqnarray}
where   $x^{\star}_{i_l}=(\xm_{i_l})_{q_{i_l}}$ is the last element of  $\xm_{i_l}$,
 while for the standard prior,
  \begin{eqnarray} \label{Citpostb}
  C_{i_l,T}^{0} =  C_{i_l,T}  ^{1} + \frac{1}{2 } (x^{\star}_{i_l})^2.
\end{eqnarray}
 Furthermore,
  \begin{eqnarray} \label{Citpostc}
  0.5 \log(|\bV_{i_l,T} ^{1} |/|\bV_{i_l,T}^0|)= -\log L^{\star}_{i_l},
\end{eqnarray}
 where $L^{\star}_{i_l}=(\Lv_i)_{q_{i_l},q_{i_l}}$ is the last diagonal element of the submatrix $\Lv_{i_l}$.
 Therefore,
$$D_{ij}=  -\log L^{\star}_{i_l} - 0.5 \log  B_{i0,jj}. $$
  \end{itemize}
\end{alg}
%\vspace*{1mm}

\noindent
{\em Derivation  of Step~(c).}  When we switch from a model  where all indicator $\delta_{i_1,j}= \ldots = \delta_{i_n,j}=1$ are equal to one
to a model where all   indicators $\delta_{i_1,j}= \ldots = \delta_{i_n,j}=0$ are zero, then the information matrix $\Omegav^0$
 and the covector $\cv^0$ of the   joint
posterior of  the remaining  nonzero factor loadings is obtained from $\Omegav$ and $\cv$ simply by deleting all
rows and columns corresponding to $\delta_{i_1,j}, \ldots, \delta_{i_n,j}$, and
the Cholesky decomposition $\Lv^0$ of  $\Omegav^0$ is  obtained from
 $\Lv$ in the same way. Also the  vector $\xm^0$
 solving
$ \Lv ^0 \xm^0 = \cv^0$   is obtained from $\xm$ simply by deleting the
rows corresponding to $\delta_{i_1,j}, \ldots, \delta_{i_n,j}$.  This last result is easily seen by considering the subsystem $\Lv_{i_l} \xm_{i_l} = \cv_{i_l,T} ^{\deltav}$ corresponding to the
$i_l$th row. Because
\begin{eqnarray} \label{deflio}
&& \Lv_{i_l}  = \left(\begin{array}{cc}
                      \Lv_{i_l} ^0 & \bfzmat  \\
                      \mathbf{l}_{i_l} & (\Lv_i)_{q_{i_l},q_{i_l}}  \\
                    \end{array} \right) =   \left(\begin{array}{cc}
                      \Lv_{i_l} ^0 & \bfzmat  \\
                      \mathbf{l}_{i_l} & L^{\star}_{i_l}  \\
                    \end{array} \right) , \qquad
\end{eqnarray}
we obtain $\Lv_{i_l} ^0 \xm_{i_l}^0= \cv^0_{i_l}$, where  $\xm_{i_l}^0$ is obtained from $\xm_{i_l}$ by deleting the  $q_{i_l}$th
element $x^{\star}_{i_l}=(\xm_{i_l})_{q_{i_l}}$.
 Hence, $\xm_{i_l}^0$ defines the desired subvector of  $\xm^0$  to compute  $C_{i_l,T}^{0}$     as in  (\ref{citfrac}).
 Since  $ \trans{(\xm_{{i_l}}^0)} \xm_{i_l}^0 = \trans{\xm_{i_l}} \xm_{i_l}- (\xm_{i_l})_{q_{i_l}}^2$  we obtain  from (\ref{idiopos}) that  (\ref{Citposta}) and (\ref{Citpostb}) hold. Note, however, that this simple relationship would not hold
 without reordering the columns as described above.

Finally, to compute the log   likelihood ratio for a standard prior,   the ratio of the determinants $|\bV^{1}_{i_l,T}|/|\bV_{i_l,T}^{0}|$ is required. Since
  the lower triangular matrices $\Lv_{i_l}$ and $\Lv_{i_l} ^0$ are, respectively, the Cholesky decomposition of $(\bV_{i_l,T}^{1})^{-1}$ and $(\bV_{i_l,T}^{0})^{-1}$,   we obtain:% the required determinants from the product of the diagonal elements of $\Lv_{i_l}$ and $\Lv_{i_l} ^0$:
 \begin{eqnarray} \label{detvchol}
&&  1/|\bV_{i_l,T} ^{1} |^{1/2} = |(\bV_{i_l,T}^{1} )^{-1} |^{1/2} = |\Lv_{i_l} |,
\end{eqnarray}
 where   $|\Lv_{i_l} |$  is   the product of the diagonal elements of  $\Lv_{i_l}$.  Computing $|\bV_{i_l,T}^{0}|$ in the same way and
    using   (\ref{deflio})  proves  (\ref{Citpostc}).

\subsection{Designing MCMC steps for econometric identification}   \label{mcmcide}

GLT structure are an example of a sparse confirmatory factor model, where a structure is imposed on the unknown indicator matrix apriori  in order to resolve rotational invariance up to trivial rotations.  The designer MCMC  scheme introduced in Algorithm~\ref{Algo3}   includes  a number of steps that are highly relevant to achieve identification for a GLT structure with an unknown number of factors.  This subsection  provides full details for these steps.

   \subsubsection{Forcing an  unordered GLT structure during MCMC sampling} \label{GLTMCMC}

   Assume that an indicator matrix $\deltav$ with $\dimy$ (not necessarily nonzero) rows and
   $\nfacr$ nonzero columns is given. Both Step~(L) and Step~(R)  as well as initialisation of Algorithm~\ref{Algo3}  discussed in Subsection~\ref{init}
   involve choosing
  a leading index $l_j$ in a particular column $j$ of   $\deltav$,  conditional on holding the leading indices $\lm_{-j}$ outside of column $j$ fixed.
 An obvious requirement is  that  $\lm_{-j}$ itself defines an unordered   GLT structure with  $\nfacr -1$ columns.

 The  leading index  $l_j$ cannot  be  chosen arbitrarily, % from  the row indices $\{1,\ldots,\leadsetn \}$,
  but  is constrained  to a subset  of $\{1,\ldots,\leadsetn \}$ that depends on $\lm_{-j}$.
 A  minimum requirement  is that  $l_j$ is different from the leading indices in  $\lm_{-j}$. This would lead to choosing
 $l_j$ from the set  $\{i:  1 \leq i\leq \dimy,  i \neq \lm_{-j} \}$.
 %\begin{eqnarray*} \label{LsetGLT}
 %\leadset{\leadsetn}{\lm_{-j}} :=\{i:  1 \leq i\leq \leadsetn,  i \neq \lm_{-j} \},
 % \end{eqnarray*}
%with cardinality $|\leadset{\leadsetn}{\lm_{-j}}| = \leadsetn - \nfacr + 1$.
%  $\leadset{\leadsetn}{\lm_{-j}}\subseteq \{1,\ldots,\leadsetn \}$ that depends on $\lm_{-j}$. % , whenever if $\nfacr > 1$.
  While Algorithm~\ref{Algo3}   could be based on this choice, for $\nfacr>1$  this  leads to indicator matrices $\deltav$ with leading indices $\lm=(l_j,\lm_{-j})$ that never can satisfy   the row deletion property \AR\  with  $\nfactrue=\nfacr$ or   the more general condition  \TS\  for a given $\tumS$, regardless of  what values are assigned to
 the remaining indicators, see Subsection~\ref{secGLT},

To avoid such indicator matrices $\deltav$,  the  stronger constraint is introduced that the leading indices $\lm=(l_j,\lm_{-j})$   satisfy condition  \GLTTS\   given in (\ref{condlj}) for a given $\tumS$ with $\nfactrue=\nfacr $:
 \begin{eqnarray} \label{LsetUnconstr}
  \leadset{\tumS}{\lm_{-j}}:=\{i:  1 \leq i\leq \dimy- \tumS -2, \lm= (i, \lm_{-j}) \mbox{ satisfies \GLTTS\  for $ \tumS$}  \}.
  \end{eqnarray}
 If we check \GLTAR\ for known  number of factors  $\nfactrue$, then $\tumS=0$. If we check   \GLTTS\ for a matrix with  $\nfacr$ nonzero columns, where  $\nfactrue$   is unknown, then  $S$ is the maximum degree of overfitting. Note that  $S$ reduces the number of available measurements for extended variance identification.

 It is possible to derive the elements of  $\leadset{\tumS}{\lm_{-j}}$ explicitly, as explained in the following.  A necessary condition for definition (\ref{LsetUnconstr})  is that
  $\lm_{-j}$ satisfies  \GLTTS\ with  $\nfactrue=\nfacr-1 $,  hence for every   $l_k \in \lm_{-j}$:
  \begin{eqnarray}
      %  l_k  \leq \leadsetn - 2 + 2(\nfacr-1-\rankl_k) = \leadsetn - 2 + 2(\nfacr-(\rankl_k +1)) ,    \label{uppldnew}
      \dimy- \tumS - 2  -    l_k  \geq   2(\nfacr-1-\rankl_k) = 2(\nfacr-(\rankl_k +1)) ,    \label{uppldnew}
\end{eqnarray}
where   $\rankl_k $  is the rank of $l_k $ in the ordered sequence
$l _{(1)}< \ldots < l _{(\nfacr-1)}$.
When adding $ l_j$, we have to ensure that $\lm= (i, \lm_{-j})$ %, including$ l_j$,
obeys condition \GLTTS\ with  $\nfactrue=\nfacr $, i.e.
 \begin{eqnarray}
 \dimy- \tumS - 2  -    l_{k}  \geq   2(\nfacr-\rankl\new_{k}).  \label{upplGLL}
\end{eqnarray}
If $ l_j  < l_k$, then  $\rankl\new _k =\rankl_k +1$ increases and    (\ref{uppldnew}) implies that (\ref{upplGLL})  holds.
However, if $ l_j  >  l_k$, then  $\rankl\new _k =\rankl_k $   and  condition  (\ref{upplGLL})
 %  $\leadsetn - 2  -    l_{k_0}  \geq   2(\nfacr-\rankl_{k_0})$
might be violated because the number of columns increases.
 This implies an upper limit $l_{\max}$ for  the position of $l_j$. If we determine the largest
leading index $l_{k_0}  \in \lm_{-j}$  for which   (\ref{upplGLL})
  holds without changing the rank (i.e.  $\rankl\new _{k_0} =\rankl_{k_0} $),
then the rank of   $l_j$  can be at most   $\rankl_{k_0}+1$, hence    $l_{\max}= \dimy- \tumS - 2(\nfacr- \rankl_{k_0} + 2)$.
% If this it satisfied for all indices, then $l_{\max}=\leadsetn - 2$.
 %
 The  elements of $ \leadset{\tumS}{\lm_{-j}}$ are then given by all rows
  between $\{1,\ldots, l_{\max}\}$  which are not occupied by any other leading index.
  This set has cardinality $|\leadset{\tumS}{\lm_{-j}}| = l_{\max} - \rankl_{k_0}$. %CHECK

  \subsubsection{Details on split and merge moves for overfitting models}  \label{RJdetails}

This subsection provides more details  concerning the  split and merge move implemented in  Step~(R) of Algorithm~\ref{Algo3}.
Let $\tumS$ be the maximum degree of overfitting and let $k$ be the  maximum number of factors.

\paragraph*{Proposing split or merge moves.}

Let $\nfactrue =\sum_{j=1}^k \indic{d_j > 1  }$
be  the current number of
\lq\lq active\rq\rq\  columns with at least two nonzero in the indicator matrix.
% If  $j_0=\nfacsp=0$, then no split/merge move is possible.
If  $\nfactrue=\nfac$, then no split/merge move is possible;
 otherwise,   a split or a merge move  that leaves $\nfactrue$ unchanged is selected.
 Let $j_0$  and  $\nfacsp$ %=\sum_{j=1}^k \indic{d_j = 0  }$  \lq\lq non-active\rq\rq\
be, respectively,   the current number of zero  and of spurious
columns, which are related through $j_0=\nfac - \nfactrue - \nfacsp$ and let
 $\nfac_0 = \min (j_0,\tumS - \nfacsp)= \min (\nfac - \nfactrue,\tumS) - \nfacsp$ be the maximum number of additional spurious columns that could be introduced.
 The probability $\psplit (\nfacsp)$ of a split move  %depends on $\nfacsp$ and $\nfac_0$  and
is zero for $\nfac_0=0$,  equal to a tuning parameter $p_0 \in (0,1)$ for $\nfacsp=0$, and  equal to a tuning parameter  $p_s \in (0,1) $ for  $\nfacsp>0$.
 Note that $p_0$  is the probability of introducing
a  spurious column for a loading matrix without spurious columns, while
 $p_s$  is the probability of introducing
a  spurious column, if one already exists.
%The split probability $p_s$ is typically a small number such as $p_s=0.01$.
The probability $\pmerge (\nfacsp)$ of a merge move is  zero for $\nfacsp=0$,  equal to one for $\nfac_0=0$, and  equal to $1-p_s$ for  $\nfac_0>0$.
Table~\ref{choosesm} expresses  $\psplit (\nfacsp)$  and  $\pmerge (\nfacsp)$  as a function of  $\nfacsp$ and $\min (\nfac - \nfactrue,\tumS)$.

\begin{table}[t!]
  \centering
   \caption{Probabilities $\psplit (\nfacsp)$  and  $\pmerge (\nfacsp)$   to propose, respectively, a split and  a merge move
   as a function of  $\nfacsp$ and $\min (\nfac - \nfactrue,\tumS)$.}\label{choosesm}
 \begin{tabular}{lcc}
    \hline
    &  $\psplit (\nfacsp)$ &  $\pmerge (\nfacsp)$ \\
    \hline
   $\nfacsp=0, \min (\nfac - \nfactrue,\tumS)=0$ & 0 & 0 \\
    $\nfacsp=0, \min (\nfac - \nfactrue,\tumS)>0$ & $p_0$ & 0 \\
    $\nfacsp = \min (\nfac - \nfactrue,\tumS)>0$  & 0 & 1 \\
  $0< \nfacsp < \min (\nfac - \nfactrue,\tumS)$  & $p_s$ & $1-p_s$ \\
     \hline
  \end{tabular}
\end{table}

\paragraph{Designing the split move.}
Let $\lm$  be the  leading indices of all  $\nfacr=\nfactrue+\nfacsp$ nonzero  columns. Updating is based on the assumption that the $\nfacr$ nonzero  columns of the current indicator matrix satisfy \GLTTS.   In  a split move,    one of the $j_0$  zero columns is chosen randomly and turned into a
spurious column.  With  $j$ being the corresponding column index,  a  leading index $l _j$  is selected randomly from the set
$\leadset{\tumS}{\lm}$ introduced in  Subsection~\ref{GLTMCMC}.
This guarantees that the proposed leading indices $ \lm \newsp = (l_j, \lm)$ satisfy condition   \GLTTS\   with  $\nfacr= \nfactrue+\nfacsp +1 $  and avoids proposing GLT structures that never can satisfy conditon \TS\ which is essential for identifying spurious columns.

% While $\delta_{l_j,j}=0$ and $\load _{l_j,j}=0$,
 The indicator $\delta\newsp_{l_j,j}=1$ is  the only nonzero element in  column $j$ and the  corresponding spurious factor loading  $\load \newsp_{l_j,j}$ is obtained by  splitting the variance $\idiov_{l_j}$ between    $(\idiov_{l_j}) \newsp$ and  $\load \newsp_{l_j,j}$ as explained in Subsection~\ref{RJMCMC}.
This  is achieved by sampling $U$
 from a distribution with support [-1,1] and defining:
 %Usually, $U$ is sampled from a uniform distribution. Alternative choices are discussed in Subsection~\ref{RJdetails}.} and defining:
\begin{eqnarray} \label{prorjitA}
\load \newsp_{l_j,j} = U \sqrt{\idiov_{l_j}}   , \qquad (\idiov_{l_j}) \newsp= (1-U^2) \idiov_{l_j} .
\end{eqnarray}
 Given $\load \newsp_{l_j,j}$ and  $(\idiov_{l_j}) \newsp$, new factors $\fac_{jt} \newsp$ are
proposed for the spurious column $j$, independently for  $t=1, \ldots,T$,  from the conditional density $ p(\fac_{jt}\newsp| \facm_{t,-j},\facload_{l_j,\cdot} \newsp, (\idiov_{l_j}) \newsp, y_{l_j,t})$ which takes a simple form, see (\ref{propconfsp}).
In addition,   a new hyperparameter $ \tau \newsp _j$ is sampled from $\tau \newsp _j | \deltav \newsp \sim \Betadis{a_0+1, b_0 + \dimy-1}$.

 % All other coefficients  remain unchainged.
  %

\paragraph{Designing the merge move.}

 The  merge move is  obtained by  reversing the split move. Let $j$ be one of the $\nfacsp $  spurious columns,
  with a single nonzero factor loading  $\load \newsp_{l_j,j}$  in row $l_j$.
   Deleting the spurious column   determines the  values of $\idiov_{l_j}$ and $U$  in the following way:
\begin{eqnarray*}
 \idiov_{l_j} =  (\load _{l_j,j}\newsp) ^2 + (\idiov_{l_j}) \newsp,\qquad
 U=\frac{\load _{l_j,j}\newsp }{\sqrt{(\load _{l_j,j}\newsp)^2 + (\idiov_{l_j}) \newsp}} ,
\end{eqnarray*}
while  $\load  _{l_j,j}=0$   and  $\delta_{l_j,j}=0$. % All other coefficients of  $\deltav$ and $\facload$ remain unchainged.
Since column $j$ is turned into a zero column,
  new factors $\fac_{jt} \sim \Normal{0,1}$ are  proposed  from the prior for  all $t=1, \ldots,T$
and  a new hyperparameter $\tau_j$ is sampled from $\tau_j | \deltav \sim \Betadis{a_0, b_0 + \dimy}$.

\paragraph*{Proposing  factors in a spurious column.}

Whenever a new spurious column $j$ is proposed,
   new factors $\facm_{j,\cdot} \newsp =
(\fac_{j1} \newsp, \ldots, \fac_{jT}\newsp) $ are proposed at the same time, while holding the factors $\facm_{t,-j}, t=1, \ldots,T,$ in all other columns  fixed.
 Draws of   $\fac \newsp_{jt}$  are available within our MCMC scheme,   however,  they were obtained from the prior $\fac_{jt}  \sim \Normal{0,1}$, as
 column $j$ was a zero column before splitting.   Since  $y_{l_j,t}$  is a measurement that contains information about $f_{jt} \newsp$ in a spurious column, its likelihood can be combined with the prior to define  the  conditional posterior density $ p(\fac_{jt}\newsp| \facm_{t,-j},\facload_{l_j,\cdot} \newsp, (\idiov_{l_j}) \newsp, y_{l_j,t})$ of $\fac_{jt}\newsp$ given $y_{l_j,t}$. This density is then used as a proposal for $f_{jt} \newsp$.

 It is easy to verify from the filter given in (\ref{filtPXsim}) that for a spurious column $j$ with leading element
 $\load _{l_j,j}\newsp$,  the conditional density $ p(\fac_{jt}\newsp| \facm_{t,-j},\facload_{l_j,\cdot} \newsp, (\idiov_{l_j}) \newsp, y_{l_j,t})$ of $\fac_{jt}\newsp$  is given by:
 \begin{eqnarray} \label{propconfsp}
&\fac_{jt} \newsp | y_{l_j,t}, \cdot  \sim \Normal{E_{jt} \newsp,V_{j} \newsp}, &\\
&\displaystyle V_{j} \newsp = \left(1+ \frac{(\load_{l_j,j}\newsp )^2 }{(\idiov_{l_j}) \newsp}\right)^{-1} =
\frac{(\idiov_{l_j}) \newsp}{(\idiov_{l_j}) \newsp + (\load_{l_j,j}\newsp )^2 }, \quad
 E_{jt} \newsp= \frac{V_{j}  \newsp \load_{l_j,j} \newsp }{(\idiov_{l_j}) \newsp} \tilde{y}_{l_j,t} =
 \frac{\load_{l_j,j} \newsp }{(\idiov_{l_j}) \newsp + (\load_{l_j,j}\newsp )^2 } \tilde{y}_{l_j,t} , &\nonumber
\end{eqnarray}
where  the pseudo outcome $\tilde{y}_{l_j,t}$ is given by
$\tilde{y}_{l_j,t}= y_{l_j,t}- \facload_{l_j,-j} \facm_{t,-j}$.
 %For a split move, we exploit (\ref{propconfsp})  to  propose  factors $\fac_{jt}\newsp$ from $\fac_{jt} | \facm_{-j,t}, \facload_{l_j,\cdot} \newsp ,(\idiov_{l_j})\newsp ,y_{l_j,t} \sim \Normal{E_{jt},V_j}$  for all $t=1, \ldots,T$,  in which case the prior/proposal ratio in (\ref{propfac}) is equal to 1.
  Using (\ref{prorjitA}), we obtain
 the  simple expressions for the posterior moments in (\ref{propconfsp})  in terms of $\idiov_{l_j}$ and $U$:
  \begin{eqnarray*}
 V_j  \newsp = \frac{(\idiov_{l_j})\newsp}{(\idiov_{l_j})\newsp + (\load_{l_j,j} \newsp)^2} = 1- U^2, \quad
 E_{jt}  \newsp= \frac{\load_{l_j,j} \newsp}{(\idiov_{l_j})\newsp + (\load_{l_j,j} \newsp)^2} \tilde{y}_{l_j,t} =
 \frac{U}{\sqrt{\idiov_{l_j}}} \tilde{y}_{l_j,t}  . \nonumber
\end{eqnarray*}

\paragraph*{Computing the acceptance ratio.}
 Suppose that the current indicator matrix  has $\nfacsp $ spurious columns and a split or a merge move has been used to change column $j$.
The acceptance probability   for  a split move reads $\min(1,A_{\mbox{\rm \footnotesize split}})$, where:
 \begin{eqnarray*}  \nonumber
A_{\mbox{\rm \footnotesize split}}=  \mbox{prior ratio} \times  \mbox{likelihood ratio}  \times \mbox{proposal ratio} \times \mbox{$|$Jacobian$|$}.
\end{eqnarray*}
Since split and merge moves are a reversible pair, this  also determines the acceptance rate  for a merge move.

The Jacobian of the transformation from $(\idiov_{l_j},U)$ to $((\idiov _{l_j})\newsp,\load \newsp_{l_j,j} )$  in
(\ref{prorjitA}) is surpsingly simple and is given by: % $ J=\sqrt{\idiov_{l_j}}$:
\begin{eqnarray*}
 | \mbox{Jacobian}|  =
 \left| \frac{\partial ((\idiov_{l_j}) \newsp,\load \newsp_{l_j,j})}{\partial (\idiov_{l_j}, U)} \right|
 = \left| \begin{array}{cc}
           1-U^2  & -2 \idiov_{l_j}\cdot U\\
           \frac{U}{2 \sqrt{\idiov_{l_j}}} & \sqrt{\idiov_{l_j}}
          \end{array}
 \right| = \sqrt{\idiov_{l_j}}.
\end{eqnarray*}
The proposal  ratio reads:
\begin{eqnarray} \label{proprat}
\mbox{proposal ratio}=  \frac{1}{g(U)}   \times
\frac{q_{\mbox{\rm \footnotesize merge}}(\deltav|\deltav \newsp)}
{q_{\mbox{\rm \footnotesize split}}(\deltav \newsp|\deltav)}  \times
 %\frac{q(\fac_{jt} |\delta_{l_j,j}=0 ) }{q(\fac_{jt} \newsp |\delta\newsp_{l_j,j}=1) }
  \prod_{t=1}^T \frac{p(\fac_{jt}) }{p(\fac_{jt}\newsp| \facm_{t,-j},\facload_{l_j,\cdot} \newsp, (\idiov_{l_j}) \newsp, y_{l_j,t})}
  \times  \frac{ p(\tau_{j} |\deltav)}{p(\tau_{j} \newsp |\deltav  \newsp ) },
 \end{eqnarray}
where $\fac_{jt} \newsp $  is  proposed  from
$q(\fac_{jt} \newsp|\delta\newsp_{l_j,j}=1)= p(\fac_{jt}\newsp| \facm_{t,-j},\facload_{l_j,\cdot} \newsp, (\idiov_{l_j}) \newsp, y_{l_j,t})$
and  $\tau \newsp_{j}$  is  proposed  from $ q(\tau_{j} \newsp |\delta\newsp_{l_j,j}=1) = p(\tau_{j} \newsp |\deltav  \newsp )$ in a split
 move. In the reverse merge move,   $\fac_{jt}$   is  proposed  from
$q(\fac_{jt} |\delta_{l_j,j}=0)=p(\fac_{jt})$   and $\tau_{j}$
 is  proposed  from   $ q(\tau_{j} |\delta_{l_j,j}=0) = p(\tau_{j} |\deltav)$.

  The  proposal  density for  $\deltav \newsp$ given $\deltav$ in a split move reads:
\begin{eqnarray*}
   q_{\mbox{\rm \footnotesize split}}(\deltav \newsp|\deltav)=
  \frac{\psplit (\nfacsp)}{ |\leadset{\tumS}{\lm}|   ( \min (\nfac - \nfactrue,\tumS) - \nfacsp)} ,
\end{eqnarray*}
where  % $\nfacsp$ is the number of  spurious columns {\em before}  splitting and
$|\leadset{\tumS}{\lm}| $ is the cardinality of   $\leadset{\tumS}{\lm}$  whereas
the proposal density for $\deltav$  given $\deltav \newsp$ in the reverse merge move simplifies to:
 %  \begin{eqnarray*} \label{prosplit}
 %q_{\mbox{\rm \footnotesize merge}}(\deltav | \deltav \newsp)=  \frac{\pmerge (\nfacsp)}{\nfacsp} ,
% \end{eqnarray*}
%where  $\nfacsp$ is the number of  spurious columns {\em before}  merging, \comment{OLD; NEW:}
  \begin{eqnarray*}
 q_{\mbox{\rm \footnotesize merge}}(\deltav | \deltav \newsp)=  \frac{\pmerge (\nfacsp+1)}{\nfacsp+1},
\end{eqnarray*}
see  Table~\ref{choosesm} for the definition  of $\psplit (\nfacsp)$  and  $\pmerge (\nfacsp+1)$.

 When deriving the likelihood  ratio and the  prior ratio, one has to keep in mind that split and merge  moves
 operate between the factor models  (\ref{spurious1})  and (\ref{spurious2}) discussed in Subsection~\ref{RJMCMC},
  conditional on  the entire indicator matrix $\deltav_{-(l_j,j)}$ except element  $\delta_{l_j,j}$,  all factor loadings $ \facload_{l_j,-j} ^{\deltav}$ except element $\load_{l_j,j}$, all idiosyncratic variances except $\idiov_{l_j}$,  and all factors $\facm_{t,-j}$ outside of column $j$, while  we marginalize over  $\tau_j$.
   Both the prior ratio  and the likelihood ratio have to be derived conditional on this information set.

Prior  (\ref{prialt}) is marginalized over  $\tau_j$ to determine the  prior ratio
   of $\delta \newsp _{l_j,j}=1$ versus $\delta_{l_j,j}=0$ without conditioning  on the hyperparameters, which are then proposed as described above.
   Hence, the prior  ratio reads:
\begin{eqnarray} \label{priorrat}
%  \mbox{prior ratio}=
  \mbox{prior ratio}_\beta \times   \mbox{prior ratio}_\sigma  \times
  \frac{\Prob{\delta \newsp _{l_j,j} =1| \deltav_{-(l_j,j)} =0}}{\Prob{\delta_{l_j,j} =0|  \deltav_{-(l_j,j)} =0}}
  \times \frac{ p(\tau_j \newsp| \delta \newsp _{l_j,j}=1, \deltav_{-(l_j,j)} =0) } {p(\tau_j| \delta_{l_j,j}=0, \deltav_{-(l_j,j)} =0 )}.
 \end{eqnarray}
The conditional priors   $p(\tau_j|\cdot )$ and $p(\tau \newsp_j |\cdot )$ in (\ref{priorrat})  cancel  against the  corresponding proposals in (\ref{proprat}).
The (marginalized) prior odds ratio is equal to:
   \begin{eqnarray*}
  \frac{\Prob{\delta \newsp _{l_j,j} =1| \deltav_{-(l_j,j)} =0}}{\Prob{\delta_{l_j,j} =0|  \deltav_{-(l_j,j)} =0}}
 = \frac{a_0}{b_0 + \dimy-1 }.
  \end{eqnarray*}
  Based on the inverted Gamma prior $p_{\footnotesize IG}(\idiov_{l_j})$ given by (\ref{priorsiidg}),  %the prior ratio
     $\mbox{prior ratio}_\sigma $ in (\ref{priorrat}) reads:
    \begin{eqnarray} \label{priorratsig}
\mbox{prior ratio}_\sigma=   \frac{p_{\footnotesize IG}((\idiov_{l_j})\newsp)}
 {p_{\footnotesize IG}(\idiov_{l_j})}.
 \end{eqnarray}
For a fractional prior (\ref{priorfrac}) , $\mbox{prior ratio}_\beta$  reads  %\footcomment{Work out details.}
\begin{eqnarray*} \label{priorrafrac}
 \mbox{prior ratio}_\beta=  \frac{p((\facload ^{\deltav}_{l_j,\cdot}) \newsp |(\idiov_{l_j})\newsp , \facm \newsp ,b)}
 {p(\facload_{l_j,\cdot} ^{\deltav}  |\idiov_{l_j}, \facm ,b) } =
  \frac{p(\tilde{\ym}_{l_j}| \deltav_{l_j,\cdot} \newsp, \facload_{l_j,\cdot} \newsp, (\idiov_{l_j}) \newsp, \facm \newsp) ^b c_{l_j} (\idiov_{l_j}, \facm ,b)}
  {p(\tilde{\ym}_{l_j}| \deltav_{l_j,\cdot}, \facload_{l_j,\cdot} , (\idiov_{l_j}) , \facm ) ^b c_{l_j} ((\idiov_{l_j}) \newsp, \facm \newsp ,b)}
 \end{eqnarray*}
and  involves the normalising constants of the fractional prior defined in  (\ref{normfrac}).
 %$p((\facload ^{\deltav}_{l_j,\cdot}) \newsp |(\idiov_{l_j})\newsp, \facm \newsp ,b)$ as well as  for $p(\facload_{l_j,\cdot} ^{\deltav}  |\idiov_{l_j}, \facm ,b)$:
   For the  standard prior  (\ref{prior1}), $\mbox{prior ratio}_\beta$  simplifies  to:
\begin{eqnarray} \label{priorratbet}
 \mbox{prior ratio}_\beta=
   p(\load \newsp_{l_j,j} | \facload_{l_j,-j} ^{\deltav}  ,(\idiov_{l_j})\newsp)  \times \frac{p(\facload_{l_j,-j}  ^{\deltav} |(\idiov_{l_j})\newsp )}
 {p(\facload_{l_j,-j}  ^{\deltav}  |\idiov_{l_j}) }.
\end{eqnarray}
Since split and merge  moves operate in the latent variable formulation of the factor model, and move  between the factor models  (\ref{spurious1})  and (\ref{spurious2}), the corresponding complete data likelihood ratio has to be considered which reads:
 \begin{eqnarray} \label{likratio}
 \mbox{likelihood ratio} =
% \prod_{t=1}^T
%  \frac{p(y_{l_j,t}| \deltav_{l_j,\cdot} \newsp, \facload_{l_j,\cdot} \newsp, (\idiov_{l_j}) \newsp, \facm \newsp) ^{(1-b^\star)} p(\fac_{jt}\newsp| \facm_{-j,t},\facload_{l_j,\cdot} \newsp, (\idiov_{l_j}) \newsp)}
%  {p(y_{l_j,t}| \deltav_{l_j,\cdot}, \facload_{l_j,\cdot} , (\idiov_{l_j}) , \facm ) ^{(1-b^\star)} p(\fac_{jt}| \facm_{-j,t})}
 \frac{p(\tilde{\ym}_{l_j}| \deltav_{l_j,\cdot} \newsp, \facload_{l_j,\cdot} \newsp, (\idiov_{l_j}) \newsp, \facm \newsp) ^{(1-b^\star)} }
 {p(\tilde{\ym}_{l_j}| \deltav_{l_j,\cdot}, \facload_{l_j,\cdot} , (\idiov_{l_j}) , \facm ) ^{(1-b^\star)}}
   \prod_{t=1}^T  \frac{p(\fac_{jt}\newsp| \facm_{t,-j},\facload_{l_j,\cdot} \newsp, (\idiov_{l_j}) \newsp)} { p(\fac_{jt}| \facm_{t,-j})},
\end{eqnarray}
where $b^\star=0$ for the standard prior. For the fractional prior, the missing fraction $b^\star=b$ appears in   $\mbox{prior ratio}_\beta$
% see  (\ref{priorrafrac}),
and  can be moved to the likelihood ratio, while changing $\mbox{prior ratio}_\beta$ to:
\begin{eqnarray} \label{priorfracnew}
 \mbox{prior ratio}_\beta=   \frac{ c_{l_j} (\idiov_{l_j}, \facm ,b)} {c_{l_j} ((\idiov_{l_j}) \newsp, \facm \newsp ,b)}.
 \end{eqnarray}
Hence, we can set   $b^\star=0$ in (\ref{likratio}) for both priors.  Although not obvious at first sight, the likelihood ratio  cancels against the  proposal ratio for the factors and therefore both terms drop from the acceptance rate. This can be verified by applying a well-known
identity to the denominator:\footnote{Apply following identity with $x=y_{l_j,t}$, $z=\fac_{jt}\newsp$,  and $\theta$ being the remaining parameters: \begin{eqnarray*}\frac{p(x|z,\theta) p(z|\theta) }{ p(x|\theta,y) }=p(x|\theta) \end{eqnarray*}}
\begin{eqnarray*} \label{priproonew}
&  \displaystyle \prod_{t=1}^T  \frac{p(y_{l_j,t}|\fac_{jt}\newsp, \delta _{l_j,j}=1,   \facload_{l_j,\cdot} \newsp, (\idiov_{l_j}) \newsp, \facm_{t,-j}, \deltav_{l_j,-j})  p(\fac_{jt}\newsp| \delta \newsp_{l_j,j}=1, \facload_{l_j,\cdot} \newsp, (\idiov_{l_j}) \newsp, \facm_{t,-j},\deltav_{-l_j,j} )}
  {p(y_{l_j,t}| \delta _{l_j,j}=0 , \facload_{l_j,\cdot} , (\idiov_{l_j}) , \facm_{t,-j} , \deltav_{l_j,-j})  p(\fac_{jt}| \delta _{l_j,-j}=0, \facm_{t,-j}, \deltav_{-l_j,j})}= &  \\
&   \displaystyle  \prod_{t=1}^T  \frac{p(y_{l_j,t}| \delta _{l_j,j}=1, \facload_{l_j,\cdot} \newsp, (\idiov_{l_j}) \newsp, \facm_{t,-j}, \deltav_{l_j,-j} ) p(\fac_{jt}\newsp| y_{l_j,t},  \delta \newsp_{l_j,j}=1, \facload_{l_j,\cdot} \newsp, (\idiov_{l_j}) \newsp, \facm_{t,-j},\deltav_{-l_j,j} ) }
{p(y_{l_j,t}| \delta _{l_j,j}=0 , \facload_{l_j,\cdot} ,   (\idiov_{l_j}) ,  \facm_{t,-j} , \deltav_{l_j,-j} ) p(\fac_{jt})}= &\\
& \displaystyle \prod_{t=1}^T  \frac{p(\fac_{jt}\newsp| y_{l_j,t},  \delta \newsp_{l_j,j}=1, \facload_{l_j,\cdot} \newsp, (\idiov_{l_j}) \newsp, \facm_{t,-j},\deltav_{-l_j,j} ) }{ p(\fac_{jt})},&
  \end{eqnarray*}
  since  the distribution $ p(y_{l_j,t}| \delta _{l_j,j}=0 , \facload_{l_j,\cdot} ,   (\idiov_{l_j}) , \facm_{t,-j} , \deltav_{l_j,-j})$,  obtained from
   (\ref{spurious1}),  is identical to the marginal distribution $  p(y_{l_j,t}| \delta _{l_j,j}=1 , \facload_{l_j,\cdot} ,   (\idiov_{l_j}) ,  \facm_{t,-j} , \deltav_{l_j,-j})$   which is  obtained from (\ref{spurious2})  after integrating over  $\fac_{jt}\newsp$   as discussed in Subsection~\ref{RJMCMC}.

Collecting all  terms together, a split move is accepted  with probability
   $\min(1,A_{\mbox{\rm \footnotesize split}}(\nfacsp))$, where: % $A_{\mbox{\rm \footnotesize split}} (\nfacsp)$ is given by:
 \begin{eqnarray}  \nonumber
A_{\mbox{\rm \footnotesize split}} (\nfacsp) &=& \frac{\sqrt{\idiov_{l_j}} \pmerge (\nfacsp+1) \cdot |\leadset{\tumS}{\lm}|  (\min (\nfac - \nfactrue,\tumS) - \nfacsp ) a_0}{g(U) (\nfacsp+1) \psplit (\nfacsp)(b_0 + \dimy-1)  }
 \times   \mbox{prior ratio}_\beta  \times   \mbox{prior ratio}_\sigma  ,
% \cdot \mbox{prior ratio}\cdot \frac{q_{\mbox{\rm \footnotesize merge}}(\deltav|\deltav \newsp)}{q_{\mbox{\rm \footnotesize split}}(\deltav \newsp|\deltav)}\\ \label{propA}&= & \frac{J}{g(U)}
% \cdot \mbox{prior ratio}\cdot \frac{ \pmerge (\nfacsp+1) \cdot |\leadset{\leadsetn}{\lm}|  (\nfac - \nfactrue -\nfacsp )  }{ (\nfacsp+1) \psplit (\nfacsp)  }.
\end{eqnarray}
where  $\lm$ are the leading indices in  $\deltav $ (before splitting).
On the other hand,   a merge move   is accepted  with probability
   $\min(1,A_{\mbox{\rm \footnotesize merge}}(\nfacsp))$,  where %$A_{\mbox{\rm \footnotesize merge}} (\nfacsp)$ is related to  $ A_{\mbox{\rm \footnotesize split}}(\nfacsp-1)$: %(\ref{propA}):
    \begin{eqnarray} \nonumber
A_{\mbox{\rm \footnotesize merge}}(\nfacsp) & = & \frac{1}{A_{\mbox{\rm \footnotesize split}}(\nfacsp-1)} ,
% \frac{g(U)}{J} \cdot  \frac{1}{\mbox{prior ratio}}   \cdot
% \frac{q_{\mbox{\rm \footnotesize split}}(\deltav \newsp|\deltav)}{q_{\mbox{\rm \footnotesize merge}}(\deltav|\deltav \newsp)}\\  \label{propB}
% & = &   \frac{g(U)}{J} \cdot  \frac{1}{\mbox{prior ratio}}  \cdot \frac{   \psplit (\nfacsp-1) \cdot  \nfacsp  }{| \leadset{\leadsetn}{\lm \newsp _{-j}} |  (\nfac - \nfactrue -\nfacsp+1 )  \pmerge (\nfacsp)},
\end{eqnarray}
and  $\lm = \lm \newsp_{-j}$ are the leading indices in all nonzero columns of  $\deltav $ , except column $j$ (before merging).

\paragraph*{Simplifications for the standard prior.}

For the standard prior, $\mbox{prior ratio}_\beta$  given in (\ref{priorratbet}) simplifies  to:%\footcomment{CAN BE SIMPLIFIED FURTHER}
\begin{eqnarray*} \label{priorrbchat}
 \mbox{prior ratio}_\beta= \frac{p_N(\load \newsp_{l_j,j}; 0 ,  A_j (\idiov_{l_j})\newsp )}{(1-U^2)^{q_{l_j}/2}}
 \exp\left( - \frac{  U^2  \|\facload_{l_j,-j}  ^{ \deltav}\|_2^2} {2  \idiov _{l_j} A_j  (1-U^2)} \right),
 %{2  \idiov _{l_j} (1-U^2)} \sum_{n=1}^\nfac  \indic{\delta_{l_j,n}=1} \frac{\load_{l_j,n}^2}{A_j}\right),
\end{eqnarray*}
where  $A_j=B_{l_j,0,jj}$   is $j$th diagonal element  of the prior covariance  matrix $\bV_{l_j,0}$
and  $q_{l_j}$ is the number of nonzero elements in  $\facload_{l_j,-j}  ^{\deltav}$, since   $\facload_{l_j,\cdot} ^{\deltav} =  \facload_{l_j,-j}^{\deltav}$.
 Furthermore, based on the inverted Gamma prior $p_{\footnotesize IG}(\idiov_{l_j})$ given by (\ref{priorsiidg}),  %the prior ratio
     $\mbox{prior ratio}_\sigma $ given in (\ref{priorratsig}) simplifies to:
    \begin{eqnarray*} \label{priorratscha}
\mbox{prior ratio}_\sigma=   \frac{1}{(1-U^2)^{c_0+1}}
 \exp\left\{ -\frac{C_{l_j,0} U^2}{\idiov_{l_j}(1-U^2)}  \right\}.
 \end{eqnarray*}

\paragraph*{Choosing a  proposal density for $U$.}

Given the fact that  the sign of $U$ is not relevant, it makes sense to choose proposals that are symmetric around 0. We investigated the following proposals for   $U$:
\begin{itemize}
  \item[(a)] $U \sim \Uniform{-1,1} $ follows a uniform distribution on $[-1,1]$.
  \item[(b)] $U=-1+2  Z$, where $Z \sim \Betadis{u_0,v_0}$, with  the corresponding density $g(u)= (1+ u)^{u_0-1} (1-u)^{v_0-1}/(2^{u_0+v_0-1} \Betafun{u_0,v_0})$. Choosing $u_0=v_0$ leads to a density that is symmetric around 0 and  $u_0=v_0=1$ leads to the  uniform distribution $U \sim \Uniform{-1,1} $.
  \item[(c)]  $U^2$ follows a uniform distribution on $[0,1]$, with  the corresponding density $g(u)= |u| $ defined on  $[-1,1]$.
     \item[(d)]  $U^2 \sim \Betadis{u_0,v_0} $ with  the corresponding density $g(u)= (u^2)^{u_0-1/2} (1-u^2)^{v_0-1}/\Betafun{u_0,v_0}$ defined on $[-1,1]$.
      Note that  $u_0=1/2,v_0=1$ leads to  the uniform distribution  $ \pm U \sim \Uniform{-1,1}$.
\end{itemize}
We found that proposal (c) and (d) led to higher acceptance rates than the other proposals and worked with
   $U^2 \sim \Betadis{3, 1.5}$  as proposal for our case studies  which implies a mode at $ \pm 0.9$ for the  proposal density  $g(u)$.

\subsubsection{Updating the leading indices in an unordered GLT structure}    \label{updatelead}

%\leadset{\leadsetn}{\lm_{-j}}

%$\leadset{\leadsetn}{\lm}$ with $\leadsetn=\dimy-\tumS$,

%%%%%%%%%%%%%%%%%%%%  {updatelead}

This subsection provides details on Step~(L) of Algorithm~\ref{Algo3}, which was shortly discussed in  Subsection~\ref{movelead}. All steps assume that the current indicator matrix $\deltav$ satisfies \GLTTS . Four local moves are applied which are illustrated  in Figure~\ref{figStepL}  in  Subsection~\ref{movelead}.

\paragraph{Shifting the leading index.}   A  shift move  is selected with probability $\pshift$.
Let   $l_\star$ denote the index of the first nonzero row below $l_j$, i.e. $\delta_{l_\star,j}= 1$, $\delta_{ij}= 0, l_j< i < l_\star$  (define  $l_\star:=\dimy+1$ for a  spurious column with  a single nonzero element).
If $l_\star>2$, then  it is proposed to move the leading index $l_j$ upwards or downwards,
% to any row that   is not occupied by any  other leading indices.
by  proposing  $l \new _j$ randomly from
the set ${\cal M} (  l_\star , \lm_{-j} )=\{1, \ldots, l_\star-1\} \cap \leadset{\tumS}{\lm_{-j}}$,
see Subsection~\ref{GLTMCMC} for a definition of   $\leadset{\tumS}{\lm_{-j}}$.  % and turned into the new leading index,
  Sampling $l_j \new$ from the set ${\cal M} (  l_\star , \lm_{-j} )$ guarantes that  \GLTTS\  holds for the new sequence of leading indices.  If  the set ${\cal M} (  l_\star , \lm_{-j} )$ is empty, then no shift  move is performed.
Otherwise, given $l_j \new$, two indicators in column $j$ are changed, namely
 $\delta_{l_j \new,j} $ from zero to one and  $\delta \new_{l_j ,j}$ from one to zero,  while the remaining elements of $\deltav$ are unchanged.
  %
%  This move is accepted  with probability   $\min(1,\alpha_{\mbox{\rm \footnotesize shift}})$, where $ \log \alpha_{\mbox{\rm \footnotesize shift}} = \odd_{l_j \new,j} - \odd_{l_j,j}$, with $\odd_{ij}$  being the log likelihood odd of $\delta_{ij}=1$ versus $\delta_{ij}=0$  defined in (\ref{samplpost}), see Subsection~\ref{updatelead} for further details.
%
 The new indicator matrix $\deltav \new $ is accepted
   with probability   $\min(1,\alpha_{\mbox{\rm \footnotesize shift}})$, where
\begin{eqnarray*} \label{propoGDdd}
\alpha_{\mbox{\rm \footnotesize shift}} =
\frac{\Prob{\delta_{l_j \new,j}=1, \delta_{l_j,j}= 0|\tilde{\ym}_i, \facm, \deltav_{l_j, -j},  \deltav_{{l_j \new}, -j} ,\tau_{j} } q(l  _j| l  \new _j,  l_\star, \lm_{-j}) }{\Prob{\delta_{l_j \new,j}=0, \delta_{l_j,j}= 1| \tilde{\ym}_i, \facm, \deltav_{l_j, -j},  \deltav_{{l_j \new}, -j} , \tau_{j}} q(l \new _j| l  _j,  l_\star, \lm_{-j}) }.
\end{eqnarray*}
Since $l \new _j$ is sampled from a set ${\cal M} (  l_\star , \lm_{-j} )$
that does not depend on $l_j$,
the proposal density is symmetric, i.e. $q(l \new _j|   l_\star, \lm_{-j})=
   q(l  _j|    l_\star, \lm_{-j})$, and  cancels from $\alpha_{\mbox{\rm \footnotesize shift}}$.
   Furthermore,   the prior ratio cancels, since the indicators in different rows are conditionally independent  given $\tau_{j}$:
      \begin{eqnarray*}
\frac{\Prob{\delta_{l_j \new,j}=1 |\tau_{j}  }}
{\Prob{\delta_{l_j \new,j}=0 | \tau_{j} }}
\frac{ \Prob{\delta_{l_j,j}= 0|\tau_{j}  }}
{\Prob{\delta_{l_j,j}= 1|\tau_{j} }}
= \frac{\tau_{j} (1-\tau_{j})}{(1-\tau_{j}) \tau_{j}} = 1.
\end{eqnarray*}
Therefore:
    \begin{eqnarray*}
\alpha_{\mbox{\rm \footnotesize shift}}
%\frac{\Prob{\delta_{l_j \new,j}=1 |\ym, \facm, \deltav_{l_j, -j},  \deltav_{{l_j \new}, -j} ,\tau_{j} } }
%{\Prob{\delta_{l_j \new,j}=0 | \ym, \facm, \deltav_{l_j, -j},  \deltav_{{l_j \new}, -j} , \tau_{j}} }
%\frac{ \Prob{\delta_{l_j,j}= 0|\ym, \facm, \deltav_{l_j, -j},  \deltav_{{l_j \new}, -j} ,\tau_{j} } }
%{\Prob{\delta_{l_j,j}= 1| \ym, \facm, \deltav_{l_j, -j},  \deltav_{{l_j \new}, -j} , \tau_{j}} }
%\frac{\Prob{\delta_{l_j \new,j}=1 |\cdot }}
%{\Prob{\delta_{l_j \new,j}=0 | \cdot}}
%\frac{ \Prob{\delta_{l_j,j}= 0|\cdot }}
%{\Prob{\delta_{l_j,j}= 1|\cdot }}
= \exp(\odd_{l_j \new,j} - \odd_{l_j,j}),
\end{eqnarray*}
with $\odd_{ij}$  being the log likelihood  ratio of $\delta_{ij}=1$ versus $\delta_{ij}=0$  defined in (\ref{samplpost}).
This move is a local move that does not change the number of nonzero elements $d_j$ in column $j$.

 \paragraph{Switching  leading indices.}

 This move is selected with probability $p_{\mbox{\rm \footnotesize switch}}$.
A nonzero column $\jtwo \neq j$ is selected randomly and  all indicators between (and including) row  $l_j$  and $l_ \jtwo$
%(or $\{l _j  , \ldots,  l_\jtwo\}$, if  $l _j >  l_\jtwo$)
that are different  are switched between the two columns, i.e. $\delta_{ij}\new=1-\delta_{ij}$
and $\delta_{i \jtwo}\new=1-\delta_{i \jtwo}$ for all $i \in  {\cal S}_{j,\jtwo} =\{i: \min (l _\jtwo ,  l_j) \leq i \leq  \max (l _\jtwo ,  l_j), \delta_{ij} \neq \delta_{i \jtwo}\}$.   %Since   $\delta_{l_j, j} \neq \delta_{l \jtwo,\jtwo}$,

 This move, which  is performed only if  $\nfacr>1$,    switches the leading elements between the two columns  and preserves condition  \GLTTS . Since the corresponding proposal density satisfies $q(\deltav \new|\deltav )=
   q(\deltav | \deltav \new)$,  $\deltav \new $  is accepted
   with probability   $\min(1,\alpha_{\mbox{\rm \footnotesize switch}})$, where
 %\begin{eqnarray} \label{propospdd}
%\alpha_{\mbox{\rm \footnotesize  switch}} =
%\prod _{i \in {\cal S}_{j,\jtwo}} \frac{\Prob{\delta_{ij}\new,\delta_{i\jtwo}\new | \deltav_{i,-(j,\jtwo)}, \tau_{j} ,\tau_{\jtwo},  \tilde{\ym}_i , \facm}}
%{\Prob{\delta_{ij} ,\delta_{i\jtwo}  | \deltav_{i,-(j,\jtwo)}, \tau_{j} ,\tau_{\jtwo}, \tilde{\ym}_i , \facm}} =
%\exp \left(  \sum_{i \in {\cal S}_{j,\jtwo}: \delta_{ij}=0 } ( \oddpost_{ij}  - \oddpost_{i\jtwo}) +   \sum_{i \in {\cal S}_{j,\jtwo}: \delta_{ij}=1 }   (\oddpost_{i\jtwo} - \oddpost_{ij}  ) \right),
%\end{eqnarray}
%and $\oddpost_{ij}$ is the log posterior odd given  in (\ref{samdelmult})  against a model where both indicators are 0;
% %, i.e. $\delta_{ij}=0$ and $ \delta_{i \jtwo}=0$:
 %\begin{eqnarray*}
 %\oddpost_{ij} = \log  \frac{\Prob{\delta_{ij}\new=1  | \delta_{i\jtwo}=0 ,  \cdot} }{\Prob{\delta_{ij} =0  | \delta_{i\jtwo}  =0  , \cdot}},
 %\quad\oddpost_{i\jtwo}= \log  \frac{\ \Prob{\delta_{i \jtwo}  \new=1  | \delta_{ij} =0 , \cdot}}{\Prob{ \delta_{i\jtwo}  =0| \delta_{ij} =0 ,\cdot}}.
%\end{eqnarray*}
\begin{eqnarray*}
\alpha_{\mbox{\rm \footnotesize  switch}} =
\prod _{i \in {\cal S}_{j,\jtwo}} \frac{p (\delta_{ij}\new,\delta_{i\jtwo}\new | \deltav_{i,-(j,\jtwo)}, \tau_{j} ,\tau_{\jtwo},  \tilde{\ym}_i , \facm)}
{p(\delta_{ij} ,\delta_{i\jtwo}  | \deltav_{i,-(j,\jtwo)}, \tau_{j} ,\tau_{\jtwo}, \tilde{\ym}_i , \facm)}.
\end{eqnarray*}
To simplify notation, subsequently, we omit the conditioning arguments. If $\delta_{ij}=0$ (and consequently $ \delta_{i \jtwo}=1$), then we obtain: %  the following log ratio $ \oddpost_{i,(j \leftrightarrow j_2)}$:
 \begin{eqnarray*}
 && \frac{\Prob{\delta_{ij}\new=1 ,\delta_{i \jtwo} \new=0  | \cdot}}
{\Prob{\delta_{ij} =0,\delta_{i \jtwo}  =1 | \cdot}} =
 \frac{\Prob{\delta_{ij}\new=1 ,\delta_{i \jtwo} \new=0  |  \cdot} \Prob{\delta_{ij} =0,\delta_{i \jtwo}  \new=0 |  \cdot}}
{\Prob{\delta_{ij} =0,\delta_{i \jtwo}  \new=0 | \cdot}  \Prob{\delta_{ij} =0,\delta_{i \jtwo} =1 | \cdot}} = \\
&&  \frac{\Prob{\delta_{ij}\new=1  | \delta_{i \jtwo}=0 ,  \cdot} \Prob{\delta_{i \jtwo}  \new=0  | \delta_{ij} =0 , \cdot}}
{\Prob{\delta_{ij} =0  | \delta_{i \jtwo}  =0  , \cdot}  \Prob{ \delta_{i \jtwo}  =1| \delta_{ij} =0 ,\cdot}} =
\exp( \oddpost_{ij|\jtwo}  - \oddpost_{i\jtwo|j}),
 \end{eqnarray*}
 where $\oddpost_{i,  j_1| j_2}$ is the log posterior odd  of $\delta_{i,j_1}=1$ versus $\delta_{i,j_1} =0$  provided that the indicator $ \delta_{i,j_2}=0$.
It can be obtained  as  the posterior  odd $\oddpost_{i , j_1 }$ given in (\ref{samdelmult}),  with  $ \delta_{i,j_2}=0$   for both models.
Therefore:
 \begin{eqnarray*}
 \oddpost_{ij|\jtwo} = \log  \frac{\Prob{\delta_{ij}=1  | \delta_{i\jtwo}=0 ,  \cdot} }{\Prob{\delta_{ij} =0  | \delta_{i\jtwo}  =0  , \cdot}},
 \quad
\oddpost_{i\jtwo|j}= \log  \frac{\ \Prob{\delta_{i \jtwo}  =1  | \delta_{ij} =0 , \cdot}}{\Prob{ \delta_{i\jtwo}  =0| \delta_{ij} =0 ,\cdot}}.
\end{eqnarray*}
 On the other hand, if $\delta_{ij}=1$ (and consequently $\delta_{i  \jtwo}=0$), then
 \begin{eqnarray*} \label{ratioj1j2}
&&  \frac{\Prob{\delta_{ij}\new=0 ,\delta_{i \jtwo} \new=1  | \cdot}}
{\Prob{\delta_{ij} =1,\delta_{i \jtwo}  =0 | \cdot}} =
 \frac{\Prob{\delta_{ij}\new=0 ,\delta_{i \jtwo} \new=1  |  \cdot} \Prob{\delta_{ij}\new =0,\delta_{i \jtwo}  =0 |  \cdot}}
{\Prob{\delta_{ij}  \new =0,\delta_{i \jtwo} =0 | \cdot}  \Prob{\delta_{ij} =1,\delta_{i \jtwo} =0 | \cdot}} = \\
&&  \frac{\Prob{\delta_{i \jtwo}  \new=1  | \delta_{ij} =0 , \cdot} \Prob{\delta_{ij}\new=0  | \delta_{i \jtwo}=0 ,  \cdot} }
{ \Prob{ \delta_{i \jtwo}  =0| \delta_{ij} =0 ,\cdot} \Prob{\delta_{ij} =1  | \delta_{i \jtwo}  =0  , \cdot} } =
\exp(  \oddpost_{i\jtwo|j} -  \oddpost_{ij| \jtwo}).
 \end{eqnarray*}
 Therefore
 \begin{eqnarray} \label{propospdd}
\alpha_{\mbox{\rm \footnotesize  switch}} =
\exp \left(  \sum_{i \in {\cal S}_{j,\jtwo}: \delta_{ij}=0 } ( \oddpost_{ij | \jtwo}  - \oddpost_{i\jtwo|j}) +   \sum_{i \in {\cal S}_{j,\jtwo}: \delta_{ij}=1 }   (\oddpost_{i\jtwo|j} - \oddpost_{ij| \jtwo}  ) \right).
\end{eqnarray}
Since the indicators   in column $j$ and $\jtwo$ are  independent  given $\tau_{j}$ and $\tau_{\jtwo}$,
  joint computation of  the log posterior odds $\oddpost_{ij}$ and $\oddpost_{i \jtwo}$  for all  rows $i \in {\cal S}_{j,\jtwo}$ is easily vectorized as in Subsection~\ref{mcmcsmodi}.

 This move allows  changes in $d_j$ and $d_\jtwo$, but leaves the overall number $d$ of nonzero elements unchanged.

 \paragraph{Adding or deleting a leading index.}
 Finally, a reversible pair of moves is selected  with probability $1- \pshift - p_{\mbox{\rm \footnotesize switch}}$. The add move
introduces  a new leading index  $l_j \new $ in a row
   above the current leading index $l_j$ which is not occupied by the leading indices
   of the other columns.
   Hence,  $l \new _j$ is selected randomly from
the set ${\cal A} (l_j,\lm_{-j} )=\{1, \ldots, l_j-1\} \cap \leadset{\tumS}{\lm_{-j}}$,  i.e. $\delta_{l_j \new,j} \new= 1$,  while the remaining elements of   $\deltav$ are unchanged (in particular    $\delta_{l_j,j} \new=\delta_{l_j,j}=1$). An add move is only possible, if   $|{\cal A} (l_j,\lm_{-j} )| >0$.\footnote{The number of  rows in ${\cal A} (l_j,\lm_{-j} )$ is equal to % $L_j^{\deltav}$, i.e. $L_j^{\deltav}=l_j- \sum_{j'=1}^r \indic{l_{j'} < l_j}$.
$l_j- \rankl_{j}$, where  $\rankl_{j}=\Count{l_{j'} \in \lm: l_{j'} \leq l_j}$  is the rank of $l_j$ among the leading indices.
Hence, an add move is possible, whenever  $l_j > \rankl_{j}$.} %\comment{CHECK!}}

The corresponding reverse move  is deterministic and  deletes the current leading index $l_j$,  making
$l_j \new=l_\star$  the new leading index where $l_\star$ is the row index of the first nonzero element in $\deltav$ below $l_j$. Hence,  $\delta_{l_j,j} \new=0$, while all other elements  of $\deltav$ remain unchanged.
 A delete move is only possible,  if   column $j$ is not spurious, $l_\star$ is not leading in any other column (that is if   % $ {\cal D}_j (\deltacol{j}) =
$ \{l_\star\} \in  \leadset{\tumS}{\lm_{-j}}$) and for $d_j=2$   the current number of spurious columns is smaller than $\tumS$. %$\nsp < \tumS$.

 If  for the current $\deltacol{j}$ neither an add nor a delete move is possible, then $l_j$ remains unchanged.
 Otherwise,   either an add or a delete move is selected with probability $p_{\mbox{\rm \footnotesize add}}(\deltav)$ and $1-p_{\mbox{\rm \footnotesize add}}(\deltav)$.
 If both add and delete moves are possible, then $p_{\mbox{\rm \footnotesize add}}(\deltav)=p_a$, with $p_a$ being a tuning parameter;
 if only an add move is possible,  then   $p_{\mbox{\rm \footnotesize add}}(\deltav)=1$, whereas
   $p_{\mbox{\rm \footnotesize add}}(\deltav)=0$, if only a delete move is possible.

   Note that whenever an add move is selected, the reverse delete move is always possible; and similarly,  the reverse add move is always possible,  whenever a delete move is selected.  The acceptance probability for an add  move is equal to $\min(1,\alpha_{\mbox{\rm \footnotesize add}})$,
with
 \begin{eqnarray*} \label{propadddd}
\alpha_{\mbox{\rm \footnotesize add}} =
\frac{\Prob{\delta_{l_j \new,j}=1| \cdot} q_{\mbox{\rm \footnotesize del}}(\deltav|\deltav \new)}
{\Prob{\delta_{l_j \new,j}=0| \cdot} q_{\mbox{\rm \footnotesize add}}(\deltav \new|\deltav)}
 =  \exp(\oddpost_{l_j \new,j})
 %\frac{(l_j- \rankl_{j}) (1-  p_{\mbox{\rm \footnotesize add}}(\deltav \new ) )}{p_{\mbox{\rm \footnotesize add}}(\deltav)},
 \frac{|{\cal A} (l_j,\lm_{-j} ) | (1-  p_{\mbox{\rm \footnotesize add}}(\deltav \new ) )}{p_{\mbox{\rm \footnotesize add}}(\deltav)},
\end{eqnarray*}
where $\oddpost_{ij}$ is the log posterior odd given  in  (\ref{samdelmult}).
    The acceptance probability for the  delete  move is equal to
    $\min(1,\alpha_{\mbox{\rm \footnotesize del}})$,
  with
 \begin{eqnarray*} \label{propospad}
\alpha_{\mbox{\rm \footnotesize del}} = \frac{\Prob{\delta_{l_j,j}=0| \cdot} %\deltav_{l_j,-j}, \tau_j,  \tilde{\ym}_{l_j}, \facm}
q_{\mbox{\rm \footnotesize add}}(\deltav|\deltav \new)}
{\Prob{\delta_{l_j,j}=1|\cdot }q_{\mbox{\rm \footnotesize del}}(\deltav \new|\deltav)}
% = \exp(-\oddpost_{l_j,j}) \frac{p_{\mbox{\rm \footnotesize add}}(\deltav\new)}{ (l_j\new - \rankl_{j}\new)
= \exp(-\oddpost_{l_j,j}) \frac{p_{\mbox{\rm \footnotesize add}}(\deltav\new)}{  |{\cal A} (l_j \new ,\lm_{-j} )| (1-p_{\mbox{\rm \footnotesize add}}(\deltav)) }.
\end{eqnarray*}
This move changes $d_j$ and increases or decreases the overall number of nonzero elements by one.

\paragraph{Tuning parameters.}
  These four moves involve three tuning probabilities, namely  $\pshift$, $p_{\mbox{\rm \footnotesize switch}}$, and $p_a$, with $1- \pshift - p_{\mbox{\rm \footnotesize switch}}>0$ and $0 < p_a <1$.

\subsubsection{Initialising  Algorithm~\ref{Algo3}}    \label{init}

To check the mixing of the MCMC chain,   two (or more) independent runs with different initial values are performed.
 First, an initial values for the number of nonzero columns $\nfacr$  of the factor loading  matrix  is chosen,
 by starting both  with small values  as well as with large  values close to the maximum number of factors  $\nfac$. Alternatively, a random initial value
   $\nfacr$  can be sampled uniformly from the set  $ \{1, \ldots,\nfac \}$ or
 as  $ \min(\max(\Poi{r_0},1),\nfac)$, where $\Poi{r_0}$ is a Poisson distribution with mean $r_0$.

Based on $\nfacr$, an initial value for the $\nfacr$ nonzero columns of the indicator matrix $\deltav$  has to be selected.
We use random initialization by  allowing factor loadings below the leading element to be zero with positive probability $p_0$, e.g. $p_0=0.5$.
 Initial  values for the leading indices   $l_1, \ldots,  l_{\nfacr}$ are chosen  by first sampling  $ l_1$ from $\{1, \ldots, u_1 \}$, where $u_1$ is a small number, e.g. 5. Then  for $j=2, \ldots, \nfacr$, we  sample  $ l_j  $  from the set $\leadset{\tumS_0}{\lm_{-j}}$ with  $\lm_{-j}=\{l_1, \ldots,  l_{j-1}\}$ and $\tumS_0=\max(p_0 \cdot \dimy, \tumS)$.\footnote{Alternatively, we could  start from a factor model  obeying the triangular constraint  $(l_1, \ldots,  l_{\nfacr})=(1,\ldots,\nfacr) $.}

 We draw at most 100  initial values $\deltav$ (including the leading indices) in this way, until a matrix  $\deltav$ is obtained where the nonzero columns satisfy condition \NC\ with $\nfactrue=\nfacr$. If no such indicator matrix is found, then we add enough nonzero elements in each nonzero column (e.g., by setting  $\delta_{l_j+1,j}=1,
 \ldots, \delta_{l_j+3,j}=1$) to ensure  variance identification for the initial value.

Given the initial value $\deltav$, an initial  value for $\hypv$ is obtained by sampling  $\tau_1, \ldots, \tau_\nfac$  as in Step~(H) of  Algorithm~\ref{Algo3}.
% To initialize  factor loadings and idiosyncratic variances,
 Finally, we perform a few (say 100) MCMC iterations in the confirmatory factor model corresponding to   $\deltav$, which is initialized  by sampling the factors $\facm_1,\ldots,\facm_T$ from the prior:  $\fac_{jt} \sim \Normal{0,1}$.
While holding   $\deltav$ fixed, we iterate between sampling  the model parameters  $\facload$ and $\idiov_1,\ldots,\idiov_{\dimy}$
as in Step~(P)  and  sampling  the factors $\facm_1,\ldots,\facm_T$  as in Step~(F) of  Algorithm~\ref{Algo3}.
The resulting model parameters  $\facload$ and $\idiov_1,\ldots,\idiov_{\dimy}$  serve as starting values for the full-blown MCMC scheme described in
Algorithm~\ref{Algo3}.

 \subsection{Details on boosting MCMC for sparse Bayesian factor models}  \label{accelerate_App}

This subsection provides further details for  boosting MCMC in Step~(A)  of Algorithm~\ref{Algo3}.
 Boosting is based on moving from  model  (\ref{fac1reg})  to  following  expanded model,
\begin{eqnarray}
&& \tilde{\facm}_t   \sim  \Normult{\nfac}{\bfz,\Psiv}, \label{fac2px} \\
 && \ym_t =  \tilde{\facload}   \tilde{\facm}_t + \errorm_t, \quad \errorm_t \sim \Normult{\dimy}{\bfz,\Vare}, \label{fac1px}
\end{eqnarray}
where $\Psiv=\Diag{\Psi_1,\ldots,\Psi_{\nfac}}$ is diagonal. The relation between the two systems is given by following transformation:
\begin{eqnarray}
% \tilde{\facm}_t = (\Psiv \old)^{1/2} \facm_t ,  \quad  \tilde{\facload} = \facload  (\Psiv \old)^{-1/2}. \label{fac5px}
 \tilde{\facm}_t = (\Psiv)^{1/2} \facm_t ,  \quad  \tilde{\facload} = \facload  (\Psiv)^{-1/2}. \label{fac5px}
\end{eqnarray}
  Algorithm~\ref{AlgoA} summarizes the boosting step.\\[1mm]

  \begin{alg}[\textbf{Implementing Step~(A)}] \label{AlgoA}
  Step~(A) in Algorithm~\ref{Algo3} is implemented in three steps:
  \begin{itemize}
   \item[(A-a)]    Choose a  (current) value $\Psiv $  %=\Diag{\Psi_1  \old, \ldots, \Psi_{\nfac}  \old }$,
and  move from system (\ref{fac1}) and (\ref{fac1reg}) to the expandend system (\ref{fac2px}) and (\ref{fac1px})  using
transformation (\ref{fac5px}).

\item[(A-b)] Sample a new value   $\Psiv \new$ in the expanded system  conditional on   $\tilde{\facm}_1, \ldots, \tilde{\facm}_T$ and   $\tilde{\facload}$ for all nonzero columns from the conditional posterior $p(\Psiv | \tilde{\facm}, \tilde{\facload} , \Vare) $ given below in (\ref{fac6px}).   Set  $\Psi_j =\Psi_j  \new=1$ for all zero columns.

\item[(A-c)] $\Psiv  \new$ % =\Diag{\Psi_1   \new, \ldots, \Psi_{\nfac}   \new}$
 is used to move from the expandend model (\ref{fac2px}) and (\ref{fac1px}) back to the original model (\ref{fac1}) and (\ref{fac1reg}),
  by means of the inverse transformation of (\ref{fac5px}).
This acceleration step   affects the  factor loadings $\facload$ and all factors $\facm_t$ in the following way for all nonzero columns $j$:
\begin{eqnarray*}
 \load_{ij}\new  = \load_{ij}  \sqrt{\Psi_j \new/\Psi_j }, % \sqrt{\frac{ \Psi_j \new}{\Psi_j } },
 \quad i=1,\ldots,\dimy, %\\
 \quad \fac_{jt}\new = \fac_{jt}  \sqrt{ \Psi_j /\Psi_j \new} % \sqrt{\frac{\Psi_j }{ \Psi_j \new} },
  \quad t=1,\ldots,T.
\end{eqnarray*}

 \end{itemize}
\end{alg}

\vspace*{2mm} \noindent
When determining the full conditional posterior  $\Psiv | \tilde{\facm}, \tilde{\facload} , \Vare $  in Step~(A-b)
in the expanded model,  it is important to account for the dependence of the prior scale of  $\tilde{\facload}$  on $\Psiv$  according to (\ref{fac5px}):
\begin{eqnarray}
 p(\Psiv | \tilde{\facm}, \tilde{\facload} , \Vare)
 % \propto p(\Psiv) p(\tilde{\facload} ^{\deltav} | \Psiv, \Vare ) p(\tilde{\facm}| \Psiv )
 \propto p(\Psiv) p(\tilde{\facload} ^{\deltav} | \Psiv, \Vare )
  \prod_{j: d_j>0}   \Psi_j ^{-T/2} \exp \left\{ - \frac{1}{2 \Psi_j} \sum_{t=1}^T \tilde{\fac}_{jt}^2 \right\} ,   \label{fac6px}
\end{eqnarray}
where $d_j=\sum_{i=1} ^\dimy \delta_{ij}$ is the number of nonzero elements in column $j$,

  The main difference between ASIS and MDA  lies in the choice of  the current value of $\Psiv$ in Step~(A-a), leading to
different priors  $p(\Psiv)$  and $ p(\tilde{\facload} ^{\deltav} | \Psiv, \Vare )$ in (\ref{fac6px}). As shown in
Subsections~\ref{ASISdat} and \ref{MDARAUG},  the conditional posterior  $p(\Psiv | \tilde{\facm}, \tilde{\facload} , \Vare) $  factors into independent  conditional posteriors for each  $\Psi_j $, arising from an inverted Gamma  distribution for the fractional prior and from a generalized inverse Gaussian distribution  for the standard prior, see Subsection~\ref{geninvgau} for details on the generalized inverse Gaussian distribution.

\subsubsection{ASIS} \label{ASISdat}

 In ASIS,  a nonzero factor loading $\load_{n_j,j}$  is chosen in each nonzero  column $j$, to define
the current value of  $\Psi_j$ as $\sqrt{\Psi_j}=\load_{n_j,j}$. This creates a factor loading matrix $\tilde{\facload}$ in the expanded system, where
 $\tilde{\load}_{n_j,j}=1$, whereas $\tilde{\load}_{i,j}= \load_{ij}/\load_{n_j,j}$ for $i\neq n_j$ in the nonzero columns.
 $\load_{n_j,j}$ can be chosen as the leading element in each columns, i.e. $n_j=l_j$, or
 such that $|\load_{n_j,j}|$ is maximized for all loadings in column $j$.  Apart from this choice, ASIS requires no further tuning.

For the fractional prior,   the  prior of $\Psi_j=  \load_{n_j,j}^2$ is given by  $p(\Psi_j) \propto  \Psi_j^{-1/2}$, whereas
\begin{eqnarray*}
  p(\tilde{\facload} ^{\deltav} | \Psiv, \Vare ) \propto
 \prod_{j: d_j>0} \Psi_j ^{(d_j-1)/2} ,
\end{eqnarray*}
%where $d_j=\sum_{i=1} ^\dimy \delta_{ij}$ is the number of nonzero elements in column $j$,
since  $p(\tilde{\load_{ij}} |\idiov_i, \Psi_j,\delta_{ij}=1) \propto  \Psi_j^{1/2}$
for all $i\neq n_j$ with $\delta_{ij}=1$. Hence,
 $\Psi_j |  \facm, \facload, \Vare $ defined in (\ref{fac6px}) reduces to an inverted Gamma distribution for all $j = 1, \ldots,k$ where
 $d_j>0$:
 \begin{eqnarray}
 \Psi_j |  \facm, \facload  \sim \Gammainv{\frac{T-d_j}{2}, \frac{\load_{n_j,j}^2}{2}  \sum_{t=1}^T \fac_{jt}^2 }. \label{fac7pxfrac}
\end{eqnarray}
 Under the standard prior,
 where the unrestricted loadings follow the prior $\load_{ij} | \idiov_i,\delta_{ij}=1  \sim \Normal{0, B_{i0,jj} \idiov_i}$,
 where   $B_{i0,jj}$ is $j$th diagonal element
 of  $\bV_{i0}$,
we obtain the prior $\Psi_j=  \load_{n_j,j}^2 \sim \Gammad{1/2,1/(2B_{i0,jj}\idiov_{n_j})}$,
whereas $\tilde{\load_{ij}} |\idiov_i, \Psi_j,\delta_{ij}=1
 \sim \Normal{0, B_{i0,jj} \idiov_{i}/\Psi_j}$, for $i\neq n_j$. Hence,
\begin{eqnarray*}
  p(\tilde{\facload} ^{\deltav} | \Psiv, \Vare ) \propto
 \prod_{j: d_j>0} \Psi_j ^{(d_j-1)/2}  \exp\left\{ - \frac{\Psi_j B_{i0,jj}^{-1} }{2} \sum_{i: \delta_{ij} =1, i\neq  n_j} \frac{\tilde{\load_{ij}}^2}{\idiov_i } \right\}.
 \end{eqnarray*}
% where $d_j=\sum_{i=1} ^\dimy \delta_{ij}$ is the number of nonzero elements in column $j$.
Hence,  $\Psi_j |  \facm, \facload, \Vare  $  defined in (\ref{fac6px}) follows  a generalized inverse Gaussian posterior   for all $j=1, \ldots, \nfac$ where $d_j>0$:
 \begin{eqnarray}
 \Psi_j |  \facm, \facload, \Vare   \sim \GIG{ \frac{d_j-T}{2},  \frac{B_{i0,jj}^{-1}}{\load_{n_j,j}^2}  \left( % \frac{1}{\idiov_{n_j}} +
 \sum_{i: \delta_{ij} =1}  \frac{\load_{ij}^2}{\idiov_i} \right),
 \load_{n_j,j}^2  \sum_{t=1}^T \fac_{jt}^2}. \label{posasisx}
\end{eqnarray}
 For spurious columns $j$ (i.e. $d_j=1$),  the second  parameter is equal to $B_{i0,jj}^{-1}/\idiov_{n_j}$, where $n_j=l_j$ is equal to the leading index.

\subsubsection{Marginal data augmentation} \label{MDARAUG}

In marginal data augmentation,  the current value $\Psiv \sim p(\Psiv) $  is drawn from a working prior $p(\Psiv)$ which is independent  both  of $\facload$ and $\Vare$.

For the fractional prior,  $p(\tilde{\facload} ^{\deltav} | \Psiv, \Vare ) $
 takes the following form,
\begin{eqnarray*}
  p(\tilde{\facload} ^{\deltav} | \Psiv, \Vare ) \propto
 \prod_{j: d_j>0} \Psi_j ^{d_j/2} ,
\end{eqnarray*}
% where $d_j=\sum_{i=1} ^\dimy \delta_{ij}$ is the number of nonzero elements in column $j$,
since  $p(\tilde{\load_{ij}} |\idiov_i, \Psi_j,\delta_{ij}=1) \propto  \Psi_j^{1/2}$
for all $i$ with $\delta_{ij}=1$.   The two last terms  in (\ref{fac6px}) factor  into a product of   independent inverted Gamma distributions.
Hence, for a fractional prior,
 the inverted Gamma working prior $\Psi_j  \sim \Gammainv{\nu_j,q_j}$ is  applied, which leads to an inverted Gamma posterior for each $\Psi_j $  given by:
  \begin{eqnarray}
 \Psi_j | \Psi_j \old, \facm, \facload, \Vare   \sim \Gammainv{\nu_j - d_j/2 + \frac{T}{2}, q_j +  \Psi_j \old /2   \sum_{t=1}^T \fac_{tj}^2}. \label{postpinv}
\end{eqnarray}
Under the standard prior, where   the unrestricted loadings  follow the prior $\load_{ij} | \idiov_i,\delta_{ij}=1  \sim \Normal{0, B_{i0,jj} \idiov_i}$,
 where   $B_{i0,jj}$ is $j$th diagonal element
 of  $\bV_{i0}$,   we obtain  $\tilde{\load_{ij}} |\idiov_i, \Psi_j, \delta_{ij}=1
 \sim \Normal{0, B_{i0,jj} \idiov_i/\Psi_j}$, hence
\begin{eqnarray*}
 p(\tilde{\facload} ^{\deltav} | \Psiv, \Vare ) \propto % \prod_{j=1}^{\nfac} %
 \prod_{j: d_j>0} \Psi_j ^{d_j/2}  \exp\left\{ - \frac{\Psi_j B_{i0,jj}^{-1} }{2} \sum_{i: \delta_{ij} =1} \frac{\tilde{\load_{ij}}^2}{\idiov_i } \right\}.
\end{eqnarray*}
 Since $ p(\tilde{\facload} ^{\deltav}| \Psiv, \Vare)$   is proportional to the kernel of a gamma density for each $\Psi_j$,
the two last terms  in (\ref{fac6px}) factor  into a product of   independent generalized inverse Gaussian distributions for each $\Psi_j $.
% (see Subsection~\ref{geninvgau} for details on the generalized inverse Gaussian distribution).
Different working priors $p(\Psi_j)$  are conditionally conjugate priors in (\ref{fac6px}), including the inverted gamma prior $\Psi_j  \sim \Gammainv{\nu_j,q_j}$ and the Gamma prior $\Psi_j  \sim \Gammad{\nu_j,q_j}$. However, we obtained the most
stable results with the  generalized inverse Gaussian working prior  $\Psi_j  \sim \GIG{p_j,a_j,b_j}$
which leads  to a generalized inverse Gaussian posterior for $\Psi_j | \Psi_j \old, \facm, \facload, \Vare  $ for all $j=1, \ldots, \nfac$:
 \begin{eqnarray}
 \Psi_j | \Psi_j \old, \facm, \facload, \Vare   \sim \GIG{p_j+ d_j/2 - \frac{T}{2}, a_j + \frac{B_{i0,jj}^{-1}}{\Psi_j  \old}  \sum_{i: \delta_{ij} =1} \frac{\load_{ij}^2}{\idiov_i},
 b_j +  \Psi_j  \old \sum_{t=1}^T \fac_{jt}^2}. \label{postpsipx}
\end{eqnarray}
The inverted  Gamma and the Gamma working prior result as those special cases where, respectively,
 $p_j=-\nu_j, a_j=0, b_j=2q_j$  and $p_j=\nu_j, a_j=2q_j, b_j=0$.
 %\footcomment{Previous choices of the working prior:  in version July 2010, a Gamma prior was used with  $\nu_j=q_j=?$; in all versions since 2011,  an inverted Gamma prior was used with $\nu_j=q_j=1$ before Oct  2014  and $\nu_j=q_j=1.5 $  after Nov  2014. However, none of these previous versions took the dependence of the prior scale of  $\tilde{\facload}$  on $\Psiv$ into account. This error was detected and corrected in October 2015.}
  In particular for columns with few positive loadings, choosing  $a_j>0$ stabilizes the algorithm. Recommended choices for the parameters of the GIG-working prior are  $p_j=p_\psi$, and $a_j=b_j=a_\psi$ with $p_\psi=1.5$ and $a_\psi=2$ (or 3).\footnote{In this case, the expected value of each $\psi_j$ is given by $\Ew{\psi_j}=K_{p_\psi+1}(a_\psi) /K_{p_\psi}(a_\psi)$, while the variance is equal to $\V{\psi_j}=K_{p_\psi+2}(a_\psi) /K_{p_\psi}(a_\psi) - \Ew{\psi_j}^2$, see Subsection~\ref{geninvgau}.}

 Note that a GIG-working prior  could also be applied for a fractional prior, which leads  a GIG-posterior as  in  (\ref{postpsipx}) with
 the second parameter  being equal to $a_j$ and independent of any actual information. However, we could not find any gain in using such an extended
  working prior.

 %\subsubsection{Example illustrating efficiency gain}

\subsubsection{The Generalized Inverse Gaussian  Distribution} \label{geninvgau}

The inverse Gaussian distribution, $\rvY \sim \GIG{p,a,b}$ is a three-parameter family of probability distribution with support  $y \in \Real^+$. The  density is given by
\begin{eqnarray*}
\displaystyle  f(y) =
 \frac{(a/b)^{p/2}}{2 K_p(\sqrt{ab})} y^{p-1} e^{-(a/2)y} e^{ -b/(2y)},
\end{eqnarray*}
where $K_p(z)$ is the modified Bessel function of the second kind, $a > 0$, $b > 0$ and $p$ is a real parameter.
The first two moments are given by:
\begin{eqnarray*}
 &&	\Ew{\rvY}= \left(\frac{b}{a}\right) ^{1/2} \frac{K_{p+1}(\sqrt{a b}) }{ K_{p}(\sqrt{a b})},\\
&&  \V{\rvY}= \left(\frac{b}{a}\right)\left[\frac{K_{p+2}(\sqrt{ab})}{K_p(\sqrt{ab})}-\left(\frac{K_{p+1}(\sqrt{ab})}{K_p(\sqrt{ab})}\right)^2\right].
 \end{eqnarray*}

%\clearpage

%\newpage

%\section{More details on the applications}   \label{appApp}

%\subsection{Exchange rate data}   \label{appApp:ex22}

%\subsection{Returns from NYSE100}   \label{appApp:NasDaq}

\end{document}